\documentclass[11pt]{article}
\usepackage[body={6.5in,9in}]{geometry}
\usepackage{graphics,graphicx,subfigure}
\usepackage{color}
\usepackage{amssymb}
\usepackage[round]{natbib}

\newtheorem{thm}{Theorem}
\newtheorem{prop}{Proposition}
\newtheorem{cor}{Corollary}
\newtheorem{exmp}{Example}
\newtheorem{rem}{Remark}
\newtheorem{lem}{Lemma}

\newcommand{\B}{\mathcal{B}}
\newcommand{\M}{\mathcal{M}}
\newcommand{\F}{\mathcal{F}}
\newcommand{\FP}{\mathcal{P}}

\newcommand{\ben}{\begin{enumerate}}
\newcommand{\een}{\end{enumerate}}
\newcommand{\bit}{\begin{itemize}}
\newcommand{\eit}{\end{itemize}}
\newcommand{\be}{\begin{equation}}
\newcommand{\ee}{\end{equation}}
\newcommand{\bdm}{\begin{displaymath}}
\newcommand{\edm}{\end{displaymath}}
\newcommand{\bea}{\begin{eqnarray}}
\newcommand{\eea}{\end{eqnarray}}
\newcommand{\f}[1]{\fbox}

\newcommand{\realnos}{\mbox{{\bf R}}}

\newcommand{\naturalnos}{\mbox{{\bf N}}}

\newcommand{\dfrac}[2]{\displaystyle{\frac{#1}{#2}}}

\begin{document}

\makeatletter      
\renewcommand{\ps@plain}{%
     \renewcommand{\@oddhead}{\textrm{}\hfil\textrm{\thepage}}%
     \renewcommand{\@evenhead}{\@oddhead}%
     \renewcommand{\@oddfoot}{}
     \renewcommand{\@evenfoot}{\@oddfoot}}
\makeatother     

\title{Deterministic Brownian Motion:\\The Effects of Perturbing a Dynamical System by a Chaotic
Semi-Dynamical System}
\author{Michael C. Mackey\thanks{e-mail: {\tt mackey@cnd.mcgill.ca},
Departments of Physiology, Physics \& Mathematics and Centre for
Nonlinear Dynamics, McGill University, 3655 Promenade Sir William
Osler, Montreal, QC, CANADA, H3G 1Y6} and Marta Tyran-Kami\'nska
\thanks{Corresponding author, email: {\tt mtyran@us.edu.pl}, Institute of Mathematics,
Silesian University, ul. Bankowa 14, 40-007 Katowice, POLAND} }
\date{\today}
\maketitle

%
%
\begin{abstract}{Here we review and extend central limit theorems for highly
chaotic but deterministic semi-dynamical discrete time systems. We
then apply these results  show how Brownian motion-like results
are recovered, and how an Ornstein-Uhlenbeck process results
within a totally deterministic framework. These results
illustrate that the contamination of experimental data by ``noise"
may, under certain circumstances, be alternately interpreted as
the signature of an underlying chaotic process.}
\end{abstract}

\tableofcontents

\section{Introduction}\label{intro}


Almost anyone who has ever looked through a microscope  at a drop of
water has been intrigued by the seemingly erratic and unpredictable
movement of small particles suspended in the water, {\it e.g.} dust
or pollen particles. This phenomena, noticed shortly after the
invention of the microscope by many individuals, now carries the
name of ``Brownian motion" after the English botanist Robert Brown
who wrote about his observations in 1828. Almost three-quarters of a
century later, \citet{einstein05} gave a theoretical (and
essentially molecular) treatment of this macroscopic motion  that
predicted the phenomenology of Brownian motion.  (A very nice
English translation of this, and other, work of Einstein's on
Brownian motion  can be found in \citet{furth}.) The contribution of
Einstein led to the development of much of the field of stochastic
processes, and  to the notion that Brownian movement is due to the
summated effect of a very large number of tiny impulsive forces
delivered to the macroscopic particle being observed.  This was also
one of the most definitive arguments of the time for an atomistic
picture of the microscopic world.

\begin{figure}
\begin{center}  
\includegraphics[width=6in]{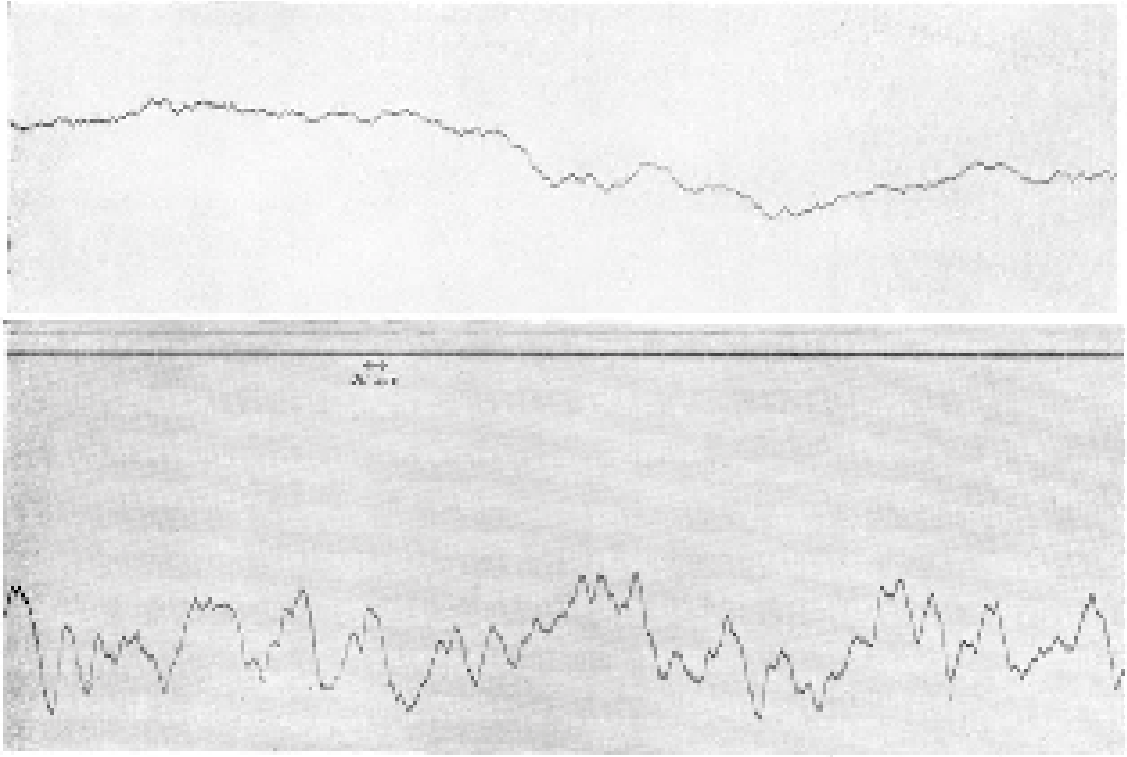}\\
  \end{center}
  \caption{In the upper panel is shown a recording of the movement of the mirror in the \citet{kappler} experiment over a period of about 30
  minutes at atmospheric pressure ($760$ mm Hg).
  The bottom panel shows the same experiment at a pressure of $4 \times 10^{-3}$ mm
  Hg. Both figures are from \citet{kappler}. See the text for more detail.}
  \label{kappler}
\end{figure}

Other ingenious experimentalists used  this conceptual idea to
explore the macroscopic effects of microscopic influences.  One of
the more interesting is due to \citet{kappler}, who devised an
experiment in which a small mirror was suspended by a quartz fiber
(c.f \citet{mazo} for an analysis of this experimental setup). Any
rotational movement of the mirror would tend to be counterbalanced
by a restoring torsional force due to the quartz fiber. The position
of the mirror was monitored by shining a light on it and recording
the reflected image some distance away (so  small changes in the
rotational position of the mirror were magnified). Air molecules
striking the mirror caused a transient deflection that could thus be
monitored, and the frequency of these collisions was controlled by
changing the air pressure. Figure \ref{kappler}, taken from
\citet{kappler}, shows two sets of data taken using this arrangement
and offers a vivid depiction of the macroscopic effects of
microscopic influences.

In trying to understand theoretically the  basis for  complicated
and irreversible experimental observations, a number of physicists
have supplemented the reversible laws of physics with various
hypotheses about the irregularity of the physical world.  One of
the first of these, and arguably one of the most well known, is
the so-called ``molecular chaos" hypothesis of \cite{boltzmann96}.
This hypothesis, which postulated a lack of correlation between
the movement of molecules in a small collision volume, allowed the
derivation of the Boltzmann equation from the Liouville equation
and led to the celebrated H theorem.  The origin of the loss of
correlations was never specified. In an effort to understand the
nature of turbulent flow, \citet{ruelle78,ruelle79,ruelle80}
postulated a type of mixing dynamics to be necessary. More
recently, several authors have made {\it chaotic hypotheses} about
the  nature of dynamics at the microscopic level. The most
prominent of these is \citet{gallavotti}, and virtually the entire
book of \citet{dorfman99} is predicated on the implicit assumption
that microscopic dynamics have a chaotic (loosely defined, but
usually taken to be mixing) nature.  All of these hypotheses have
been made in spite of the fact that none of the microscopic
dynamics that we write down in physics actually display such
properties.

\begin{figure}
\begin{center}  
\includegraphics[width=0.9\linewidth,bb=297 259 549 441,clip]{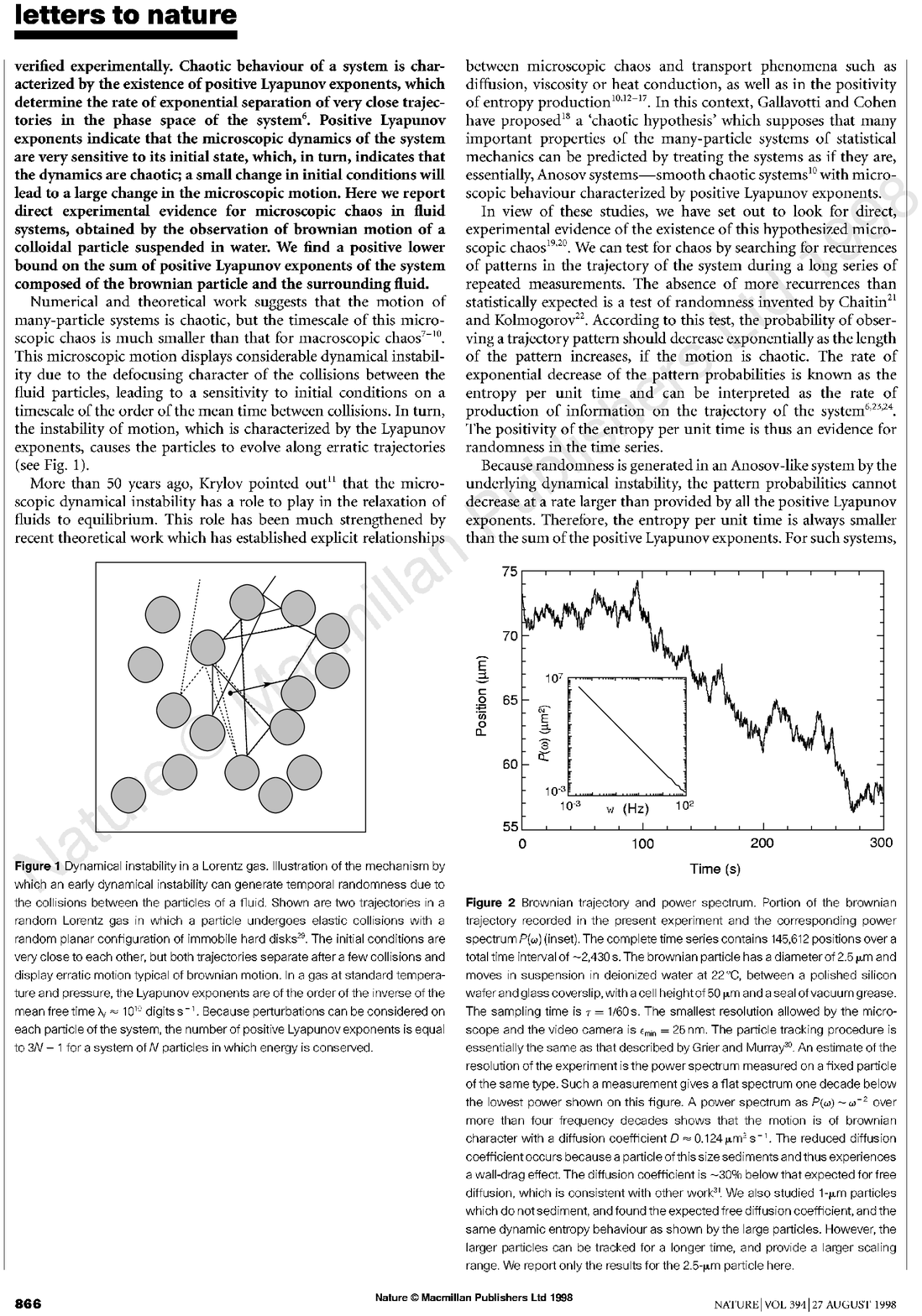}
  \end{center}
  \caption{The data shown here, taken from \cite{gaspard98}, show the
position of a $2.5 \,\mu\mbox{m}$ Brownian particle in water over a 300 second period with a sampling interval
of $\frac 1{60}$ sec (see \cite{gaspard98,briggs01} for the experimental details). The inset figure shows the
power spectrum, which displays a typical decay (for Brownian motion) with $\omega^{-2}$.}\label{gaspard}
\end{figure}

Others  have taken this suggestion (chaotic hypothesis) quite
seriously, and attempted an experimental confirmation. Figure
\ref{gaspard} shows a portion of the data, taken from
\cite{gaspard98}, that was obtained in an examination of a
microscopic system for the presence of chaotic behavior. Their
data analysis showed a positive lower bound on the sum of 
Lyapunov exponents of the system composed of a macroscopic
Brownian particle and the surrounding fluid.  From their analysis,
they argued that the Brownian  motion was due to (or the signature
of) deterministic microscopic chaos.  However, \citet{briggs01}
were more cautious in their interpretation, and \citet[Chapter
18]{mazo} has explored the possible interpretations of experiments
like these in some detail.

If true, the existence of deterministic chaos  (whatever that
means) would be an intriguing possibility since, if generated by a
non-invertible dynamics (semi-dynamical system), it could serve as
an explanation of a host of unresolved problems in the sciences.
Most notably, it could serve as an explanation for the manifest
irreversibility of our physical and biological world in the face
of physical laws that fail to encompass irreversibility without
the most incredulous of assumptions.  In particular, it would
clarify the foundations of irreversible statistical mechanics,
{\it e.g.} the operation of the second law of thermodynamics
\citep{dorfman99,gallavotti,mcm89rmp,mcmtdbk,schulman97}, and the
implications of the second law for the physical and biological
sciences.

\begin{figure}
 \includegraphics[width=6in]{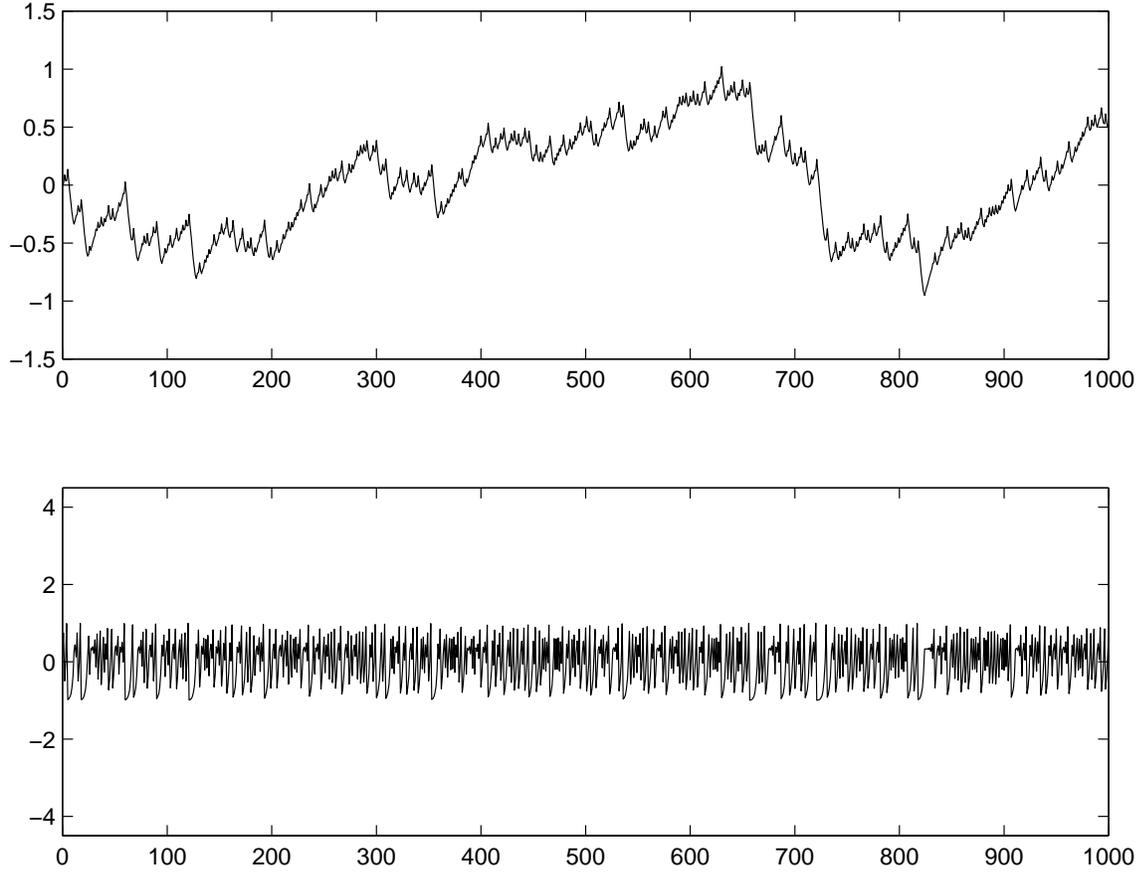}\\
  \caption{The top panel shows the simulated position of a particle obeying
  Equations \ref{vel} through \ref{hat} using Equation \ref{xnplus1}, while the bottom panel shows the velocity of
  the same particle computed with Equation \ref{finalx}.  The parameters used were:
  $\gamma = \Gamma/m =  10$,
$\kappa = 1$,
  and $\tau = -\frac{1}{10}
  \ln (9 \times 10^{-4})\simeq 0.932$ so $\lambda \equiv e^{-\gamma \tau} = 9 \times 10^{-4}$.  The initial condition on the hat map given by Equation \ref{hat} was $y_0=0.12562568$.
  }
  \label{s0009}
\end{figure}

\begin{figure}
 \includegraphics[width=6in]{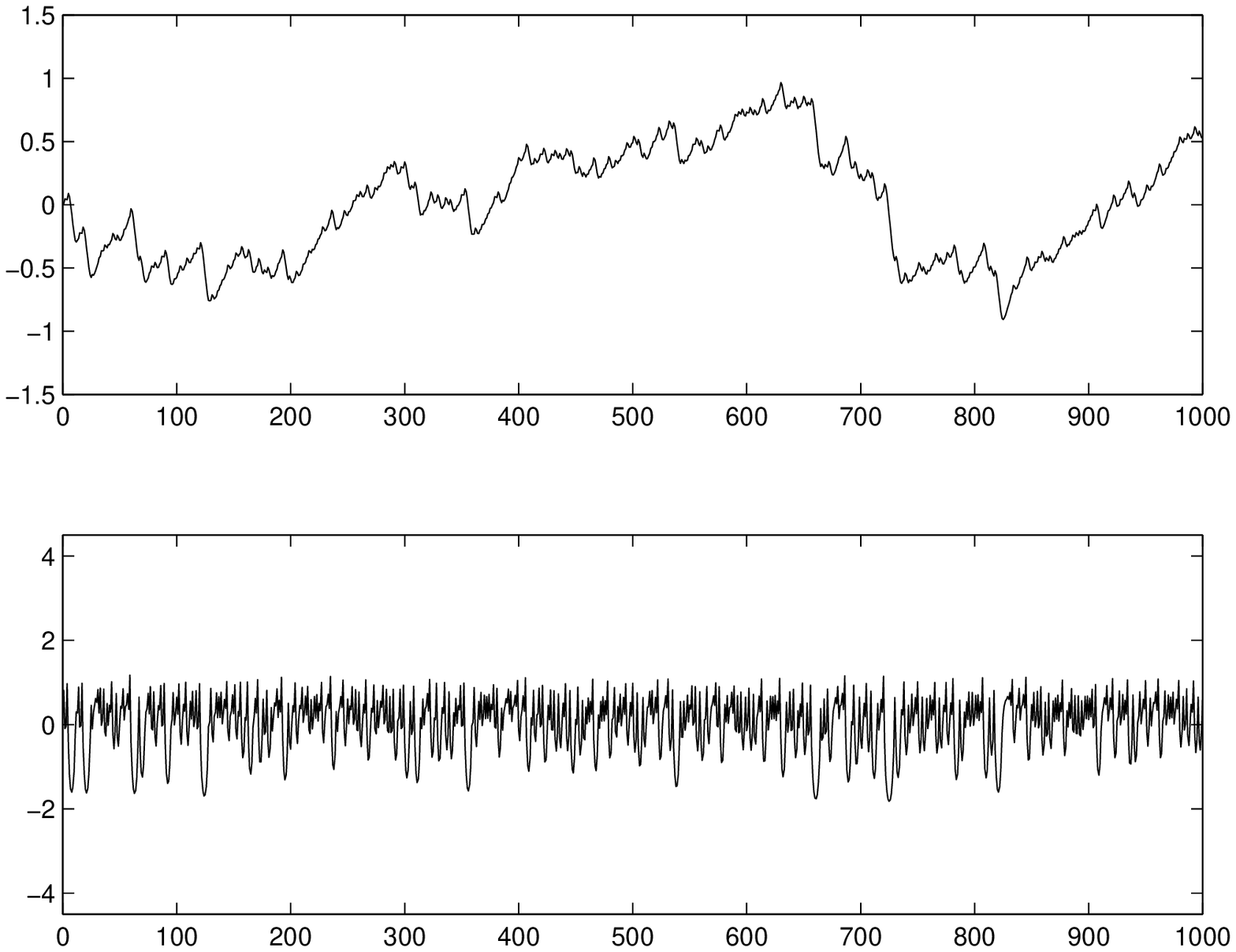}\\
  \caption{As in Figure \ref{s0009} except that
  $\tau = \frac {1}{10}\ln (2)
  \simeq 0.069$ so $\lambda = \frac 12$.
  }

  \label{simulation}
\end{figure}

\begin{figure}
 \includegraphics[width=6in]{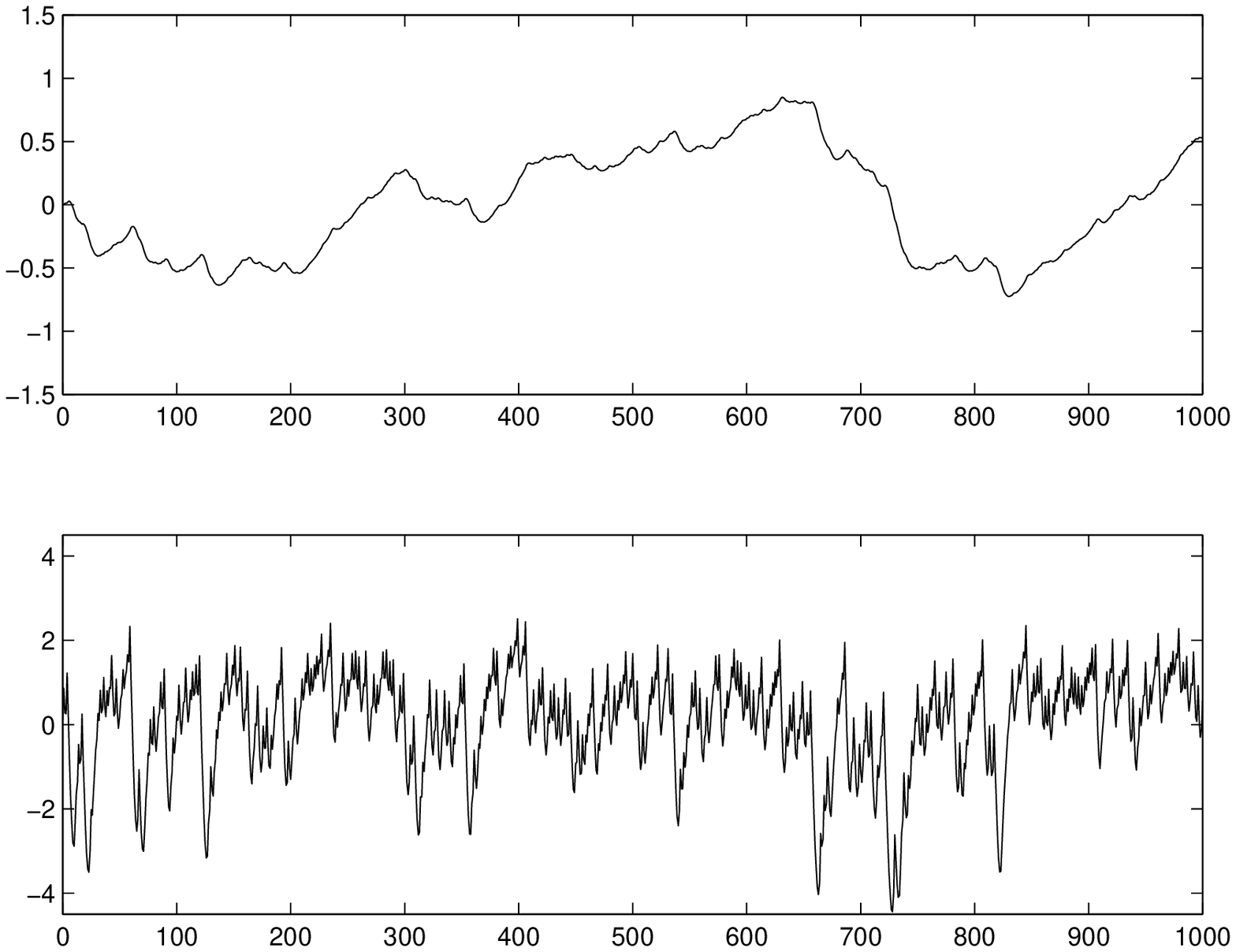}\\
  \caption{ As in Figure \ref{simulation} except that
$\tau = -\frac{1}{10}
  \ln (0.9 ) \simeq 0.011$ so $\lambda = 0.9$.  
  }
  \label{s9}
\end{figure}

\begin{figure}
 \includegraphics[width=6in]{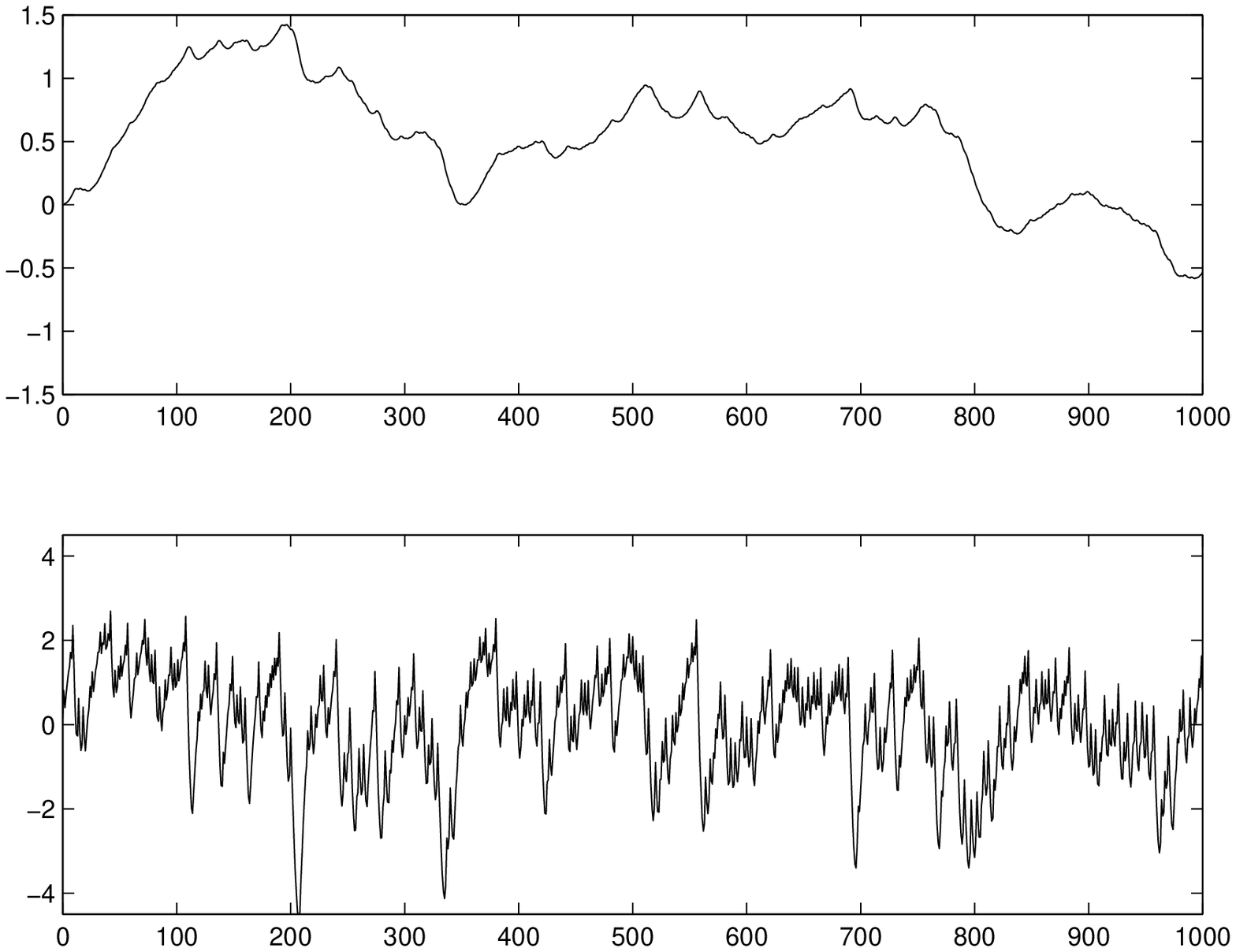}\\
  \caption{As in Figure \ref{s0009}, with the parameters of Figure \ref{s9} and
  an initial condition on the hat map (\ref{hat}) of $y_0=0.1678549321$.
  }
  \label{s9spd}
\end{figure}

In this paper, we have a  rather more modest goal. We address a
different facet of this chaotic hypothesis by studying  how and
when the characteristics of Brownian motion can be reproduced by
deterministic  systems. To motivate this, in Figures \ref{s0009}
through \ref{s9spd}
we show the position $(x)$  and velocity $(v)$ of a particle of mass
$m$ whose dynamics are described by
\begin{eqnarray}
\dfrac {dx}{dt}&=&v \label{vel}\\
m\dfrac{dv}{dt} &=& -\Gamma v + {\mathcal{F}}(t) \label{simul-eqns}.
\end{eqnarray} In Equations \ref{vel} and \ref{simul-eqns}, $\mathcal{F}$ is a
fluctuating ``force" consisting of a sequence of delta-function like
impulses given by \be \mathcal{F}(t) = m \kappa \sum_{n=0}^\infty
\xi(t) \delta(t-n\tau), \label{force}\ee and $\xi$ is a ``highly
chaotic" (exact, see Section \ref{sdsys}) deterministic variable
generated by $\xi_{t+\tau} = T(\xi_t)$ where $T$ is the hat map on
$[-1,1]$ defined by: \be T(y) = \left\{
\begin{array}{ll}
2 \left ( y + \frac 12 \right ) & \qquad \mbox{for} \quad y \in
\left [ -1, 0 \right ) \\
2 \left (\frac 12  - y   \right ) & \qquad \mbox{for} \quad y \in \left [ 0, 1 \right ).
\end{array}
\right. \label{hat} \ee

In this paper we examine the  behavior  of systems described by
equations like (\ref{vel}) through (\ref{hat}) and establish,
analytically, the eventual limiting behavior of ensembles. In
particular, we address the question of how Brownian-like motion
can arise from a purely deterministic dynamics. We do this by
studying the dynamics from a statistical, or ergodic theory,
standpoint.

The outline of the paper is as  follows. Section  \ref{sdsys}
gives required background and new material including the
definitions of a heirarchy of chaotic behaviours (ergodic, mixing,
and exact) with a discussion of their different behaviours in
terms of the evolution of densities under the action of transfer
operators such as the Frobenius-Perron  operators.  We then go on
to treat Central Limit theorems and Functional Central Limit
Theorems for non-invertible dynamical systems.
Section \ref{anal} returns to the specific problem that Equations
\ref{vel} through \ref{hat} illustrate. We show how the particle
velocity distribution may converge and how the particle position
may become asymptotically Gaussian, but for a more general class
of maps than given by (\ref{hat}).
In Section \ref{identify} we illustrate the application
of the results from Section \ref{anal} using a specific chaotic
map  (the dyadic map, Equation \ref{hat}) to
 act as a surrogate noise source.
Section \ref{tau} considers the question when one can obtain
 Gaussian processes by studying appropriate scaling limits of the
velocity and position variables, and the convergence of the
velocity process to an Ornstein-Uhlenbeck process as the interval
$\tau$ between chaotic perturbations approaches $0$. The paper
concludes with a brief discussion in Section \ref{disc}. The
Appendix collects and extends general central limit theorems from
probability theory that are used in the main results of the paper.

\setcounter{equation}{0} \setcounter{figure}{0}
\section{Semi-dynamical systems}\label{sdsys}

We are going to examine the behavior illustrated in Section
\ref{intro} using techniques from ergodic theory, and closely
related concepts from probability theory, applied to the dynamics
of semi-dynamical (non-invertible) systems.  In this section we
collect together the necessary machinery to do so.  Much of this
background material can be found in \cite{almcmbk94}.

\subsection{Density evolution operators}

Let $(Y_1,\B_1,\nu_1)$ and $(Y_2,\B_2,\nu_2)$ be two $\sigma$-finite measure spaces and let the transformation $T:Y_1\to Y_2$ be measurable,
{\it i.e.} $T^{-1}(\B_2)\subseteq \B_1$ where $T^{-1}(\B_2)=\{T^{-1}(A): A\in\B_2\}$. Then we say that $T$ is {\it nonsingular} (with respect to
$\nu_1$ and $\nu_2$) if $\nu_1(T^{-1}(A))=0$ for all $A\in\B_2$ with $\nu_2(A)=0$. Associated with the transformation $T$ we have the {\bf
Koopman operator} $U_T$ defined by
\[
U_T g=g\circ T
\]
for every measurable function $g:Y_2\to\realnos$. We define the
{\bf transfer} operator $P_T:L^1(Y_1,\B_1,\nu_1)\to
L^1(Y_2,\B_2,\nu_2)$ as follows.  For any $f\in
L^1(Y_1,\B_1,\nu_1)$, there is a unique element $P_T f$  in
$L^1(Y_2,\B_2,\nu_2)$ such that \be \int_{A}P_T
f(y)\nu_2(dy)=\int_{T^{-1}(A)}f(y)\nu_1(dy). \label{fpoper}\ee
Equation \ref{fpoper} simply gives an implicit relation between an
initial density of states $(f)$ and that density after the action
of the map $T$, {\it i.e.} $P_Tf$.  The Koopman operator
$U_T:L^\infty(Y_2,\B_2,\nu_2)\to L^\infty(Y_1,\B_1,\nu_1)$ and the
transfer operator $P_T$ are adjoint, so
\[
\int_{Y_2} g(y) P_T f(y) \nu_2(dy)=\int_{Y_1} f(y) U_T g(y)
\nu_1(dy)
\]
for $g\in L^\infty(Y_2,\B_2,\nu_2)$, $f\in L^1(Y_1,\B_1,\nu_1)$.

In some special cases Equation \ref{fpoper} allows us to obtain an
explicit form for $P_T$. Let $Y_2=\realnos$, $\B_2=\B(\realnos)$
be the Borel $\sigma$-algebra, and $\nu_2$ be the Lebegue measure.
Let $Y_1$ be an interval $[a,b]$ on the real line $\realnos$,
$\B_1=[a,b]\cap \B(\realnos)$ and $\nu_1$ be the Lebesgue measure
restricted to $[a,b]$. We will simply write $L^1([a,b])$
when the underlying measure is the Lebesgue measure.

The transformation $T:[a,b]\to\realnos$ is called {\it piecewise
monotonic} if \bit
\item[(i)]  there is a partition
$a=a_0<a_1<...<a_l=b$ of $[a,b]$ such that for each integer
$i=1,...,l$ the restriction of $T$ to $(a_{i-1},a_i)$ has a $C^1$
extension to $[a_{i-1},a_i]$ and \item[(ii)] $|T'(x)|>0$ for $
x\in(a_{i-1},a_i), i=1,...,l$.\eit
If a transformation
$T:[a,b]\to\realnos$ is  piecewise monotonic, then  for $f\in
L^1([a,b])$ we have
\[
P_T
f(y)=\sum_{i=1}^{l}\frac{f(T_{(i)}^{-1}(y))}{|T'(T_{(i)}^{-1}(y))|}1_{T((a_{i-1},a_i))}(y),
\]
where $T_{(i)}^{-1}$ is the inverse function for the restriction
of $T$ to $(a_{i-1},a_i)$. Note that we have equivalently \be P_T
f(y)=\sum_{x\in T^{-1}(\{y\})}\frac{f(x)}{|T'(x)|}.
\label{formulaPF}\ee Of course these formulas hold almost
everywhere with respect to the Lebesque measure.

Let $(Y,\B)$ be a measurable space and let $T:Y\to Y$ be a
measurable transformation. The definition of the transfer operator
for $T$ depends on a given  $\sigma$-finite measure on $\B$, which
in turn gives rise to different operators for different underlying
measures on $\B$. If $\nu$ is a probability measure on $\B$ which is
{\it invariant} for $T$, {\it i.e.} $\nu(T^{-1}(A))=\nu(A)$ for all
$A\in \B$, then $T$ is nonsingular. The transfer operator
$P_T:L^1(Y,\B,\nu)\to L^1(Y,\B,\nu)$ is well defined and when we
want to emphasize that the underlying measure $\nu$ in the transfer
operator is invariant under the transformation $T$ we will write
$\FP_{T,\nu}$.  The Koopman operator $U_T$ is also well defined for
$f\in L^1(Y,\B,\nu)$ and is an isometry of $L^1(Y,\B,\nu)$ into
$L^1(Y,\B,\nu)$, {\it i.e.} $||U_T f||_1=||f||_1$ for all $f\in
L^1(Y,\B,\nu)$. The following relation holds between the operators
$U_T,\FP_{T,\nu}:L^1(Y,\B,\nu)\to L^1(Y,\B,\nu)$
 \be
\FP_{T,\nu} U_T f=f\;\;\mbox{and}\;\;U_T \FP_{T,\nu}
f=E(f|T^{-1}(\B))\label{fpcond}
 \ee for $f\in L^1(Y,\B,\nu),$ where
$E(\cdot|T^{-1}(\B)):L^1(Y,\B,\nu)\to L^1(Y,T^{-1}(\B),\nu)$ denotes the operator of conditional expectation
(see Appendix). 
Both of these equations are based on the following change of
variables \citep[Theorem 16.13]{billingsley95}: $f\in
L^1(Y,\B,\nu)$ if and only if $f\circ T\in L^1(Y,\B,\nu)$, in
which case the following holds \be \int_{T^{-1}(A)} f \circ T(y)
\nu(dy)= \int_A f(y)  \nu(dy) ,\;\;A\in\B.\ee If the measure $\nu$
is finite, we have $L^p(Y,\B,\nu)\subset L^1(Y,\B,\nu)$ for $p\ge
1$. The operator $U_T:L^p(Y,\B,\nu)\to L^p(Y,\B,\nu)$ is also an
isometry in this case. Note that if the conditional expectation
operator
$E(\cdot|T^{-1}(\B)):L^1(Y,\B,\nu)\to L^1(Y,\B,\nu)$ 
is restricted to $L^2(Y,\B,\nu)$, then this is the orthogonal projection of $L^2(Y,\B,\nu)$ onto
$L^2(Y,T^{-1}(\B),\nu).$

One can also consider any $\sigma$-finite measure $m$ on $\B$ with
respect to which $T$ is nonsingular and the corresponding transfer
operator $P_T: L^1(Y,\B, m)\to L^1(Y,\B, m)$. To be specific, let
$Y$ be a Borel subset of $\realnos^k$ with Lebesque measure $m$
and $\B=\B(Y)$ be the $\sigma-$algebra of Borel subsets of $Y$.
Throughout this paper $m$ will denote Lebesque measure and
$L^1(Y)$ will denote $L^1(Y,\B,m)$. The transfer operator
$P_T:L^1(Y)\to L^1(Y)$ is usually known as the {\bf
Frobenius-Perron} operator. A measure $\nu$ (on $Y$) is said to
have a {\it density} $g_*$ if $\nu(A)=\int_{A} g_*(y) dy $ for all
$A\in\B$, where $g_*\in L^1(Y)$ is nonnegative and
$\int_{Y}g_*(y)dy=1$. A measure $\nu$ is called {\it absolutely
continuous} if it has a density. If the Frobenius-Perron operator
$P_T$ has a nontrivial fixed point in $L^1(Y)$, {\it i.e.} the
equation $P_Tf=f$ has a nonzero solution in $L^1(Y)$, then the
transformation $T$ has an absolutely continuous invariant measure
$\nu$, its density $g_*$ is a fixed point of $P_T$, and we call
$g_*$ an {\it invariant density} under the transformation $T$.
The following relation holds between the operators $P_T$ and
$\FP_{T,\nu}$ \be P_T(fg_*)=g_*\FP_{T,\nu}f\;\;\mbox{for}\;\;f\in
L^1(Y,\B,\nu).\label{dfpo} \ee In particular, if the density $g_*$
is strictly positive, {\it i.e.} $g_*(y)>0$ a.e. for $y\in Y$, then
the measures $m$ and $\nu$ are equivalent and we also have
\[
P_T(f)=g_* \FP_{T,\nu}
\left(\frac{f}{g_*}\right)\;\;\mbox{for}\;\;f\in L^1(Y).
\]
The notion of a piecewise monotonic transformation on an interval
can be extended to ``piecewise smooth" transformations $T:Y\to Y$
  with  $Y\subset\realnos^k$. Therefore if $T$ has, for example, a
finitely many inverse branches and the Jacobian matrix $DT(x)$ of
$T$ at $x$ exists and $\det DT(x)\neq 0$ for almost every $x$,
then the Frobenius-Perron operator is given by
\[
P_T f(y)=\sum_{x\in T^{-1}(\{y\})}\frac{f(x)}{|\det
DT(x)|}\qquad\mbox{a.e.}
\]
for $f\in L^1(Y).$ If $T$ is invertible then we have
$
P_T f(y)=f(T^{-1}(y))|\det DT^{-1}(y)|.
$

Finally, we briefly mention Ruelle's transfer operator. Let $Y$ be a
compact metric space, $T:Y\to Y$ be a continuous map such that
$T^{-1}(\{y\})$ is finite for each $y\in Y$ and let
$\phi:Y\to\realnos$ be a  function (typically continuous or
H{\"o}lder continuous). The so called Ruelle operator $
\mathcal{L}_{\psi}$
 acts on functions  rather than on $L^1(Y)$ elements and
is defined by
\[
 \mathcal{L}_{\psi}f(y)=\sum_{x\in T^{-1}(\{y\})}e^{\psi(x)}f(x)
\]
 for every $y\in Y$. The function
$\psi$ is a so called potential. If we take $\psi(x)=-\log |\det
DT(x)|$, provided it makes sense, then we arrive at the
representation for the Frobenius-Perron operator.

\subsection{Probabilistic and ergodic
properties of density evolution}\label{proband}

Let $(Y,\B,\nu)$ be a normalized measure space and let $T:Y\to Y$ be
a measurable transformation preserving the measure $\nu$. We can
discuss the ergodic properties of $T$ in terms of the convergence
behavior of its transfer operator $\FP_{T,\nu}:L^1(Y,\B,\nu)\to
L^1(Y,\B,\nu)$. To this end, we note that the transformation $T$ is

\bit \item[(i)] {\it Ergodic} (with respect to $\nu$) if and only if
every invariant set $A \in \B$ is such that $\nu(A) = 0$ or $\nu (Y
\setminus A)=0$.  This is equivalent to: $T$ is  {\it ergodic} (with
respect to $\nu$) if and only if for each $f\in L^1(Y,\B,\nu)$ the
sequence $\frac{1}{n}\sum_{k=0}^{n-1}\FP_{T,\nu}^k f$ is weakly
convergent in $L^1(Y,\B,\nu)$ to $\int f(y)\nu(dy)$, {\it i.e.} for
all $g\in L^\infty(Y,\B,\nu)$
\[
\lim_{n\to\infty}\int\frac{1}{n}\sum_{k=0}^{n-1}\FP_{T,\nu}^k f(y)g(y) \nu(dy)=\int
f(y)\nu(dy)\int g(y) \nu(dy);
\]
\item[(ii)] {\it Mixing} (with respect to $\nu$) if and only if
\[
\lim_{n\to\infty} \nu(A \cap T^{-n}(B)) = \nu(A) \nu(B) \qquad
\mbox{for}\quad A,B \in \B.
\]
Mixing is equivalent to: For each  $f\in L^1(Y,\B,\nu)$ the
sequence $P^n_T f$ is weakly convergent in $L^1(Y,\B,\nu)$ to
$\int f(y)\nu(dy)$, {\it i.e.}
\[
\lim_{n\to \infty}\int \FP_{T,\nu}^n f(y)g(y)\nu(dy) = \int
f(y)\nu(dy) \int g(y)\nu(dy) \quad\mbox{for}\quad g\in
L^\infty(Y,\B,\nu).
\]
\item[(iii)] {\it Exact} (with respect to $\nu$) if and only if
\[
\lim_{n \to \infty} \nu (T^n(A))=1 \qquad\mbox{for}\quad A \in \B
\quad\mbox{with}\quad T(A)\in\B,\; \nu(A) > 0.
\]
Exactness  is equivalent to: For each $f\in L^1(Y,\B,\nu)$ the
sequence $P^n_{T,\nu} f$ is strongly convergent in $L^1(Y,\B,\nu)$
to $\int f(y)\nu(dy)$, {\it i.e.}
\[
\lim_{n\to\infty}\int |\FP_{T,\nu}^nf(y)-\int f(y)\nu(dy)|
\nu(dy)=0.
\]

 \eit

The characterization of the ergodic properties of transformations
through the properties of the evolution of densities requires that
we know an invariant measure $\nu$ for $T$. Examples of ergodic,
mixing, and exact
transformations are
given in the following.

\begin{exmp}\label{ergod}
{\em The transformation on $[0,1]$
\[
T(y) = y + \phi \qquad \mbox{mod} \,1,
\]
known as rotation on the circle,  is ergodic with respect to the
Lebesgue measure when $\phi$ is irrational. The associated
Frobenius-Perron operator is given by
\[
P_Tf(y) = f(y-\phi). \]}
\end{exmp}

\begin{exmp}\label{baker}
{\em The baker map on $[0,1]\times[0,1]$
\[
T(y,z)= \left\{
\begin{array}{ll}
(2y,\frac {1}{2} z) &0 \leq z \leq \frac {1}{2} \\
(2y-1, \frac 12 + \frac {1}{2} z ) &\frac {1}{2} < z \leq 1
\end{array}
\right.
\]
is mixing with respect to the Lebesgue measure.  The Frobenius-Perron operator is given by
\[
P_Tf(y,z)= \left\{
\begin{array}{ll}
f(\frac 12 y, 2 z) &0 \leq z \leq \frac {1}{2} \\
f(\frac 12 + \frac 12 y , 2z-1 ) &\frac {1}{2} < z \leq 1
\end{array}
\right.
\]
}
\end{exmp}

\begin{exmp}\label{exp:hat}
{\em The hat map on $[-1,1]$ defined by Equation \ref{hat} is exact
with respect to the Lebesgue measure, and has a Frobenius-Perron
operator given by
\[
P_Tf(y)= \frac 12 \left [f\left( \frac 12 y -\frac 12 \right ) +
f\left ( \frac 12 - \frac 12 y \right ) \right ].
\]
}
\end{exmp}

\begin{exmp}\label{exp:genhat}
{\em A class of piecewise linear transformations on $[0,1]$ are given by \be T_N(y) = \left\{
\begin{array}{ll}
N \left ( y - \frac {2n}{N} \right ) & \qquad \mbox{for} \quad y \in
\left [ \frac {2n}{N}, \frac {2n+1}{N} \right ) \\
N \left (\frac {2n+2}{N} - y   \right ) & \qquad \mbox{for} \quad y \in \left [\frac {2n+1}{N}, \frac {2n+2}{N}
\right ),
\end{array}
\right. \label{genhat} \ee where $n = 0,1,\ldots,[(N-1)/2]$ and
$[z]$ denotes the integer part of $z$. For $N \geq 2$, these
piecewise linear maps generalize the hat map, are exact with
respect to the Lebesgue measure,  and have the invariant density
 \be g_*(y) =  1_{[0,1]}(y). \ee }
\end{exmp}

\begin{exmp}\label{cheby}
{\em 
 The Chebyshev maps (\cite{adler64}) on
$[-1,1]$ studied by \citet{beck87}, \citet{beck96} and
\citet{beck99} are given by \be S_N(y) = \cos (N \arccos y),
\qquad N=0,1,\cdots \label{chebyeqn}\ee with $S_0(y) = 1$ and
$S_1(y) = y$. They are conjugate to the transformation of Example
\ref{exp:genhat}, and satisfy the recurrence relation $S_{N+1}(y)
= 2 y S_N(y) - S_{N-1}(y)$.  For $N \geq 2$ they are exact with
respect to the measure with the density
\[
g_*(y) = \dfrac{1}{\pi \sqrt{1-y^2}}.
\]
For $N = 2$ the Frobenius-Perron operator is given by
\[
P_{S_2} f(y) = \frac{1}{2 \sqrt {2y+2}}\left [
f\left(\sqrt{\frac{1}{2} y + \frac{1}{2}}\right) +
f\left(-\sqrt{\frac{1}{2} y + \frac{1}{2}}\right) \right ]
\]
and the transfer operator by
\[
\FP_{S_2,\nu} f(y) = \frac{1}{2
}\left [ f\left(\sqrt{\frac{1}{2} y + \frac{1}{2}}\right) +
f\left(-\sqrt{\frac{1}{2} y + \frac{1}{2}}\right) \right ].
\]
}
\end{exmp}

The construction of a transfer operator for a conjugate map is
presented in the next theorem from
\cite{almcmbk94}.

\begin{thm}(\citet[Theorem 6.5.2]{almcmbk94})\label{conjugate}
Let $T:[0,1]\to[0,1]$ be a measurable and nonsingular (with
respect to the Lebesgue measure) transformation. Let
$\nu:\B([a,b])\to [0,\infty)$ be a probability measure with a
strictly positive density $g_*$, that is $g_*(y)>0$ a.e. Let a
second transformation $S:[a,b]\to [a,b]$ be given by
$S=G^{-1}\circ T\circ G$, where
\[
G(x)=\int_{a}^x g_*(y)\,dy, \qquad a\le x\le b.
\]
Then the transfer operator $\FP_{S,\nu}$ is given by \be \FP_{S,\nu}
f=U_G P_T U_{G^{-1}} f, \qquad \mbox{for}\;\;f\in
L^1([a,b],\B([a,b]),\nu), \ee where $U_G, U_{G^{-1}}$ are  Koopman
operators for $G$ and $G^{-1}$, respectively, and $P_T$ is the
Frobenius-Perron operator for $T$. As a consequence, $\nu$ is
invariant for $S$ if and only if the Lebesgue measure is invariant
for $T$.
\end{thm}

\begin{exmp}\label{exp:dyadic}
{\em The dyadic map on $[-1,1]$ is given by \be T(y)= \left\{
\begin{array}{ll}
2y +1, & y \in \left [-1,0\right ]\\ 2y -1, & y \in \left (0, 1\right ],
\end{array}
\right. \label{dyadic} \ee and has the uniform invariant density
\[
g_*(y) = \dfrac 12 1_{[-1,1]}(y).
\]
Like the hat map, it is exact with respect to the normalized Lebesgue measure on $[-1,1].$  It has a
Frobenius-Perron operator given by
\[
P_Tf(y) = \dfrac12 \left [f \left
(\dfrac{1}{2}y-\dfrac{1}{2}\right) +
f\left(\dfrac{1}{2}y+\dfrac{1}{2}\right ) \right ].
\]
}
\end{exmp}

\begin{exmp}\label{fatbaker} {\em \cite{alexyorke} defined a generalized baker
transformation (also known as a fat/skiny baker transformation)
$S_\beta:[-1,1]\times[-1,1]\to [-1,1]\times[-1,1]$ by
\[
S_\beta(x,y)=(\beta x + (1-\beta) h(y), T(y))
\]
where $0<\beta<1$, $T$ is the dyadic map on $[-1,1]$, and
\[
h(y)=\left\{\begin{array}{ll}1, &  y\ge 0,\\ -1, & y<0.\end{array}\right.
\]
For every $\beta\in(0,1)$ the transformation $S_\beta$ has an
invariant probability measure on $[-1,1]\times [-1,1]$ and is
mixing. The invariant measure is the product of a so called
infinitely convolved Bernoulli measure (see Section \ref{identify})
and the normalized Lebesgue measure on $[-1,1].$ If
$\beta=\frac{1}{2}$, the transformation $S_\beta$ is conjugated
through a linear transform of the plane to the baker map of Example
\ref{baker}. If $\beta<\frac{1}{2}$,  the transformation $S_\beta$
does not have an invariant density (with respect to the planar
Lebesgue measure). }\end{exmp}

\begin{exmp}\label{confr}
{\em The continued fraction map \be T(y) = \dfrac 1y \qquad \mbox{mod} \;1 \quad y \in (0,1] \ee has an
invariant density \be g_*(y) = \dfrac 1{(1+y)\ln {2}}\ee and is exact. The Frobenius-Perron operator is given by
\[
P_T f(y) = \sum_{k=1}^\infty \dfrac{1}{(y+k)^2} f\left(\dfrac{1}{y+k}\right)
\]
and the transfer operator by
\[
\FP_{T,\nu} f(y) = \sum_{k=1}^\infty \dfrac{y+1}{(y+k)(y+k+1)}
f\left(\dfrac{1}{y+k}\right).
\]
}
\end{exmp}

\begin{exmp}\label{quadratic}
{\em The quadratic map is given by
\[
T_\beta(y)=1-\beta y^2, \;\;y\in\left[-1,1\right]
\]
where $0<\beta\le 2.$ It is known that there exists a positive Lebesgue measure set of parameter values $\beta$
such that the map $T_\beta$ has an absolutely continuous (with respect to Lebesgue measure) invariant measure
$\nu_\beta$ (\cite{jakobson} and \cite{benedics}). Let $\alpha>0$ be a very small number and let
\[
\Delta_\epsilon=\{\beta\in[2-\epsilon,2]: |T_\beta^n(0)|\ge e^{-\alpha n}\;\;\mathrm{and}\;\;
|(T_\beta^n)'(T_\beta(0))|\ge (1.9)^n\;\; \forall n\ge 0\}
\]
for $\epsilon>0$. \cite{young} proved that for sufficiently small
$\epsilon$ and for every $\beta\in\Delta_\epsilon$ the
transformation $T_\beta$ is exact with respect to $\nu_\beta$ and
this measure is supported on $[T_\beta^2(0),T_\beta(0)]$. }
\end{exmp}

\begin{exmp}\label{pomeau} {\em The Manneville-Pomeau map $T_\beta:[0,1]\to[0,1]$
is given by
\[
T_\beta(y)=y+y^{1+\beta} \qquad \mbox{mod} \,1,
\]
 where $\beta\in (0,1)$. The map has an absolutely continuous invariant probability measure $\nu_\beta$ with density
satisfying
\[
\frac{c_1}{y^\beta}\le g_*(y)\le \frac{c_2}{y^\beta}
\]
for some constants $c_2\ge c_1>0$ (cf. \cite{thaler}), and is
exact.
 }
\end{exmp}

Finally, we discuss the notion of {\bf Sinai-Ruelle-Bowen measure}
or {\bf SRB measure} of $T$ which was first conceived in the
setting of Axiom A diffeomorphisms on compact Riemannian
manifolds. This notion varies from author to author
\citep{alexyorke, eckmannruelle85,tsujii96,young02,hunt02}. Let
$(Y,\rho)$ be a compact metric space with a reference measure $m$,
e.g. a compact subset of $\realnos^k$ and $m$ the Lebesque or a
compact Riemannian manifold and $m$ the Riemannian measure on $Y$.
If $T:Y\to Y$ is a continuous map, then by the Bogolyubov-Krylov
theorem there always exists at least one invariant probability
measure for $T$.
 When there is more than one measure, the question arises which
invariant measure is ``interesting", and has led to  attempts to
give a good definition of ``physically" relevant invariant
measures.  Though this seems to be a rather vague and poorly
defined concept, loosely speaking one would expect that one
criteria for a physically relevant invariant measure would be
whether or not it was observable in the context of some laboratory
or numerical experiment.

An invariant measure $\nu$ for $T$ is called a {\it natural} or
{\it physical measure} if there is a positive Lebesque measure set
$Y_0\subset Y$ such that for every $y\in Y_0$ and for every
continuous observable $f:Y\to\realnos$ \be
\lim_{n\to\infty}\frac{1}{n}\sum_{i=0}^{n-1}f(T^i(y))= \int
f(z)\nu(dz) \label{c:physicalmeasure}.\ee  In other words the
average of $f$ along the trajectory of $T$ starting in  $Y_0$  is
equal to the average of $f$ over the space $Y$. 
We can also say that for each $y\in Y_0$ the measures
$\frac{1}{n}\sum_{i=0}^{n-1}\delta_{T^i(y)}$ are weakly convergent
to $\nu$
\[
\frac{1}{n}\sum_{i=0}^{n-1}\delta_{T^i(y)}\to^d \nu
\]
(see the next
section for this notation).

Observe that if $\nu$ is ergodic then from the individual Birkhoff
ergodic theorem it follows that for every $f\in L^1(Y,\B,\nu)$
Condition \ref{c:physicalmeasure} holds for almost all $y\in Y$,
i.e. except for a subset of $Y$ of $(\nu)$ measure zero. Thus, if
$T$ has an ergodic absolutely continuous invariant measure $\nu$
with density $g_*$ then every continuous function $f$ is
integrable with respect $\nu$ and Condition
\ref{c:physicalmeasure} holds for  almost every point from the set
$\{y\in Y: g_*(y)>0\}$, i.e. except for a subset of Lebesque
measure zero. Therefore such $\nu$ is a physical measure for $T$.
Not only absolutely continuous measures are physical measures.
Consider, for example, the generalized baker transformation
$S_\beta$ of Example \ref{fatbaker}. \cite{alexyorke} showed that
there is a unique physical measure $\nu_\beta$ for each $\beta\in
(0,1)$. This measure is mixing and hence ergodic. Although for
$\beta>\frac{1}{2}$ the transformation expands areas, the measure
$\nu_\beta$  might not be absolutely continuous for certain values
of the parameter $\beta$ (e.g. $\beta=\dfrac{-1+\sqrt{5}}{2}$) in
which case the Birkhoff ergodic theorem only implies Condition
\ref{c:physicalmeasure} on a zero Lebesque measure set. Therefore
a completely different argument was needed in the proof of the
physical property.

In the context of smooth invertible maps having an Axiom A
attractor the existence of a unique physical measure on the
attractor was first proved for Anosov diffeomorhisms by
\cite{sinai72} and later generalized by \cite{ruelle76} and
\cite{bowen75}. Roughly speaking, these are maps having uniformly
expanding and contracting directions and their physical invariant
measures have densities with respect to the Lebesque measure in
the expanding directions (being usually singular in the
contracting directions). This property lead then to the
characterization of a Sinai-Ruelle-Bowen measure. In a recent
attempt to go beyond maps having an Axiom A attractor,
\cite{young02} additionally requires that $T$ has a positive
Lyapunow exponent a.e. The precise definition strongly relies on
the smoothness and invertibility of the map $T$. Note that the
generalized baker transformation $S_\beta$ has Lyapunov exponents
equal to $\ln 2$ and $\ln \beta$ and the measure $\nu_\beta$ is
absolutely continuous along all vertical directions.

%
%
%
%


\subsection{Brownian motion from deterministic
perturbations}\label{s:brownian}

\subsubsection{Central limit theorems}\label{s:prob}

We follow the terminology of \cite{billingsley68}. If $\zeta$ is a measurable mapping from a probability space
$(\Omega,\F,\Pr)$ into a measurable space $(Z,\mathcal{A})$, we call $\zeta$ a $Z$-valued random variable. The
distribution of $\zeta$ is the normalized measure $\mu=\Pr\circ \zeta^{-1}$ on $(Z,\mathcal{A})$, {\it i.e.}
\[
\mu(A)=\Pr(\zeta^{-1}(A))=\Pr\{\omega:\xi(\omega)\in A\}=\Pr\{\xi\in A\},\;\;A\in\mathcal{A}.
\]
If $Z=\realnos^k$, we also have the associated distribution function of $\zeta$ or $\mu$, defined by
\[
F(x)=\mu\{y:y\leq x\}=\Pr\{\zeta\leq x\},\;\;x\in\realnos^k,
\]
where $\{y:y\le x\}=\{y:y_i\le x_i, i=1,\ldots,k\}$ for $x=(x_1,\ldots,x_k)$. The random variables $\zeta$ and
$\xi$ are, by definition, {\it (statistically) independent} if
\[
\Pr\{\zeta\in A, \xi\in B\}=\Pr\{\zeta\in A\}\Pr\{\xi\in B\},
\]
{\it i.e.} the distribution of the pair $(\zeta,\xi)$ is the product of the distribution of $\zeta$ with that of
$\xi$.

Let  $(Z,\rho)$ be a  metric space and $\B(Z)$ be the $\sigma$-algebra of Borel subsets of $Z$. A sequence
$(\mu_{n})$ of normalized measures on $(Z,\B(Z))$ is said to {\it converge weakly} to a normalized measure $\mu$
if
\[
 \lim\limits_{n\to\infty} \int_Z f(z) \mu_{n}(dz) =
  \int_Z f(z)\mu(dz)
\]
for every continuous bounded function $f:Z\to\realnos$. Note that
the integrals $\int_Z f(z)\mu(dz)$ completely determine $\mu$,
thus the sequence $(\mu_{n})$ cannot converge weakly to two
different limits. Note also that weak convergence depends only on
the topology of $Z$, not on the specific metric that generates it;
thus two equivalent metrics give rise to the same notion of weak
convergence. If we have a family $\{\mu_\tau:\tau\ge 0\}$ of
normalized measures instead of a sequence, we can also speak of
weak convergence of $\mu_\tau$ to $\mu$ when $\tau$ goes to
$\infty$ or some finite value $\tau_0$ in a continuous manner.
This then means that $\mu_\tau$ converges weakly to $\mu$ as
$\tau\to\tau_0$ if and only if $\mu_{\tau_n}$ converges weakly to
$\mu$ for each sequence $(\tau_n)$ such that  $\tau_n\to\tau_0$ as
$n\to\infty$.

If $Z=\realnos^k$ and $F$ and $F_n$ are, respectively, the distribution functions of $\mu$ and $\mu_n$, then
$(\mu_n)$ converges weakly to $\mu$ if and only if
\[
\lim_{n\to\infty}F_n(z)=F(z)\;\;\mathrm{at\;continuity\; points}\;z\;\;\mathrm{of}\;F.
\]
The characteristic function
$\varphi_\mu$ of a normalized measure $\mu$ on $\realnos^k$ is
defined by
\[
\varphi_\mu(r)=\int \exp{(i<r,z>)}\mu(dz),
\]
where $i=\sqrt{-1}$ and $<r,z>=\sum_{j=1}^kr_j z_j$ denotes the inner product in $\realnos^k$. The {\bf continuity theorem} \citep[Theorem
7.6]{billingsley68} gives us the following: $(\mu_n)$ converges weakly to $\mu$ if and only if
\[
\lim_{n\to\infty}\varphi_{\mu_n}(r)=\varphi_\mu(r) \;\;\mbox{for each}\;\;r\in\realnos^k.
\]
A sequence $\zeta_n$ of $Z$-valued random variables converges {\it in distribution}, or {\it weakly}, to a
normalized measure $\mu$ on $(Z,\B(Z))$, if the corresponding distributions of $\zeta_n$ converge weakly to
$\mu$. This is denoted by
\[
\zeta_n\to^{d} \mu.
\]
If $\mu$ is the
distribution of a random variable $\zeta$, we write $\zeta_n\to^{d} \zeta.$ Note that the underlying probability
spaces for the random variables $\zeta, \zeta_1, \zeta_2...$ may be all distinct.

A sequence $\zeta_n$ of $Z$-valued random variables converges {\it
in probability} to a $Z$-valued random variable $\zeta$  if \be
\lim_{n\to\infty}\Pr(\rho(\zeta_n,\zeta)>\varepsilon)=0\;\;\mbox{for
all}\;\;\varepsilon>0. \label{d:conpr}\ee This is denoted by
\[
\zeta_n\to^{P}\zeta.
\]
Here all the random variables are defined on
the same probability space. Note that if Condition \ref{d:conpr}
holds then
\[
\lim_{n\to\infty}{\Pr}_{0}(\rho(\zeta_n,\zeta)>\varepsilon)=0\;\;\mbox{for all}\;\;\varepsilon>0
\]
for every probability measure $\Pr_0$ on $(\Omega,\F)$ which is
absolutely continuous with respect to $\Pr$. In other words
convergence in probability is preserved by an absolutely
continuous change of measure. We will also frequently use the
following result from \citet[Theorem 4.1]{billingsley68}: If
$(Z,\rho)$ is a separable metric space, and \be \mbox{if}\;\;
\tilde{\zeta}_n\to^d\zeta\;\; \mbox{and}\;\;
\rho(\zeta_n,\tilde{\zeta}_n)\to^P 0,\;\; \mbox{then}
\;\;\zeta_n\to^d \zeta.\label{slutsky}\ee

We will write $N(0,\sigma^2)$ for either a real-valued random
variable which is Gaussian distributed with mean $0$ and variance
$\sigma^2$, or the measure on $(\realnos, \B(\realnos))$ with
density \be
\dfrac{1}{\sqrt{2\pi}\sigma}\exp\left({-\dfrac{x^2}{2\sigma^2}}\right).
\label{gaussian}\ee Since $\sigma N(0,1)=N(0,\sigma^2)$ when
$\sigma>0$, we can always write $\sigma N(0,1)$ for $\sigma\ge 0$,
which in the case $\sigma=0$ reduces to $0$. The characteristic
function of $N(0,1)$ is of the form
$\phi(r)=\exp{(-\frac{1}{2}r^2)},$ $r\in\realnos.$

Let $(\zeta_j)_{j\ge 1}$ be a sequence of real-valued zero-mean
random variables with finite variance. If there is $\sigma>0$ such
that \be \frac{\sum_{j=1}^n \zeta_j}{\sqrt{n}}\to^d \sigma N(0,1),
\label{clt} \ee then $(\zeta_j)_{j\ge 1}$ is said to satisfy  the
{\bf central limit theorem (CLT)}. Note that if the random
variables $\zeta_j$ are defined on the same probability space
$(\Omega,\F,\Pr)$, we have the equivalent formulation of
(\ref{clt}) in the case of $\sigma>0$
\[
\lim_{n\to\infty}\Pr\left\{\omega\in \Omega:
\frac{\sum_{j=1}^{n}\zeta_j(\omega)}{\sigma\sqrt{n}}\le
z\right\}=\Phi(z),\;\;z\in\realnos,
\]
where
 \be \Phi(z)=\frac{1}{\sqrt{2\pi}}
\int_{-\infty}^{z}\exp\left(-\frac{1}{2}t^2\right)dt\ee is the standard Gaussian distribution function.  In the
case of $\sigma=0$
\[
\frac{\sum_{j=1}^n \zeta_j}{\sqrt{n}}\to^P 0.
\]

Let $\{w(t),t\in[0,\infty)\}$ be a standard Wiener process (Brownian motion), {\it i.e.}
$\{w(t),t\in[0,\infty)\}$ is a family of real-valued random variables on some probability space
$(\Omega,\F,\Pr)$ satisfying the following properties: \bit
\item[(i)] the process starts at zero: $w(0)=0$ a.e;\item[(ii)]
for $0\le t_1<t_2$ the random variable $w(t_2)-w(t_1)$ is Gaussian distributed with mean $0$ and variance
$t_2-t_1$;
\item[(iii)] for times $t_1<t_2<\ldots<t_n$ the increments
$w(t_2)-w(t_1),\ldots,w(t_n)-w(t_{n-1})$ are independent random variables. \eit Existence of the Wiener process
$\{w(t),t\in[0,1]\}$ is equivalent to the existence of the Wiener measure $W$ on the space $C[0,1]$ of
continuous functions on $[0,1]$ with uniform convergence, in a topology which makes $C[0,1]$ a complete
separable metric space. Then, simply, $W$ is the distribution of a random variable $W:\Omega\to C[0,1]$ defined
by $W(\omega):t\mapsto w(t)(\omega)$.

Let $D[0,1]$ be the space of right continuous real valued functions on $[0,1]$ with left-hand limits. We endow
$D[0,1]$ with the Skorohod topology which is defined by the metric
\[
\rho_S(\psi,\widetilde{\psi})=\inf_{s\in
S}\left(\sup_{t\in[0,1]}|\psi(t)-\widetilde{\psi}(s(t))|+\sup_{t\in[0,1]}|t-s(t)|\right),\;\;\psi,\widetilde{\psi}\in
D[0,1],
\]
where $S$ is the family of strictly increasing, continuous
mappings $s$ of $[0,1]$ onto itself such that $s(0)=0$ and
$s(1)=1$  \citep[Section 14]{billingsley68}. The metric space
$(D[0,1],\rho_S)$ is separable and is not complete, but there is
an equivalent metric on $D[0,1]$ which turns $D[0,1]$ with the
Skorohod topology into a complete separable metric space. Since
the Skorohod topology and the uniform topology on $C[0,1]$
coincide, $W$ can be considered as a measure on $D[0,1]$.

A stronger result than the CLT is a {\it weak invariance
principle},  also called a {\bf functional central limit theorem
(FCLT)}. Let $(\zeta_j)_{j\ge 1}$ be a sequence of real-valued
zero-mean random variables with finite variance. Let $\sigma>0$
and  define the process $\{\psi_n(t),t\in[0,1]\}$ by
\[
\psi_n(t)=\frac{1}{\sigma\sqrt{n}}\sum_{j=1}^{[nt]}\zeta_j \;\;\mbox{for}\;\;t\in [0,1],
\]
(where the sum from $1$ to $0$ is set equal to $0$). Note that
$\psi_n$ is a right continuous step function, a random variable of
$D[0,1]$ and $\psi_n(0)=0$. If
\[
\psi_n\to^d w
\]
(here the convergence in distribution is in $D[0,1]$), then $(\zeta_{j})_{j\ge 1}$ is said to satisfy the FCLT.

If for every $k\ge 1$ and every vector $(t_1,\ldots,t_k)\in[0,1]^k$ with $t_1<\ldots<t_k$ the joint distribution of the  vector
$(\psi_n(t_1),\ldots,\psi_n (t_k))$ converges to the joint distribution of $(w(t_1),\ldots,w(t_k))$, then we say that the {\it finite
dimensional distributions} of $\psi_n$ converge to those of $w$. For one dimensional distribution this convergence is equivalent to the central
limit theorem.

The convergence of all finite-dimensional distributions of $\psi_n$ to those of $w$ is not sufficient to
conclude that $\psi_n\to^d w$ in $D[0,1]$. According to Theorems 15.1 and 15.5 of \cite
{billingsley68} if, additionally, 
for each positive $\epsilon$ \be \lim_{\delta\to
0}\limsup_{n\to\infty}\Pr(\sup_{|t-s|\le\delta}|\psi_n(s)-\psi_n(t)|>\epsilon)=0,\label{tight}
\ee then $\psi_n$ converges in distribution to the Wiener process
$w$.

The term {\it functional central limit theorem} comes from the mapping theorem \citep[Theorem 5.1]{billingsley68}, according to which for any
functional $f: D[0,1]\to \realnos$, measurable and continuous on a set of Wiener measure $1$, the distribution of $f(\psi_n)$ converges weakly
to the distribution of $f(w)$. This applies in particular to the functional $f(\psi)=\sup_{0\le s\le 1} \psi(s)$. Instead of real-valued
functionals one can also consider mappings with values in a metric space.  For example, this theorem applies for any $f: D[0,1]\to D[0,1]$ of
the form $f(\phi)(t)=\sup_{s\le t} \phi(s)$ or $f(\phi)(t)=\int_0^t \phi(s)ds$.

\subsubsection{FCLT for noninvertible maps}
\label{s:CLT}

How can we obtain, and in what sense, Brownian-like motion from a
(semi) dynamical system? This question is intimately connected
with Central Limit Theorems (CLT) for non-invertible systems and
various invariance principles.

Many CLT results and invariance principles for maps have been
proved, see {\it e.g.} the survey \cite{denker89}. These results
extend back over some decades, including contributions by
\citet{ratner73}, \citet{boyarsky79}, \citet{wong79},
\citet{keller80}, \citet{jab83}, \citet{jab91}, \citet{liverani}
and \citet{viana}.

First, however, remember that if we have a time series $y(j)$ and
a bounded integrable function $h:X \rightarrow R$, then the {\bf
correlation} of $h$ is defined as
\[
R_{h}(n) = \lim_{N \to \infty} \dfrac 1N \sum_{j=0}^{N-1}
h(y(j+n)) h(y(j)).
\]
If the time series is generated by a measurable transformation
$T:Y\to Y$ operating on a normalized measure space
$(Y,\mathcal{B},\nu)$, and if further $\nu$  is invariant under
$T$ and $T$ is ergodic, then we can rewrite the correlation as
\[
R_{h}(n) = \lim_{N \to \infty} \dfrac 1N \sum_{j=0}^{N-1}
h(y(j+n)) h(y(j)) = \int h(y)h(T^n(y))\nu(dy).
\]
The average $<h>$ is just
\[
<h> = \lim_{N \to \infty} \dfrac 1N \sum_{j=0}^{N-1} h(y(j))  =
\int h(y)\nu(dy).
\]

Let $(Y,\mathcal{B},\nu)$ be a normalized measure space, and
$T:Y\to Y$ be a measurable transformation such that $\nu$ is
$T$-invariant. $(Y,\mathcal{B},\nu)$ will serve as our probability
space $(\Omega,\F,\Pr)$. Let $h\in L^2(Y,\B,\nu)$ be such that
$\int h(y)\nu(dy)=0$. The random variables $\zeta_j=h\circ T^{j-1}$,
$j\ge 1$ are real-valued, have zero-mean and finite variance equal
to $||h||_2^2=\int h^2(y)\nu(dy)$. Thus the terminology from Section
\ref{s:prob} applies. The explicit formulae for the Frobenius-Perron
operators in Section \ref{proband} show that the equations $P_T h=0$
or $\FP_{T,\nu}h=0$ can be easily solved. For instance, in Example
\ref{exp:hat}, every function $h$ which is odd is a solution of
these equations. 
In particular, considering $h$ with $\FP_{T,\nu}h=$ turns out to be
very fruitful. Statistical properties of the sequence $(h\circ
T^j)_{j\ge 0}$ are summarized in the following

\begin{thm} (CLT)\label{CLT1} Let $(Y,\B,\nu)$ be a
normalized measure space and $T:Y\to Y$ be 
ergodic with respect to $\nu$. If $h\in L^2(Y,\B,\nu)$ is such that
$\FP_{T,\nu}h=0$, then \bit \item[(i)] $\int h(y) \nu(dy)=0$ and
$\int h(y)h(T^n(y))\nu(dy)=0$ for all $n\ge 1$. 
\item[(ii)] $\dfrac{\sum_{j=0}^{n-1} h\circ T^j }{\sqrt{n}}\to^d
\sigma N(0,1)$ and $\sigma=||h||_2.$ \item[(iii)] If $\sigma>0$
then $(h\circ T^j)_{j\ge 0}$ satisfies the CLT and FCLT.
\item[(iv)] If $h\in L^\infty(Y,\B,\nu)$ and $\sigma>0$ then all
moments of $\dfrac{\sum_{j=0}^{n-1} h\circ T^j }{\sqrt{n}}$
converge to the corresponding moments of $\sigma N(0,1)$, {\it
i.e.} for each $k\ge 1$ we have
\[
\lim_{n\to\infty}\int \left(\dfrac{\sum_{j=0}^{n-1} h( T^j(y))
}{\sqrt{n}}\right)^{2k}\nu(dy) = \dfrac{(2k)!\sigma^k}{k! 2^k},
\]
\[
\lim_{n\to\infty}\int \left(\dfrac{\sum_{j=0}^{n-1} h( T^j(y))
}{\sqrt{n}}\right)^{2k-1}\nu(dy) = 0.
\]
\eit
\end{thm}
{\noindent\bf Proof.} First note that the transfer operator
$\FP_{T,\nu}$ preserves the integral, {\it i.e} $\int
\FP_{T,\nu}h(y)\nu(dy)=\int h(y)\nu(dy)$. Hence $\int
h(y)\nu(dy)=0$. Now let $n\ge 1$. Since
$\nu$ is a finite measure, the Koopman and transfer 
operators are adjoint on the space $L^2(Y,\B,\nu)$. This implies
\begin{eqnarray*}
 \int h(y)h( T^n(y))\nu(dy) &=& \int  h(y) U_T^n h(y)\nu(dy) \\
   &=& \int \FP_{T,\nu}h(y)U_T^{n-1}h(y)\nu(dy)=0
\end{eqnarray*}
and completes the proof of (i). Part (ii) follows from Lemma
\ref{mdarr} and Theorem \ref{CLGen} in Appendix, since for each
$n\ge 1$ we have
\[\frac{1}{\sqrt{n}}\sum_{j=0}^{n-1}h\circ
T^j=\frac{1}{\sqrt{n}}\sum_{j=1}^{n}h\circ T^{n-j}.
\]
If $\sigma>0$, then the CLT is a consequence of part (ii), while
the FCLT follows from Lemma \ref{finitedim} and \ref{FCLGen} in
Appendix. Finally, the existence and convergence of moments follow
from Theorem 5.3 and 5.4 of \cite{billingsley68} and from Lemma
\ref{mdarr} in Appendix.


\begin{rem}Note that if $T$ is invertible then the equation
$\FP_{T,\nu}(h)=0$ has only a zero solution, so the theorem does not
apply.
\end{rem}

We now address the question of solvability of the equation
$\FP_{T,\nu}(h)=0$.

\begin{prop}
Let $(Y,\B,\nu)$ be a normalized measure space and $T:Y\to Y$ be a
measurable map preserving the measure $\nu$. Let $Y=Y_1\cup Y_2$
with $Y_1, Y_2\in\B$ and $\nu(Y_1\cap Y_2)=0$ and let a bijective
map $\varphi:Y_1\to Y_2$ be such that both $\varphi$ and
$\varphi^{-1}$ are measurable and preserve the measure $\nu$. Assume
that for every $A\in \B$ there is $B\in\B$ such that $B\subseteq
Y_1$ and \be T^{-1}(A)=B\cup \varphi(B). \label{turning}\ee If
$h(y)+h(\varphi(y))=0$ for almost every $y\in Y_1$, then
$\FP_{T,\nu}h=0$.
\end{prop}
{\noindent\bf Proof.} From Condition \ref{turning} it follows that
\[
\int_{T^{-1}(A)}h(y) \nu(dy)=\int_{B}h(y)\nu(dy)+ \int_{\varphi(B)}h(y)\nu(dy).
\]
Since
\[
\int_{\varphi(B)}h(y)\nu(dy)=\int_{\varphi(B)}h(\varphi^{-1}(\varphi(y)))\nu(dy),
\]
the last integral is equal to
\[
\int_{B}h(\varphi(y))(\nu\circ\varphi)(dy)=\int_{B}h(\varphi(y))\nu(dy)
\]
by the change of variables applied to $\varphi^{-1}$ and finally by
the invariance of $\nu$ for $\varphi$. This, with the definition of
$\FP_{T,\nu}$, completes the proof.

\begin{rem}
The above proposition can be easily generalized. For example, we can have
\[
T^{-1}(A)=B\cup \varphi_1(B)\cup \varphi_2(B)
\]
with the sets $B, \varphi_1(B), \varphi_2(B)$ pairwise disjoint.

Note that if  $Y$ is an interval then it is enough to check
Condition \ref{turning} for intervals of the form $[a,b)$.
\end{rem}

\begin{exmp}\label{exp:sol}
{\em For an even transformation $T$ on $[-1,1]$ with an  even
invariant density we can take $Y_1=[-1,0]$ and $\varphi(y)=-y$. In
this case $\FP_{T,\nu}h=0$ for every odd function on $[-1,1]$. In
particular, this applies to the tent map and to the Chebyshev maps
$S_N$ of Example \ref{cheby} with $N$ even. We also have
$\FP_{S_N,\nu}h=0$ when $h(y)=y$ and $S_N$ is the Chebyshev map with
$N$ odd. Indeed, first observe that by Theorem \ref{conjugate} we
have $\FP_{S_N,\nu}h=0$ if $P_{T_N}f=0$ where $T_N$ is given by
\ref{genhat} and $f(y)=\cos(\pi y)$ for $y\in[0,1]$ and then note
that $P_{T_N}f=0$ follows because the expression  \[f(y)+
\sum_{n=1}^{(N-1)/2}\left(f\left(\frac{2n}{N}+y\right)+f\left(\frac{2n}{N}-y\right)\right)
\] reduces to
\[ \cos(\pi
y)(1+2\sum_{n=1}^{(N-1)/2}\cos\left(\frac{2n\pi}{N}\right))\] which
is equal to $0$. For the dyadic map, we can take $\varphi(y)=y+1$.
Then any function satisfying  $h(y)+h(y+1)=0$ gives a solution to
$\FP_{T,\nu}h=0$. }
\end{exmp}

The next example shows that the assumption of ergodicity in Theorem \ref{CLT1} is in a sense essential.

\begin{exmp}
{\em Let $T:[0,1]\to[0,1]$ be defined by
\[
T(y)=\left\{
\begin{array}{ll}
2y , & y \in  [0,\frac{1}{4} )\\ 2y -\frac{1}{2}, & y
\in  [\frac{1}{4}, \frac{3}{4} ),\\
2y-1, & y\in[\frac{3}{4},1].
\end{array}
\right.
\]
The Frobenius-Perron operator is given by
\[
P_Tf(y)=\frac{1}{2}f\left(\frac{1}{2}y\right)1_{[0,\frac{1}{2})}(y) +
\frac{1}{2}f\left(\frac{1}{2}y+\frac{1}{4}\right)+ \frac{1}{2}f\left(\frac{1}{2}y +
\frac{1}{2}\right)1_{[\frac{1}{2},1]}(y).
\]
Observe that the Lebesgue measure on $([0,1],\B([0,1]))$ is invariant for $T$ and that $T$ is not ergodic since
$T^{-1}([0,\frac{1}{2}])=[0,\frac{1}{2}]$ and $T^{-1}([\frac{1}{2},1])=[\frac{1}{2},1]$.
Consider the following functions
\[
h_1(y)=\left\{
\begin{array}{ll}
1 , & y \in  [0,\frac{1}{4} )\\ -1, & y
\in  [\frac{1}{4}, \frac{3}{4} ),\\
1, & y\in[\frac{3}{4},1],
\end{array}
\right.\;\; h_2(y)=\left\{
\begin{array}{ll}
1 , & y \in  [0,\frac{1}{4} )\\ -1, & y
\in  [\frac{1}{4}, \frac{1}{2} ),\\
-2, & y
\in  [\frac{1}{2}, \frac{3}{4} ),\\
2, & y\in[\frac{3}{4},1].
\end{array}
\right.
\]
We see at once
 that $P_Th_1=P_Th_2=0$, $P_Th_1^2=h_1^2$, and $P_Th_2^2=h_2^2$. It is immediate that for every $y\in[0,1]$ we have $h_1^2(T
(y))=h_1^2(y)=1$ and $h_2^2(T(y))=h_2^2(y)$. Lemma \ref{mdarr} and
Theorem \ref{CLEgl} in Appendix show that
\[\frac{1}{\sqrt{n}}\sum_{j=0}^{n-1}h_1\circ T^j\to^d N(0,1),\] while
\[
\frac{1}{\sqrt{n}}\sum_{j=0}^{n-1}h_2\circ T^j\to^d \zeta,
\]
where $\zeta$ has the characteristic function of the form
\[
\phi_\zeta(r)=\int_{0}^{1}\exp\left(-\frac{r^2}{2}h_2^2(y)\right)dy.
\]
The density of $\zeta$ is equal to
\[
\frac{1}{2}\frac{1}{\sqrt{2\pi}}\exp \left(-\frac{x^2}{2}\right)+ \frac{1}{2}\frac{1}{\sqrt{8\pi}}\exp
\left(-\frac{x^2}{8}\right),\;\;x\in\realnos.
\]
Consequently, the sequence $(h_2\circ T^j)_{j\ge 0}$ does not satisfy the CLT.}
\end{exmp}

We now discuss the problem of changing the underlying probability space for the sequence $(h\circ T^j)_{j\ge
0}$. The random variables $h\circ T^j$ in Theorem \ref{CLT1} are defined on the probability space $(Y,\B,\nu)$.
Since $\nu$ is invariant for $T$, they have the same distribution and constitute a stationary sequence. We shall
show that the result (iii) of Theorem \ref{CLT1} remains true if the transformation $T$ is exact and $h\circ
T^j$ are random variables on $(Y,\B,\nu_0)$ where $\nu_0$ is an arbitrary normalized measure absolutely
continuous with respect to $\nu$. In other words we can consider random variables $h(T^j(\xi_0))$ with $\xi_0$
distributed according to $\nu_0$. Now these random variables are not identically distributed and constitute a
non-stationary sequence. For example, consider the hat map $T(y)=1-2|y|$ on $[-1,1]$ and $h(y)=y$. Then Theorem
\ref{CLT1} applies if $\xi_0$ is uniformly distributed. We will show that we can also consider $\xi_0$ having a
density with respect to the normalized Lebesgue measure on $[-1,1]$.

\begin{thm}\label{CLTE}
Let $(Y,\B,\nu)$ be a normalized measure space and $T:Y\to Y$ be
exact with respect to $\nu$. Let $h\in L^2(Y,\B,\nu)$ be such that
$\FP_{T,\nu}h=0$ and $\sigma=||h||_2>0$. If $\xi_0$ is distributed
according to a normalized measure $\nu_0$ on $(Y,\B)$ which is
absolutely continuous with respect to $\nu$, then $(h\circ
T^j(\xi_0))_{j\ge 0}$ satisfies both the CLT and FCLT.
\end{thm}
{\noindent \bf Proof.} Let $g_0$ be the density of the measure $\nu_0$ with respect to $\nu$. On the probability
space $(Y,\B)$ define the random variables $\zeta_n$ by
\[
\zeta_n(y)=\frac{1}{\sigma \sqrt{n}}\sum_{j=0}^{n-1} h( T^j(y)), \;\;y\in Y.
\]
To prove the CLT we shall use the continuity theorem
\citep[Theorem 7.6]{billingsley68}, and to do so we must show that
\[
\lim_{n\to\infty}\int \exp(i r \zeta_{n}(y))g_0(y)\nu(dy)=\exp \left(-\frac{r^2}{2} \right),\;\;r\in\realnos .
\]
Fix $\epsilon>0$. Since $T$ is exact, there exists $m\ge 1$ such that
\[
\int|\FP_{T,\nu}^mg_0(y)-1|\nu(dy)\le \epsilon.
\]
Define
\[\tilde{\zeta}_{n}(y)=\frac{1}{\sigma \sqrt{n}}\sum_{j=0}^{n-m-1}
h(T^j(y)),\;\;n> m\] and observe that
\[
\zeta_n=\frac{1}{\sigma \sqrt{n}}\sum_{j=0}^{m-1} h\circ T^j+ \tilde{\zeta}_{n}\circ T^m
\]
for sufficiently large $n$. Since for every $y$ the sequence $\frac{1}{\sigma \sqrt{n}}\sum_{j=0}^{m-1}
h(T^j(y))$ converges to $0$ as $n\to \infty$, we obtain
\[
\lim_{n\to\infty}\left|\int \exp(ir \zeta_n(y)) g_0(y)\nu(dy)-\int \exp(ir
 \tilde{\zeta}_n(T^m(y))) g_0(y)\nu(dy)\right|=0.
\]
The equality $\exp(ir\tilde{\zeta}_n\circ T^m)=U_{T}^m(\exp(i r \tilde{\zeta}_n))$ implies that
\[\int \exp(ir
 \tilde{\zeta}_n(T^m(y))) g_0(y)\nu(dy)=\int \exp(ir
 \tilde{\zeta}_n(y)) \FP_{T,\nu}^m g_0(y)\nu(dy),\]
since the operators $U_T$ and $\FP_{T,\nu}$  are adjoint. This gives
\[\left|\int \exp(ir
 \tilde{\zeta}_n(T^m(y)))g_0(y)\nu(dy)-\int \exp(ir
 \tilde{\zeta}_n(y))\nu(dy)\right|\le \int|\FP_{T,\nu}^mg_0(y)-1|\nu(dy)\le
 \epsilon.
\]
From Theorem \ref{CLT1} and the continuity theorem \citep[Theorem
7.6]{billingsley68}, it follows that
\[
\lim_{n\to\infty}\int \exp(ir
 \tilde{\zeta}_n(y))\nu(dy)=\exp \left (-\frac{r^2}{2}\right ).
\]
Consequently,
\[
\limsup_{n\to\infty}\left|\int \exp(ir \zeta_n(y)) g_0(y)\nu(dy)-\exp \left( -\frac{r^2}{2}\right)\right|\le \epsilon,
\]
which leads to the desired conclusion, as $\epsilon$ was arbitrary.

Similar arguments as above, the multidimensional version of the
continuity theorem, and Lemma \ref{finitedim} in Appendix allow us
to show that the finite dimensional distributions of
\[
\psi_n(t)=\frac{1}{\sigma \sqrt{n}}\sum_{j=0}^{[nt]-1} h(T^j(\xi_0))
\]
converge to the finite dimensional distributions of the Wiener
process $w$. By Lemma \ref{FCLGen} in Appendix Condition
\ref{tight} holds with $\Pr=\nu$. Since $\nu_0$ is absolutely
continuous with respect to $\nu$, it is easily seen that this
Condition also holds with $\Pr=\nu_0$, which completes the proof.


Does the CLT still hold when  $h$ does not satisfy the equation
$\FP_{T,\nu}h=0$? The answer to this question is positive provided
that $h$ can be written as a sum of two functions in which one
satisfies the assumptions of Theorem \ref{CLT1} while the other is
irrelevant for the Central Limit Theorem to hold.
This idea goes back to \cite{gordin69}. 

\begin{thm}\label{p:CLT}
Let $(Y,\B,\nu)$ be a normalized measure space, $T:Y\to Y$ be
ergodic with respect to $\nu$, and $h\in L^2(Y,\B,\nu)$ be such that
$\int h(y)\nu(dy)=0$. If there exists $\tilde{h}\in L^2(Y,\B,\nu)$
such that $\FP_{T,\nu}\tilde{h}=0$ and  the sequence
$\frac{1}{\sqrt{n}}\sum_{j=0}^{n-1}(h-\tilde{h})\circ T^j$ is
convergent in $L^2(Y,\B,\nu)$ to $0$, then \be
\lim_{n\to\infty}\frac{||\sum_{j=0}^{n-1}h\circ
T^j||_2^2}{n}=||\tilde{h}||_2^2 \label{htildelimit}\ee and
\[
\frac{1}{\sqrt{n}}\sum_{j=0}^{n-1}h\circ T^j\to^d||\tilde{h}||_2
N(0,1).
\]
Moreover, if the series $\sum_{j=1}^\infty\int
h(y)h(T^j(y))\nu(dy)$ is convergent, then \be ||\tilde{h}||_2^2=
\int h^2(y)\nu(dy)+2\sum_{n=1}^\infty\int h(y)h(T^n(y))\nu(dy).
\ee
\end{thm}

{\noindent\bf Proof.} We have \be \dfrac{\sum_{j=0}^{n-1} h\circ T^j
}{\sqrt{n}}=\dfrac{\sum_{j=0}^{n-1} \tilde{h}\circ T^{j}}{\sqrt{n}}+
\dfrac{\sum_{j=0}^{n-1}(h- \tilde{h})\circ
T^{j}}{\sqrt{n}}.\label{sumforclt} \ee Since
$\FP_{T,\nu}\tilde{h}=0$, we obtain $\int \tilde{h}(y)\nu(dy)=0$ and
$\int \tilde{h}(T^i(y))\tilde{h}(T^j(y))\nu(dy)=0$ for $i\neq j$ by
Condition (i) of Theorem \ref{CLT1}. Hence
\[||\tilde{h}||_2^2=\frac{1}{n}\sum_{j=0}^{n-1}||\tilde{h}\circ
T^j||_2^2=||\frac{1}{\sqrt{n}}\sum_{j=0}^{n-1}\tilde{h}\circ
T^j||_2^2\] and therefore Equation \ref{htildelimit} holds. Since
the sequence $\dfrac{1}{\sqrt{n}}\sum_{j=0}^{n-1}(h- \tilde{h})\circ
T^{j}$ is  convergent to $0$ in $L^2(Y,\B,\nu)$, it is also
convergent to $0$ in probability. Equation \ref{sumforclt},
Condition (ii) of Theorem \ref{CLT1} applied to $\tilde{h}$, and
property (\ref{slutsky})  complete the proof of the first part.

It remains to show that
\[\lim_{n\to\infty}\frac{||\sum_{j=0}^{n-1}h\circ
T^j||_2^2}{n}=\int h^2(y)\nu(dy)+2\sum_{n=1}^\infty\int
h(y)h(T^n(y))\nu(dy).\] Since $\nu$ is $T$-invariant, we have
\[
\frac{1}{n}\int
\left(\sum_{j=0}^{n-1}h(T^j(y))\right)^2\nu(dy)=||h||_2+2\frac{1}{n}\sum_{j=1}^{n-1}\sum_{l=1}^{j}
\int h(y) h(T^l(y))\nu(dy),
\]
but the sequence  $(\sum_{j=1}^n\int h(y)h(T^j(y))\nu(dy))_{n\ge
1}$ is convergent to $\sum_{j=1}^\infty\int h(y)h(T^j(y))\nu(dy)$,
which completes the proof.

\begin{rem}
Note that in the above proof of the CLT we only used the weaker
condition that the sequence
$\dfrac{1}{\sqrt{n}}\sum_{j=0}^{n-1}(h- \tilde{h})\circ T^{j}$
 is convergent to
$0$ in probability. The stronger assumption that this sequence is
convergent in $L^2(Y,\B,\nu)$ was  used to derive Equation
\ref{htildelimit}.  Note also that all of the computations
are useless when $||\tilde{h}||_2=0$ and the most interesting
situation is when $\tilde{h}$ is nontrivial, {\it i.e.}
$||\tilde{h}||_2>0$.
\end{rem}

Strengthening the assumptions of the last theorem leads to the
functional central limit theorem.

\begin{thm}\label{t:FCLT}
Let $(Y,\B,\nu)$ be a normalized measure space, $T:Y\to Y$ be
ergodic with respect to $\nu$, and $h\in L^2(Y,\B,\nu)$ be such that
$\int h(y)\nu(dy)=0$. If there exists a nontrivial $\tilde{h}\in
L^2(Y,\B,\nu)$ such that $\FP_{T,\nu}\tilde{h}=0$ and  the sequence
$ \frac{1}{\sqrt{n}}\sum_{j=0}^{n-1}(h-\tilde{h})\circ T^j $ is
$\nu$-a.e. convergent to $0$, then $(h\circ T^j)_{j\ge 0}$ satisfies
the FCLT.
\end{thm}

{\noindent\bf Proof.} Since every sequence convergent $\nu$ almost
everywhere is convergent in probability, the CLT follows by the
preceding Remark and Theorem \ref{p:CLT}. To derive the FCLT
define
\[
\widetilde{\psi}_n(t)=\frac{1}{\sigma\sqrt{n}}\sum_{j=0}^{[nt]-1}\tilde{h}\circ
T^j\;\;\mbox{and}\;\;\psi_n(t)=\frac{1}{\sigma\sqrt{n}}\sum_{j=0}^{[nt]-1}h\circ
T^j,\;\;t\in[0,1],
\]
where $\sigma=||\tilde{h}||_2$. Then by (iii) of Theorem
\ref{CLT1} we have
\[
\widetilde{\psi}_n\to^d w.
\]
By property (\ref{slutsky}) it remains to show that
\[
\rho_S(\psi_n,\widetilde{\psi}_n)\to^P 0.
\]
To this end observe that
\[
\rho_S(\psi_n,\widetilde{\psi}_n)  \le  \sup_{0\le t\le
1}|\psi_n(t)-\widetilde{\psi}_n(t)|\le
\frac{1}{\sigma\sqrt{n}}\max_{1\le k\le
n}|\sum_{j=0}^{k-1}(h-\tilde{h})\circ T^j|.
\]
Since the sequence $
\frac{1}{\sqrt{n}}\sum_{j=0}^{n-1}(h-\tilde{h})\circ T^j $ is
$\nu$-a.e. convergent to $0$, the same holds for the sequence
$\frac{1}{\sigma\sqrt{n}}\max_{1\le k\le
n}|\sum_{j=0}^{k-1}(h-\tilde{h})\circ T^j| $ by an elementary
analysis, which completes the proof.

\begin{rem}
With the settings and notation of Theorem \ref{p:CLT} and
respectively Theorem \ref{t:FCLT}, if $T$ is exact, then the same
conclusions hold for the sequence $(h\circ T^j(\xi_0))_{j\ge 0}$
provided that $\xi_0$ is distributed according to a normalized
measure $\nu_0$ on $(Y,\nu)$ which is absolutely continuous with
respect to $\nu$. Indeed, since convergence to zero in probability
is preserved by an absolutely continuous change of measure, we can
apply the above arguments again, with Theorem \ref{CLT1} replaced
by Theorem \ref{CLTE}.
\end{rem}

One situation when all assumptions of the two preceding theorems
are met is described in the following

\begin{thm} 
\label{FCLT1} Let $(Y,\B,\nu)$ be a normalized measure space,
$T:Y\to Y$ be ergodic with respect to $\nu$, and $h\in
L^2(Y,\B,\nu)$. Suppose that the series \[ \sum_{n=0}^\infty
\FP_{T,\nu}^n h \] is convergent in $L^2(Y,\B,\nu)$. Define $
f=\sum_{n=1}^\infty \FP_{T,\nu}^n h$ and $\tilde{h}=h+f- f\circ T$.

Then $\tilde{h}\in L^2(Y,\B,\nu)$, $\FP_{T,\nu}\tilde{h}=0$, $\int
h(y)\nu(dy)=0$, and the sequence
$(\frac{1}{\sqrt{n}}\sum_{j=0}^{n-1}(h-\tilde{h})\circ T^j)_{n\ge
1}$ is convergent to $0$ both in $L^2(Y,\B,\nu)$ and $\nu-$a.e.

In particular, $||\tilde{h}||_2=0$ if and only if $h=f\circ T-f$
for some $f\in L^2(Y,\B,\nu)$.
\end{thm}

{\noindent\bf Proof.} Since $\FP_{T,\nu}(h+f)=f$, we have by
Equation \ref{fpcond}
\[\FP_{T,\nu}\tilde{h}=\FP_{T,\nu}(h+f)-\FP_{T,\nu} (f\circ T) = f-\FP_{T,\nu} U_Tf=0.
\]
Thus it remains to study the behavior of the sequence
$\frac{1}{\sqrt{n}}\sum_{j=0}^{n-1}(h-\tilde{h})\circ T^j$, which
with our notations reduces to $\frac{1}{\sqrt{n}}(f\circ T^n-f)$.
This sequence is obviously convergent to $0$ in $L^2(Y,\B,\nu)$
because $||f\circ T^n-f||_2\le 2||f||_2$. It is also $\nu$-a.e.
convergent to $0$ which follows from the Borel-Cantelli lemma and
the fact that for every $\epsilon>0$ the series $\sum_{n=1}^\infty
\nu( f^2\circ T^n\ge n\epsilon)=\sum_{n=1}^\infty \nu( f^2\ge
n\epsilon)$ is convergent as $f\in L^2(Y,\B,\nu)$, which completes the proof. 

Summarizing our considerations for general $h$ we arrive at the
following sufficient conditions for the CLT and FCLT to hold.

\begin{cor}\label{FCLT2}
Let $(Y,\B,\nu)$ be a normalized measure space, $T:Y\to Y$ be
ergodic with respect to $\nu$, and $h\in L^2(Y,\B,\nu)$. If  \be
\sum_{n=0}^\infty ||\FP_{T,\nu}^n h||_2<\infty, \label{conclt}\ee
then $\sigma\ge 0$ given by
\[\sigma^2=\int h^2(y)\nu(dy)+2\sum_{n=1}^\infty\int
h(y)h(T^n(y))\nu(dy)\label{defsigma}\] is finite 
%
and $(h\circ T^j)_{j\ge 0}$ satisfies the CLT and FCLT provided
that $\sigma>0$.
\end{cor}
{\noindent\bf Proof.} Since the operators $\FP_{T,\nu}$ and $U_T$
are adjoint on the space $L^2(Y,\B,\nu)$, we have
\[
\int h(y)h(T^n(y))\nu(dy)=\int \FP_{T,\nu}^nh(y) h(y)\nu(dy).
\]
Thus
\[
\left|\int h(y)h(T^n(y))\nu(dy)\right|\le \int
|\FP_{T,\nu}^nh(y)h(y)|\nu(dy)\le ||\FP_{T,\nu}^nh||_2||h||_2
\]
by Schwartz's inequality. Hence
\[
\sum_{n=1}^\infty\int |h(y)h(T^n(y))|d\nu\le||h||_2\sum_{n=1}^\infty
||\FP_{T,\nu}^nh||_2,
\]
which shows that the series $\sum_{n=1}^\infty\int
h(y)h(T^n(y))\nu(dy)$ is convergent. Since assumption \ref{conclt}
implies that the series $\sum_{n=0}^\infty \FP_{T,\nu}^n h$ is
absolutely convergent in $L^2(Y,\B,\nu)$, the assertions follow from
Theorems \ref{p:CLT}, \ref{t:FCLT}, and \ref{FCLT1}.

\begin{rem}
Note that if Condition \ref{conclt} holds then
\[
\lim_{n\to\infty}||\FP_{T,\nu}^nh||_2=0.
\]
Since $\nu$ is finite, we have
\[
\lim_{n\to\infty}||\FP_{T,\nu}^nh||_1=0.
\]
Therefore the validity of Condition \ref{conclt} on a dense subset
of $\{h\in L^1(Y,\B,\nu):\int h(y)\nu(dy)=0\}$ implies that $T$ is
exact.
\end{rem}

Assume that $Y$ is an interval $[a,b]$ in $\realnos$ for some $a, b$. Recall that a function
$h:[a,b]\to\realnos$ is said to be of {\it bounded variation} if
\[
\bigvee\limits_{a}^{b}h=\sup\sum_{i=1}^{n}|h(y_{i-1})-h(y_i)|<\infty,
\]
where the supremum is taken over all finite partitions,
$a=y_0<y_1<...<y_n=b$, $n\ge 1$, of $Y$. 

Let $V([0,1])$ denote the space of all integrable functions with
bounded variation over $[0,1]$ such that $\int_{0}^1 h(y) dy=0$.
We have \be |h(y)|\le \bigvee\limits_{0}^{1}h
\;\;\mbox{for}\;\;h\in V([0,1]),\;y\in[0,1].\label{supvar} \ee

\begin{exmp}
{\em For the continued fraction map there exists a positive
constant $c<1$ such that for every function $h$ of bounded
variation over $[0,1]$ we have \citep[Corollary, p. 904]{iosif92}
\[
\bigvee\limits_{0}^{1}\FP_{T,\nu}^n h\le c^n \bigvee\limits_{0}^{1}h
\;\;\mbox{for all}\;\; n\ge 1,
\]
where $\nu$ is Gauss's measure with density $g_*$ as in Example \ref{confr}. From this and Condition
\ref{supvar} it follows that for every $h\in V([0,1])$
\[
||\FP_{T,\nu}^n h||_{2}\le \sup_{y\in [0,1]}|\FP_{T,\nu}^nh(y)|\le
c^n \bigvee\limits_{0}^{1}h.
\]
Consequently,  Condition \ref{conclt} is satisfied and Corollary \ref{FCLT2} applies. }
\end{exmp}

By definition, the Frobenius-Perron operator is a linear operator from $L^1([0,1])$ to $L^1([0,1])$, but for
sufficiently smooth piecewise monotonic maps it can be defined as a pointwise map of $V([0,1])$ into $V([0,1])$.
Since functions of bounded variation have only countably many points of discontinuity, redefining $P_T$ at those
points does not change its $L^1$ properties. If, moreover, one is able to give an estimate for the iterates of
$P_T$ in the bounded variation norm
\[
||h||_{BV}=\bigvee\limits_{0}^{1}h + \int_{0}^{1}|h(y)|dy,
\]
then obviously one is able to estimate the norm of $P_{T}^n f$ in all $L^p([0,1])$ spaces. In many cases there
exist $c_1,c_2>0$ and $r\in(0,1)$ such that \be ||P_{T}^n h||_{BV}
 \le c_1 r^n
(\bigvee\limits_{0}^{1}h +c_2 ||h||_1),\;\;h\in V([0,1]).
\label{rateofconv}\ee 

We now describe two classes of chaotic maps for which one can easily
show that Condition \ref{conclt} holds for every $h\in V([0,1])$.
Consider a transformation $T:[0,1]\to [0,1]$ having the following
properties \bit \item[(i)] there is a partition
$0=a_0<a_1<...<a_l=1$ of $[0,1]$ such that for each integer
$i=1,...,l$ the restriction of $T$ to $[a_{i-1},a_i)$ is continuous
and convex, \item[(ii)] $T(a_{i-1})=0$ and $T'(a_{i-1})>0$,
\item[(iii)] $T'(0)>1$. \eit For such transformation the
Frobenius-Perron operator has a unique fixed point $g_*$, where
$g_*$ is of bounded variation and a decreasing function of $y$
\citep{almcmbk94}. Moreover it is bounded from below when, for
example, $T([0,a_1])=[0,1]$. It is known \citep{jabmal83} that the
estimate in Equation \ref{rateofconv} holds for the Frobenius-Perron
operator $P_T$ for transformations with these three properties.
Suppose that $g_*(y)>0$ for a.a. $y\in[0,1]$. Since
$||\FP_{T,\nu}||_\infty\le ||f||_\infty$ for all $f\in L^\infty
([0,1],\B,\nu)$, we have for all $h\in f\in L^\infty ([0,1],\B,\nu)$
\be ||\FP_{T,\nu}^n h||_{2}\le||h||_\infty^{1/2} ||\FP_{T,\nu}^n
h||_{1}^{1/2} \ee If $h$ is of bounded variation with $\int
h(y)g_*(y)dy=0$, then $h g_*\in V([0,1])$ and $||\FP_{T,\nu}^n
h||_1=|| P_T^n (hg_*)||_1$. Thus
\[
||\FP_{T,\nu}^n h||_{2}= O(r^{n/2})
\]
by Equation \ref{rateofconv}
and Corollary \ref{FCLT2} applies.

Let a transformation $T:[0,1]\to [0,1]$ be piecewise monotonic,
the function $\frac{1}{|T'(y)|}$ be of bounded variation over
$[0,1]$ and $\inf_{y\in [0,1]}|T'(y)|>1$. For such transformations
the Frobenius-Perron operator has a fixed point $g_*$ and $g_*$ is
of bounded variation \citep{almcmbk94}. Suppose that $P_T$ has a
unique invariant density $g_*$ which is strictly positive. Then
the transformation $T$ is ergodic and $g_*$ is bounded from below.
There  exists $k$ such that $T^k$ is exact, the estimate in
Equation \ref{rateofconv} is valid for the Frobenius-Perron
operator $P_{T^k}$ corresponding to $T^k$, and the following holds
\citep{jabkow85}: There exists $c_1>0$ and $r\in (0,1)$ such that
\[
|P_{T^k}^n f(y)|\le c_1 r^n (\bigvee\limits_{0}^{1}f + \int_{0}^{1}|f(y)|dy),\;\;y\in [0,1],\;\;f\in V([0,1]);
\]
Hence Corollary \ref{FCLT2} applies to $T^k$ and every $h$ of bounded variation with $\int h(y)g_*(y)dy=0$. One
can relax the assumption that $g_*$ is strictly positive and have instead $g_*\ge c$ for a.e. $y\in Y_*=\{y\in
[0,1]: g_*(y)>0\}$. Then the above estimate is valid for $y\in Y_*$ and $f$ with $\mbox{supp}f=\{y\in [0,1]:
f(y)\neq 0\}\subset Y_*$ and Corollary \ref{FCLT2} still applies to $T^k$.
%

%
%
%

We now describe how to obtain the conclusions of Corollary \ref{FCLT2} for conjugated maps. If the Lebesgue
measure on $[0,1]$ is invariant with respect to $T$, then Theorem \ref{conjugate} offers the following.

\begin{cor}\label{conjugateclt}
Let $T:[0,1]\to[0,1]$ be a transformation for which the Lebesgue measure on $[0,1]$ is invariant and  for which
Equation \ref{rateofconv} holds. Let $g_*$ be a positive function and $S:[a,b]\to [a,b]$ be given by
$S=G^{-1}\circ T\circ G$, where
\[
G(x)=\int_{a}^x g_*(y)\,dy, \qquad a\le x\le b.
\]
If  $h:[a,b]\to\realnos$ is a function of bounded variation with $\int h(y)g_*(y)dy=0$, then
\[
\sum_{n=0}^\infty ||\FP_{T,\nu}^n h||_2<\infty,
\]
where $||\cdot ||_2$ denotes the norm in $L^2([a,b],\B([a,b]),\nu)$ and $\nu$ is the measure with density $g_*$.
\end{cor}
{\noindent\bf Proof.} By Theorem \ref{conjugate} we have \[
\FP_{S,\nu} f=U_G P_T U_{G^{-1}} f, \qquad \mbox{for}\;\;f\in
L^1([a,b],\B([a,b]),\nu).
\]
The operator $U_G:L^2([0,1])\to L^2([a,b],\B([a,b]),\nu)$ is an
isometry, thus \[||\FP_{S,\nu}^nh||_2=||P_T^n U_{G^{-1}}
h||_{L^2([0,1])}.\] Since $G$ is increasing, $G^{-1}$ is a function
of bounded variation, as a result $U_{G^{-1}} h=h\circ G^{-1}$ is of
bounded variation over $[0,1]$, which completes the proof.

Finally we discuss the case of quadratic maps. We follow the
formulation in \cite{viana}. Consider the quadratic map $T_\beta$,
$\beta\in(0,2)$, of Example \ref{quadratic} and assume that for
the critical point $c=0$ there are constants $\lambda_c>1$ and
$\alpha>0$ such that $\lambda_c>e^{2\alpha}$ and \bit \item[(i)]
$|(T_\beta^n)'(T(c))|\ge \lambda_c^n$ for every $n\ge 1$;
\item[(ii)] $|T_\beta^n(c)-c|\ge e^{-\alpha n}$ for every $n\ge
1$; \item[(iii)] $T_\beta$ is topologically mixing on the interval
$[T_\beta^2(c),T_\beta(c)]$, i.e. for every interval $I\subset
[-1,1]$ there exists $k$ such that $T_\beta^k(I)\supset
[T_\beta^2(c),T_\beta(c)]$.  \eit \cite{young} shows that there is
a set of parameters $\beta$ close to $2$ for which conditions (i),
(ii), (iii) hold with $\lambda_c=1.9$ and $\alpha=10^{-6}$ and
that the central limit theorem holds for $(h\circ T_\beta^j)$ with
$h$ of bounded variation. \cite{viana} (Section 5) proves that
conditions (i), (ii), (iii) imply all assumptions of our Corollary
\ref{FCLT2} for the transformation $T_\beta$ and functions $h$ of
bounded variation with $\int h(y)\nu_\beta(dy)=0$.


We now use the following generalization of Theorem \ref{FCLT1}
which allow us to have the CLT and FCLT for maps with polynomial
decay of correlations.

\begin{thm}(\cite{tyran})\label{t:FCLTEX}
Let $(Y,\B,\nu)$ be a normalized measure space, $T:Y\to Y$ be
ergodic with respect to $\nu$, and $h\in L^2(Y,\B,\nu)$ be such that
$\int h(y)\nu(dy)=0$. Suppose that \be \sum_{n=1}^\infty
n^{-\frac{3}{2}}||\sum_{k=0}^{n-1}\FP_{T,\nu}^k h||_2<\infty.
\label{concltpo}\ee Then there exists $\tilde{h}\in L^2(Y,\B,\nu)$
such that $\FP_{T,\nu}\tilde{h}=0$ and the sequence
$(\frac{1}{\sqrt{n}}\sum_{j=0}^{n-1}(h-\tilde{h})\circ T^j)_{n\ge
1}$ is convergent in $L^2(Y,\B,\nu)$ to zero and if
\[||\sum_{k=0}^{n-1}\FP_{T,\nu}^k h||_2=O(n^\alpha) \qquad\mbox{with}
\qquad\alpha<\frac{1}{2}\] then
$(\frac{1}{\sqrt{n}}\sum_{j=0}^{n-1}(h-\tilde{h})\circ T^j)_{n\ge
1}$ is convergent $\nu-$a.e. to $0$.
\end{thm}

Now we give a simple result that derives CLT and FCLT from a decay
of correlations.

\begin{cor}(\cite{tyran})\label{c:rapid}
Let $(Y,\B,\nu)$ be a normalized measure space, $T:Y\to Y$ be
ergodic with respect to $\nu$, and let $h\in L^\infty(Y,\B,\nu)$
be such that $\int h(y)\nu(dy)=0$. If there are $\alpha>1$ and
$c>0$ such that \be \left|\int h(y)g(T^n(y))\nu(dy)\right|\le
\frac{c}{n^\alpha}||g||_\infty\label{corrdecay} \ee for all $g\in
L^\infty(Y,\B,\nu)$ and $n\ge 1$, then $\sigma\ge 0$ given by
\[\sigma^2=\int h^2(y)\nu(dy)+2\sum_{n=1}^\infty\int
h(y)h(T^n(y))\nu(dy)\] is finite and 
%
$(h\circ T^j)_{j\ge 0}$ satisfies the CLT and FCLT provided that
$\sigma>0$.
\end{cor}


Only recently  the FCLT was established by \cite{pollicottsharp}
for maps such as the Manneville-Pomeau map of Example \ref{pomeau}
and for H{\"o}lder continuous functions $h$ with $\int
h(y)\nu(dy)=0$ under the hypothesis that $0<\beta<\frac{1}{3}$.
When $0<\beta<\frac{1}{2}$ the CLT was proved by \cite{young99},
where it was shown that in this case condition \ref{corrdecay}
holds. Thus our Corollary \ref{c:rapid} gives both the CLT and the
FCLT for maps satisfying the following: 
\bit
    \item[(i)] $T(0)=0$, $T'(0)=1$, $T$ is increasing and
    piecewise $C^2$ and onto $[0,1]$,
    \item[(ii)]  $\inf _{\epsilon\le y\le
    1}|T'(y)|>1$ for every $\epsilon>0$,
    \item[(iii)] $\lim_{y\to 0}y^{1-\beta}T''(y)\neq 0$.
\eit as the ``tower method" of \cite{young99} gives us the estimate
in Equation \ref{corrdecay} with $\alpha=\dfrac{1}{\beta}-1$ for all
H{\"o}lder continuous $h$ and $g\in L^\infty([0,1],\B,\nu)$ with the
constant $c$ dependent only on $h$.


\subsection{Weak convergence criteria}
\label{technical}

Let  $(X,|\cdot|)$ be a phase space which is either $\realnos^{k}$
or a separable Banach space, and denote by $\M_1$ the space of all
probability measures defined on the $\sigma$-algebra $\B(X)$ of
Borel subsets of $X$. For a real-valued measurable bounded
function $f$, and $\mu\in\M_1$, we introduce the scalar product
notation
\[
 \left\langle f,\mu \right\rangle = \int_{X}f(x)\mu(dx).
\]
One way to characterize weak convergence in $\M_1$ is to use the Fortet-Mourier metric in $\M_1$, which is
defined by
\[
d_{FM}(\mu_1,\mu_2)=\sup\{| \left\langle f,\mu_{1}\right\rangle-\left\langle f,\mu_{2} \right\rangle |:
f\in\mathcal{F}_{FM}\} \;\;\mathrm{for}\;\;\mu_{1},\mu_{2}\in\M_{1},
\]
where
\[\mathcal{F}_{FM}=\{f:X\to\realnos: \sup_{x\in X}|f(x)| \le 1, |f|_L\le
1\}\]
 and $|f|_L=\sup_{x\neq
y}\frac{|f(x)-f(y)|}{|x-y|}$. This defines a complete metric on $\M_1$, and we have $\mu_n\to\mu$ weakly if and
only if $d_{FM}(\mu_n,\mu)\to 0$ (cf. \cite{dudley} [Chapter 3])

We further introduce a distance on $\M_{1}$ by
\[
 d(\mu_{1},\mu_{2})=\sup\{| \left\langle f,\mu_{1}\right\rangle-\left\langle f,\mu_{2} \right\rangle |:
 |f|_L\le 1\}\;\;\mathrm{for}\;\;\mu_{1},\mu_{2}\in\M_{1}.
\]
This quantity is always defined, but for some measures it may be infinite. It is easy to check that the function
$d$ is finite for elements of the set
\[
 \M_{1}^{1}=\{\mu\in\M_{1}:
 \int_{X}|x|\mu(dx)<\infty\},
\]
and defines a metric on this set. Moreover, $\M_{1}^{1}$ is a dense subset of $(\M_1,d_{FM})$ and
\[
d_{FM}(\mu_1,\mu_2)\le d(\mu_1,\mu_2).
\]

Let $(Y,\mathcal{B},\nu)$ be a normalized measure space and let
 $R_{n}:X\times Y\to X$ be a measurable transformation for each
 $n\in\naturalnos$. We associate with each transformation $R_n$ an operator
$P_n:\M_1\to\M_1$ defined by \be
 P_{n}\mu(A)=\int_{X}\int_Y 1_A(R_n(x,y))\nu(dy)\mu(dx)
 \ee
for $\mu\in\M_{1}$, where \[ 1_A(x) = \left\{
\begin{array}{ll}
1 & x \in A \\ 0 & x \not\in A
\end{array}
\right. \] is the indicator function of a set $A$. Write
\[
U_nf(x)=\int_Y f(R_n(x,y))\nu(dy)
\]
for measurable functions $f:X\to\realnos$, for which the integral is defined. The operators $U_n$ and $P_n$
satisfy the identity $ <U_nf,\mu>=<f,P_n\mu>$.  Note that if $\mu=\delta_x$, where $\delta_x$ is the point
measure at $x$ defined by \be \delta_x(A) = \left\{
\begin{array}{ll}
1 & x \in A \\ 0 & x \not\in A,
\end{array}
\right.  \ee then $U_nf(x)=<f,P_n\delta_x>$.

\begin{rem}
Note that if $R_n(x,y)$ does not depend on $y$, then
$U_nf=U_{R_n}f$ where $U_{R_n}$ is the Koopman operator
corresponding to $R_n: X\to X$. The following relation holds
between the Frobenius-Perron operator $P_{R_n}$ on
$L^1(X,\B(X),m)$ and the operator $P_n$:  If $\mu$ has a density
$f$ with respect to $m$, then $P_{R_n}f$ is a density of $P_n\mu$.

On the other hand if $R_n(x,y)$ does not depend on $x$, then $U_nf$ is equal to $\int U_{R_n}f(y)\nu(dy),$ where
$U_{R_n}$ is the Koopman operator corresponding to $R_n:Y\to X$. The operator $P_n$ has the same value $\nu\circ
R_n^{-1}$ for every $\mu\in\M_1$.
\end{rem}

Assume that for each $n\in\naturalnos$ the transformation
$R_{n}:X\times Y\to X$ satisfies the following conditions: \bit
\item[(A1)] There exists a measurable function
$L_{n}:Y\to\realnos_{+}$ such that
\[
|R_{n}(x,y)-R_{n}(\bar{x},y)|\le L_{n}(y)|x-\bar{x}|\ \ \mathrm{for}\ \ x,\bar{x}\in X,\ y\in Y.
\]
\item[(A2)] The series
\[
\sum_{n=1}^{\infty}\int_{Y}|R_n(0,T(y))-R_{n+1}(0,y)|\nu(dy)
\]
is convergent, where $T:Y\to Y$ is a transformation preserving the measure $\nu$. \item[(A3)] The integral
$\int_{Y}|R_{n}(0,y)|\nu(dy) $ is finite for at least one $n$. \eit
\begin{prop}\label{proplimit}
Let the transformations $R_n$ satisfy conditions (A1)-(A3). If
\[\lim_{n\to\infty}\int_YL_n(y)\nu(dy)=0,\] then there exists a unique
measure $\mu_*\in\M_1$ such that $(P_n\mu)$ converges weakly to $\mu_*$ for each measure $\mu\in\M_1$.
\end{prop}

\noindent {\bf Proof.} Assumptions (A2) and (A3) imply  that
$P_n(\M_1^1)\subset \M_1^1$.  By the definition of the metric $d$ we
have
\[d(P_n\delta_{0},P_{n+1}\delta_{0})=\sup\{|U_nf(0)-U_{n+1}f(0)|:|f|_L\le
1\}.\] Since the transformation $T$ preserves the measure $\nu$, we
can write \[U_nf(0)=\int_Y f(R_n(0,y)\nu(dy)=\int_Y
f(R_n(0,T(y))\nu(dy)\] for any $f$ with $|f|_L\le 1$. Hence
\[|U_nf(0)-U_{n+1}f(0)|\le
\int_{Y}|R_n(0,T(y))-R_{n+1}(0,y)|\nu(dy).\] Consequently
\[
d(P_n\delta_{0},P_{n+1}\delta_{0})\le \int_{Y}|R_n(0,T(y))-R_{n+1}(0,y)|\nu(dy),
\]
and
\[
d_{FM}(P_n\delta_{0},P_{n+1}\delta_{0})\le
d(P_n\delta_{0},P_{n+1}\delta_{0}).\] From Condition (A2), the
sequence $(P_n\delta_{0})$ is a Cauchy sequence.  Since the space
$(\M_1,d_{FM})$ is complete, $(P_n\delta_{0})$ is weakly convergent
to a $\mu_*\in\M_1$. From (A1) it follows that
\[d(P_n\mu_1,P_n\mu_2)\le \int_YL_n(y)\nu(dy)d(\mu_1,\mu_2).\] Hence
$(P_n\mu)$ is weakly convergent for each $\mu\in\M_1^1$ and has the
limit $\mu_*$. Since, for sufficiently large $n$, each operator
$P_n$ satisfies \[d_{FM}(P_n\mu_1,P_n\mu_2)\le d_{FM}(\mu_1,\mu_2)\]
and the set $\M_1^1$ is dense in $(M_1,d_{FM})$, the proof is
complete.

\setcounter{equation}{0} \setcounter{figure}{0}
\section{Analysis}\label{anal}

We now return to the original problem posed in Section \ref{intro}. We consider the position $(x)$ and velocity
$(v)$ of a
dynamical system defined by \bea \dfrac{dx (t)}{dt}&=&v(t),\label{position} \\
\dfrac{dv(t)}{dt} &= & b(v(t)) + \eta(t),\label{first}  \eea  with
a perturbation $\eta$ in the velocity. We assume that $\eta(t)$
consists of a series of
 delta-function-like perturbations that occur at times $t_0,
 t_1, t_2, \cdots$.  These perturbations have an amplitude $  h(\xi(t))$,
 and $\eta(t)$
 takes the explicit form
 \be \eta(t) =  \kappa \sum_{n=0}^\infty h(\xi(t))
\delta(t-t_n). \label{pert} \ee

We assume that $\xi$ is generated by a dynamical system that at least has an invariant measure for the results
of Section \ref{weakconv} to hold, or is at least ergodic for the Central Limit Theorem to hold as in Section
\ref{linearcase}.

In practice, we will illustrate our results assuming that $\xi$ is
the trace of a  highly chaotic semidynamical system that is,
indeed, even {\bf exact} in the sense of \cite{almcmbk94} (c.f.
Section \ref{sdsys}). $\xi$ could, for example, be generated by
the differential delay equation \be \delta \dfrac{d\xi}{dt} =
-\xi(t) + T(\xi(t-1)), \ee where the nonlinearity $T$ has the
appropriate properties to generate chaotic solutions
(\citet{mcmlg77sci,adhmcm82}). The parameter $\delta$ controls the
time scale for these oscillations, and in the limit as $\delta \to
0$ we can approximate the behavior of the solutions through a map
of the form \be \xi_{n+1}=T(\xi_n). \label{ximap} \ee Thus, we can
think of the map $T$ as being generated by the sampling of a
chaotic continuous time signal $\xi(t)$ as, for example, by the
taking of a Poincar\'e section of a semi-dynamical system
operating in a high dimensional phase space.

Let $(Y,\mathcal{B},\nu)$ be a normalized measure space. Let $b:\realnos^k\to \realnos^k$, $h:Y\to \realnos^k$
be measurable transformations, and let $(t_n)_{n\ge 0}$ be an increasing sequence of real numbers. Assume that
$\xi:\realnos_{+}\times Y\to Y$ is such that $\xi(t_{n+1})=T(\xi(t_n))$ for $n\ge 0$, where $T:Y\to Y$ is a
measurable transformation preserving the measure $\nu$. Combining (\ref{first}) with (\ref{pert}) we have
%
 \be
 \frac{dv(t)}{dt} = b(v(t)) + \kappa \sum_{n=0}^\infty
 h(\xi(t)) \delta(t-t_n) \label{perturbedgen}
 \ee
We say that $v(t)$, $t\ge t_0$, is a solution of Equation \ref{perturbedgen} if, for each
 $n\ge 0$,  $v(t)$ is  a solution of the Cauchy
problem \be \left\{
\begin{array}{lll}
\dfrac{dv(t)}{dt}=& b(v(t)),\;\;t\in(t_n,t_{n+1})\\ v(t_n)= & v(t_n^-)+  \kappa h(\xi(t_n)),
\end{array}
\right. \label{cauchy} \ee where $v_0$ is an arbitrary point of
$\realnos^k$ and 
$v(t_n^-)=\lim_{t\to t_n^{-}} v(t)$ for $n\ge 1$.

Let $\pi:\realnos_{+}\times \realnos^k\to \realnos^k$ be the semigroup generated by the Cauchy problem (if $b$
is a Lipschitz map then $\pi$ is well defined)\be \left\{
\begin{array}{lll}
 \dfrac{d\tilde{v}(t)}{dt}=& b(\tilde{v}(t)),\;\;t>0\\
\tilde{v}(0)= & \tilde{v}_0,\end{array} \right. \label{e:Cauchy}\ee {\it i.e.} for every $\tilde{v}_0\in
\realnos^k$ the unique solution of (\ref{e:Cauchy}) is given by $\tilde{v}(t)=\pi(t,\tilde{v}_0)$ for $t\ge 0$.
As a result, the solution of (\ref{perturbedgen}) is given by
\[
 v(t)=\pi(t-t_{n},v(t_{n})),\;\;\mbox{for}\;\;t\in
 [t_{n},t_{n+1}), n\ge 0.
\]
After integration, for $t\in[t_{n},t_{n+1})$ we have
\[
x(t)-x(t_n)=\int_{t_n}^t v(s)ds =\int_{t_n}^t \pi(s-t_n,v(t_n))ds=\int_{0}^{t-t_n} \pi(s,v(t_n))ds.
\]
Consequently, the solutions of Equations \ref{position} and \ref{first} are given by
\bea x(t)&=& x(t_n) + \int_{0}^{t-t_n} \pi(s,v(t_n))ds, \\
v(t)&=&\pi(t-t_{n},v(t_{n})),\;\;\mbox{for}\;\;t\in
 [t_{n},t_{n+1}), n\ge 0. \eea
Observe that $x(t)$ is continuous in $t$, while $v(t)$ is only right continuous, with left-hand limits, and
$v(t_n)=\lim_{t\to t_n^+}v(t)$.

We are interested in the variables $v(t_n)$, $v_n:=v(t_n^-)$, $\xi_n:=\xi(t_n)$, and $x_n:=x(t_n)$ which appear
in the definition of the solution $v(t)$ and $x(t)$.  We have \bea
v(t_n)&=&v_n+ \kappa h(\xi_n),\label{thev+}\\
v_{n+1}&=&\pi(t_{n+1}-t_n,v(t_n)),\label{thev}\\
\xi_{n+1}&=& T(\xi_{n}),\label{thexi}\\
x_{n+1}&=&x_{n}+ \int_{0}^{t_{n+1}-t_n} \pi(s,v(t_n))ds. \eea We are going to examine the dynamics of these
variables  from a statistical point of view.


Suppose that $v_0$ has a distribution $\mu$, $\xi_0$ has a distribution $\nu$, and that the random variables are
independent. What can we say about the long-term behavior of the distribution of the random variables $v(t_n)$
or $v_n$?


\subsection{Weak convergence of $v(t_n)$ and
$v_n$.}\label{weakconv}

To simplify the presentation and easily use Proposition
\ref{proplimit} of Section \ref{technical},  assume that the
differences $t_{n+1}-t_n$ do not depend on $n$, and that $t_n=n\tau$
for $n\ge 0$. Define $\Lambda:\realnos^k\to \realnos^k$ by
\be\Lambda(v)=\pi(\tau,v), \;\; v\in \realnos^k,\label{deflambda}\ee
where $\pi$ describes the solutions of the unperturbed system as
defined by Equation \ref{e:Cauchy}. In particular, adding chaotic
deterministic perturbations to any exponentially stable system
produces a stochastically stable system as stated in the following

\begin{cor}\label{corollaryweak}
Let $\Lambda:\realnos^k\to \realnos^k$ be a Lipschitz map with a Lipschitz constant $\lambda\in (0,1)$. Let
$T:Y\to Y$ be a transformation preserving the measure $\nu$, and $h:Y\to \realnos^k$ be such that
$\int_Y|h(y)|\nu(dy)<\infty$. Assume that the random variables $v_0$ and $\xi_0$ are independent and that
$\xi_0$ has a distribution $\nu$. Then $v(n\tau)$ converges in distribution to a probability measure $\mu_*$ on
$\realnos^k$ and $\mu_*$ is independent of the distribution of the initial random variable $v_0$. Moreover,
$v_n$ converges in distribution to the probability measure $\mu_*\circ\Lambda^{-1}$.
\end{cor}
\noindent {\bf Proof.} From Equations \ref{thev+} and \ref{thev} it follows that
\[
v((n+1)\tau)=\Lambda(v(n\tau)) +  \kappa h(\xi_{n+1}),\qquad n\ge 0.
\] Define the transformation $R_n:\realnos^k\times Y\to \realnos^k$
recursively: \be \left\{
\begin{array}{lll}
R_0(v,y)&=&v+ \kappa h(y),\\
R_{n+1}(v,y)&=&\Lambda(R_n(v,y))+ \kappa h(T^{n+1}(y)), \;\;v\in \realnos^k, y\in Y, n\ge 0.
\end{array}\right.\label{thesn}
\ee Then $v(n\tau)=R_n(v_0,\xi_0)$. One can easily check by induction that all assumptions of Proposition
\ref{proplimit} are satisfied. Thus $v(n\tau)$ converges in distribution to a unique probability measure $\mu_*$
on $\realnos^k$. Since $v_{n+1}=\Lambda(v(n\tau))$ and $\Lambda$ is a continuous transformation, it follows from
the definition of weak convergence that the distribution of $v_{n+1}$ converges weakly to
$\mu_*\circ\Lambda^{-1}$.

We call the measure $\mu_*$ the \textit{limiting measure} for $v(n\tau)$. Note that $\mu_*$ may depend on $\nu$.

\begin{rem}
Although this Corollary shows that there is a unique limiting measure, we cannot conclude in general that this
measure has a density absolutely continuous with respect to the Lebesgue measure. See Example \ref{fatbaker} and
Remark \ref{r:fatbaker}.
\end{rem}

\subsection{The linear case in one
dimension}\label{linearcase}
We now consider Equation \ref{first} when $b(v)=-\gamma v$ and $\gamma\ge 0$.  In this situation, we are
considering a frictional force linear in the velocity, so Equations \ref{position} and \ref{first} become
\bea \dfrac{dx (t)}{dt}&=&v(t),\nonumber \\
\frac{dv(t)}{dt} & = & -\gamma v(t) + \kappa \sum_{n=0}^\infty h(\xi(t)) \delta(t-t_n). \label{perturbed}
 \eea

To make the computations of the previous section completely transparent, multiply Equation \ref{perturbed} by
the integrating factor $\exp(\gamma t)$, rearrange, and integrate from $(t_n-\epsilon)$ to $(t_{n+1}-\epsilon)$,
where $0 < \epsilon< \min_{n\geq0}(t_{n+1}-t_n)$, to give
\begin{eqnarray}
 \nonumber v(t_{n+1}-\epsilon) e^{\gamma (t_{n+1}-\epsilon)} -
v(t_{n}-\epsilon) e^{\gamma (t_{n}-\epsilon)} &=& \kappa \sum_{n=0}^\infty
\int_{t_{n}-\epsilon}^{t_{n+1}-\epsilon}
h(\xi(z)) \delta(z-t_n) \,dz\\
&=& \kappa  e^{\gamma (t_n-\epsilon)}  h(\xi(t_n - \epsilon))
\label{integrated}
\end{eqnarray}
Taking the $\lim_{\epsilon \to 0}$ in Equation \ref{integrated} and remembering that $v(t_n^-) = v_n$ and
$\xi(t_n) = \xi_n$, we have
 \be
 v_{n+1}
= \lambda_n v_n + \kappa \lambda_n h(\xi_n), \label{almostfinalx}
\ee where $\tau_{n} \equiv t_{n+1}-t_n$ and  $ 0 \leq \lambda_n
\equiv e^{-\gamma \tau_{n}} < 1 .$

We simplify this  formulation by taking $ t_{n+1}-t_n \equiv \tau > 0 $ so the perturbations are assumed to be
arriving periodically. As a consequence, $\lambda_n \equiv \lambda$ with \be \lambda= e^{-\gamma \tau}.\ee Then,
Equation \ref{almostfinalx} becomes
 \be
 v_{n+1} = \lambda v_n + \kappa \lambda h(\xi_n).
\label{finalx} \ee This result can also be arrived at from other
assumptions \footnote{ Alternately but, as it will turn out,
equivalently, we can think of the perturbations as constantly
applied. In this case we write an Euler approximation to the
derivative in Equation \ref{perturbed} so with an integration step
size of $\tau$ we have \be v(t+\tau) = (1-\gamma \tau) v(t) + \tau
 \kappa h(\xi(t)). \label{secondinteg}  \ee
Measuring time in units of $\tau$ so $t_{n+1} = t_n + \tau$ we then can write this in the alternate equivalent
form
 \be
 v_{n+1} = \lambda v_n +  \kappa_1\lambda h(\xi_n),
\label{secfinalx} \ee where, now, $ \lambda = 1-\gamma \tau $ and
$\kappa_1 = \kappa \tau \lambda^{-1}$.  Again, by induction we
obtain Equation \ref{firstinduc}.}.

\subsubsection{Behaviour of the velocity variable}

For a given initial $v_0$ we have, by induction, \bea v_n &=& \lambda^n v_0 + \kappa \lambda \sum_{j=0}^{n-1}
\lambda^{n-1-j}
h(\xi_{j}) \nonumber \\
&=& \lambda^n v_0 + \kappa \lambda \sum_{j=0}^{n-1} \lambda^{n-1-j} h(T^j(\xi_{0})). \label{firstinduc}
 \eea

We now  calculate the asymptotic behaviour of the variance of
$v_n$ when $\xi_0$ is distributed according to $\nu$, the
invariant measure for $T$. Assume for simplicity that $v_0=0$, set
$\sigma^2=\int h^2(y)\nu(dy)$ and assume that $\int h(y)h(
T^n(y))\nu(dy)=0$ for $n \geq 1$.  
Then we have
\[
\int v_n^2 \nu(dy) = \kappa^2 \int \left( \sum_{j=0}^{n-1} \lambda^{n-j}h(T^j(y))\right)^2  \nu(dy).
\]
Since the sequence $h\circ T^j$ is uncorrelated by our assumption,
\[
\int \left( \sum_{j=0}^{n-1} \lambda^{n-j}h(T^j(y))\right)^2\nu(dy)= \sum_{j=0}^{n-1} \lambda^{2n-2j}\int
h(T^j(y))^2\nu(dy)=\dfrac{1-\lambda^{2n}}{1-\lambda^2}\sigma^2.
\]
Thus \be \int v_n^2\nu(dy)= \kappa ^2\sigma^2 \left( \dfrac{1-\lambda^{2n}}{1-\lambda^2} \right).\ee


Since $b(v)=-\gamma v$, we have $\pi(t,v)=e^{-\gamma t}v$. Equation \ref{finalx}, in conjunction with  Equations
\ref{thev} and \ref{thev+}, leads to $v_{n+1} = \lambda v(n\tau)$ where
 $ v(n\tau) = v_n + \kappa h(\xi_n).$
Thus \be v(n\tau)= \lambda^n v_0 + \kappa \sum_{j=0}^{n}
\lambda^{n-j} h(T^j(\xi_{0}))\label{calvel}\ee


%
\noindent {\bf Case 1:} If $\lambda<1$ and $\xi_0$ is distributed according to $\nu$, then by Corollary
\ref{corollaryweak} there exists a unique limiting measure $\mu_*$ for $v(n\tau)$ provided that $h$ is
integrable with respect to $\nu$. The sequence $v_n$ also converges in distribution. Since the function
$\Lambda$ defined by Equation \ref{deflambda} is linear, $\Lambda(v)=\lambda v$, both sequences $v(n\tau)$ and
$v_n$ are either convergent or divergent in distribution.

What can happen if the random variable $\xi_0$ in Equation \ref{calvel} is distributed according to a different
measure?

\begin{prop}
Let $\lambda<1$, the transformation $T$ be exact with respect to $\nu$, and $h\in L^1(Y,\B,\nu)$. If the random
variable $\xi_0$ is distributed according to a normalized measure $\nu_0$ on $(Y,\B)$ which is absolutely
continuous with respect to $\nu$, then \be \kappa \sum_{j=0}^{n} \lambda^{n-j} h(T^j(\xi_{0}))\to^d \mu_* \ee
and $\mu_*$ does not depend on $\nu_0$.
\end{prop}
{\noindent\bf Proof.} Recall that by the continuity theorem
\[
\kappa \sum_{j=0}^{n} \lambda^{n-j} h(T^j(\xi_{0}))\to^d \mu_*
\]
if and only if for every $r\in\realnos$
\[
\lim_{n\to\infty}\int_{Y} \exp(i r \kappa \sum_{j=0}^{n} \lambda^{n-j} h(T^j(y)))\nu_0(dy)=\int_{\realnos}
\exp(i r x)\mu_*(dx).
\]
We know that the last Equation is true when $\nu_0=\nu$. Since for every $m\ge 1$ the sequence
\[
\kappa \sum_{j=0}^{m-1} \lambda^{n-j} h(T^j(y))
\]
is convergent to $0$ as $n\to\infty$ and
\[
\sum_{j=m}^n\lambda^{n-j}h\circ T^j =\sum_{j=0}^n\lambda^{n-m-j}h\circ T^j\circ T^m,
\]
an analysis similar to that in the proof of Theorem \ref{CLTE}
completes the demonstration.

\noindent {\bf Case 2:} If $\lambda = 1$ we  have $v_n=v_0 + \kappa \sum_{j=0}^{n-1}h(T^j(\xi_0))$. Since $v_0$
and $\xi_0$ are independent random variables, $v_0$ and $ \kappa \sum_{j=0}^{n-1}h(T^j(\xi_0))$ are also
independent. Hence $v_n$ converges if and only if $\kappa \sum_{j=0}^{n-1}h(T^j(\xi_0))$ does. Moreover, if
there is a limiting measure $\mu_*$ for $v_n$, then the sequence $v(n\tau)$ converges in distribution, say to
$\nu_*$, and $\mu_*$ is a convolution of the distribution of $v_0$ and $\nu_*$. As a result, $\mu_*$ depends on
the distribution of $v_0$. However, if the map $T$ and function $h$ satisfy the FCLT, then
\[
\dfrac{\sum_{j=0}^{n-1}h(T^j(\xi_0))}{\sigma\sqrt{n}}\to^d N(0,1).
\]
Hence $\sum_{j=0}^{n-1}h(T^j(\xi_0))$ is not convergent in distribution since the density is spread on the
entire real line.

\subsubsection{Behaviour of the position variable}

For the position variable we have, for $t\in[n\tau,(n+1)\tau)$,
\[
x(t)-x(n\tau)=\int_{n\tau}^t v(s)ds=\int_{n\tau}^t
e^{-\gamma(s-n\tau)}v(n\tau)ds=\dfrac{1-e^{-\gamma(t-n\tau)}}{\gamma}v(n\tau).
\]
With $x(n\tau)=x_n$ we have \be
x_{n+1}-x_{n}=\dfrac{1-e^{-\gamma\tau}}{\gamma}v(n\tau)=\dfrac{1-\lambda}{\gamma}v(n\tau).
\label{xnplus1} \ee Summing from $0$ to $n$ gives \be
x_{n+1}=x_{0}+ \dfrac{1-\lambda}{\gamma}\sum_{j=0}^n v(j\tau). \ee
From this and Equation \ref{calvel} we obtain
\begin{eqnarray*}
   x_{n+1}&=&  x_{0}+ \dfrac{1-\lambda}{\gamma}\sum_{j=0}^n\left(\lambda^j v_0 +
\kappa \sum_{i=0}^{j} \lambda^{j-i} h(T^i(\xi_{0}))\right)\\
   &=& x_{0}+ \dfrac{(1-\lambda^{n+1})}{\gamma}v_0+
\dfrac{(1-\lambda)\kappa }{\gamma}\sum_{j=0}^n\sum_{i=0}^{j} \lambda^{j-i} h(T^i(\xi_{0})).
\end{eqnarray*}
Changing the order of summation in the last term gives
\begin{eqnarray*}
  \sum_{j=0}^n\sum_{i=0}^{j} \lambda^{j-i}
h(T^i(\xi_{0}))&=&\sum_{i=0}^n\sum_{j=i}^{n} \lambda^{j-i}
h(T^i(\xi_{0}))=\sum_{i=0}^n\frac{1-\lambda^{n-i+1}}{1-\lambda}h(T^i(\xi_{0}))\\
&=&\frac{1}{1-\lambda}\sum_{i=0}^nh(T^i(\xi_{0})) -\frac{\lambda}{1-\lambda}\sum_{i=0}^n
\lambda^{n-i}h(T^i(\xi_{0})).
\end{eqnarray*}
Consequently \[ x_{n+1}=x_{0}+
\dfrac{(1-\lambda^{n+1})}{\gamma}v_0+\dfrac{\kappa
}{\gamma}\sum_{i=0}^nh(T^i(\xi_{0})) -\dfrac{\lambda \kappa
}{\gamma}\sum_{i=0}^n \lambda^{n-i}h(T^i(\xi_{0})). \] In
conjunction with Equation \ref{xnplus1}, this gives \be x_{n}=x_{0}+
\dfrac{(1-\lambda^{n})}{\gamma}v_0+\dfrac{\kappa
}{\gamma}\sum_{i=0}^nh(T^i(\xi_{0})) -\dfrac{\kappa
}{\gamma}\sum_{i=0}^n \lambda^{n-i}h(T^i(\xi_{0})).\label{posxn} \ee

Next we calculate the asymptotic behavior of the variance of $x_n$.
Assume as before that $x_0=v_0=0$, and that $\int
h(y)h(T^j(y))\nu(dy)=0$ and $\sigma^2=\int h^2(y)\nu(dy)$. We have
\begin{eqnarray*}
  \left(\sum_{i=0}^nh(T^i(\xi_{0}))
-\sum_{i=0}^n \lambda^{n-i}h(T^i(\xi_{0})\right)^2 &=& \left(\sum_{i=0}^nh(T^i(\xi_{0}))\right)^2+ \left(
\sum_{i=0}^n
\lambda^{n-i}h(T^i(\xi_{0}))\right)^2\\
   & &-
2\sum_{i=0}^nh(T^i(\xi_{0}))\sum_{i=0}^n \lambda^{n-i}h(T^i(\xi_{0}))
\end{eqnarray*}
Since, by assumption, the sequence $h\circ T^i$ is again
uncorrelated we have
\begin{eqnarray*}
\int\left(\sum_{i=0}^nh(T^i(y))\right)^2\nu(dy)&=&\sum_{i=0}^n\sum_{j=0}^n\int
h(T^i(y))h(T^j(y))\nu(dy)\\
 & = & \sum_{i=0}^n\int h(T^i(y))^2\nu(dy)=(n+1)\sigma^2.
\end{eqnarray*}
Analogous to the computation for the velocity variance
\[
\int \left( \sum_{i=0}^n \lambda^{n-i}h(T^i(y))\right)^2\nu(dy)= \sum_{i=0}^n \lambda^{2n-2i}\int
h(T^i(y))^2\nu(dy)=\dfrac{1-\lambda^{2n+2}}{1-\lambda^2}\sigma^2
\]
and
\begin{eqnarray*}
  \int \sum_{i=0}^nh(T^i(y))\sum_{j=0}^n
\lambda^{n-j}h(T^j(y))\nu(dy) &=& \sum_{i=0}^n\sum_{j=0}^n\lambda^{n-j}\int
h(T^i(y))h(T^j(y))\nu(dy) \\
   &=&\sum_{i=0}^n \lambda^{n-i}\int
h(T^i(y))^2\nu(dy)=\dfrac{1-\lambda^{n+1}}{1-\lambda}\sigma^2.
\end{eqnarray*}
Consequently, if $x_0=v_0=0$ then  \be \int x_n^2\nu(dy)= \frac{\kappa ^2\sigma^2}{\gamma^2}\left(n+1
+\dfrac{1-\lambda^{2n+2}}{1-\lambda^2}-2\dfrac{1-\lambda^{n+1}}{1-\lambda}\right).\ee

\begin{thm}\label{++}
Let $(Y,\B,\nu)$ be a normalized measure space, $T:Y\to Y$ be a measurable map such that $T$ preserves the
measure $\nu$,  let $\sigma>0$ be a constant, and $h\in L^2(Y,\B,\nu)$ be such that $\int h(y)\nu(dy)=0$.
Then \be \frac{\sum_{i=0}^nh(T^i(\xi_{0}))}{\sqrt{n}}\to^d N(0,\sigma^2)\label{clttrans}\ee
%
%
if and only if \be \frac{x_n}{\sqrt{n}}\to^d N\left(0,\frac{ \kappa^2\sigma^2}{\gamma^2}\right).
\label{cltpos}\ee
\end{thm}
{\noindent \bf Proof.} Assume that Condition \ref{clttrans} holds.
From Equation \ref{posxn} we obtain
\[
\frac{x_{n}}{\sqrt{n}}=\frac{x_{0}}{\sqrt{n}}+ \dfrac{(1-\lambda^{n})}{\gamma}\frac{v_0}{\sqrt{n}} +\dfrac{
\kappa}{\gamma \sqrt{n}}
\sum_{i=0}^nh(T^i(\xi_{0}))
-\dfrac{\kappa }{\gamma\sqrt{n}}\sum_{i=0}^n \lambda^{n-i}h(T^i(\xi_{0})).
\]
By assumption
\[
\dfrac{ \kappa}{\gamma \sqrt{n}} \sum_{i=0}^nh(T^i(\xi_{0}))\to^d \dfrac{ \kappa}{\gamma}N(0,\sigma^2).
\]
Thus the result will follow when we show that the remaining terms are convergent in probability to zero. The
first term
\[
\frac{x_{0}}{\sqrt{n}}+ \dfrac{(1-\lambda^{n})}{\gamma}\frac{v_0}{\sqrt{n}}
\]
is convergent to zero a.e. hence in probability. The sequence
$\sum_{i=0}^n\lambda^{n-i}h(T^i(\xi_{0}))$ is convergent in
distribution and $\dfrac{\kappa }{\gamma\sqrt{n}}\to 0$ as
$n\to\infty$.  Consequently, the sequence
\[
\dfrac{\kappa }{\gamma\sqrt{n}}\sum_{i=0}^n\lambda^{n-i}h(T^i(\xi_{0}))
\]
is convergent in probability to zero which completes the proof. The proof of the converse is analogous.

\begin{rem}
Observe that if the transformation $T$ is exact and $\xi_0$ is distributed according to a measure absolutely
continuous with respect to $\nu$, then the conclusion of Theorem \ref{++} still holds.
\end{rem}

Theorem \ref{++} generalizes the results of \citet{chew}. In Section \ref{s:CLT} we have discussed when
Condition \ref{clttrans} holds for a given ergodic transformation.

\begin{rem}
Note that if we multiply Gaussian distributed random variable
$N(0,1)$ by a positive constant $c$, then it becomes Gaussian
distributed $N(0,c^2)$ with variance $c^2$. Thus if we multiply
both sides of Equation \ref{cltpos} by $\dfrac{1}{\sqrt{\tau}}$,
we obtain
\[
\frac{x(n\tau)}{\sqrt{n\tau}}\to^d N(0,\frac{ \kappa^2\sigma^2}{\tau\gamma^2}).
\]
So if $\kappa=\sqrt{\gamma m \tau}$, as in \citet{chew}, then
\[
\frac{ \kappa^2\sigma^2}{\tau\gamma^2}=\frac{m\sigma^2}{\gamma}.
\]
\end{rem}

\setcounter{equation}{0} \setcounter{figure}{0}
\section{Identifying the Limiting Velocity Distribution}\label{identify}

Let $Y\subset \realnos$ be an interval, $\B=\B(Y)$, and $T:Y\to Y$
be a transformation preserving a normalized measure $\nu$ on
$(Y,\B(Y))$.  Recall from Section \ref{weakconv} that $\mu_*$ is
the limiting measure for  the sequence of random variables
$(v(n\tau))$ starting from $v_0\equiv 0$, {\it i.e.}
\[
v(n\tau)=  \kappa \sum_{i=0}^{n}\lambda^{n-i}h(\xi_i),
\]
where $h:Y\to\realnos$ is a given integrable function, $0<\lambda<1$, $\xi_i=T^i(\xi_0)$, and $\xi_0$ is
distributed according to $\nu$.

\begin{prop}\label{p:mom}
Let $Y=[a,b]$ and  $h:Y\to\realnos$ be a bounded function. Then
the limiting measure $\mu_*$ has moments of all order given by \be
\int x^k\mu_*(dx)=\lim_{n\to\infty}\int
v(n\tau)^k\nu(dy)\label{moments} \ee and the characteristic
function of $\mu_*$ is of the form
\[
\phi_*(r)=\sum_{k=0}^{\infty}\frac{(ir)^k}{k!}\int x^k\mu_*(dx),\;\;r\in\realnos.
\]
Moreover, $\mu_*([-\dfrac{\kappa c}{1-\lambda},\dfrac{\kappa
c}{1-\lambda}])=1$, where $c=\sup_{y\in Y}|h(y)|$.
\end{prop}
{\noindent \bf Proof.}  Since $h$ is bounded, we have
\[
|v(n\tau)|^k\le  \left(\frac{\kappa c}{1-\lambda}\right)^k,\;\;n,k\ge 0.
\]
The existence and convergence of moments now follow from Theorem
5.3 and 5.4 of \cite{billingsley68}. Since $v(n\tau)$ has all its
values in the interval $[-\dfrac{\kappa
c}{1-\lambda},\dfrac{\kappa c}{1-\lambda}]$ we obtain
$\mu_n([-\dfrac{\kappa c}{1-\lambda},\dfrac{\kappa
c}{1-\lambda}])=1$ where $\mu_n$ is the distribution of
$v(n\tau)$. Convergence in distribution \citep[Theorem
2.1]{billingsley68} implies that
\[
\limsup_{n\to\infty}\mu_n(F)\le \mu_*(F)
\]
for all closed sets. Therefore $\mu_*([-\dfrac{\kappa
c}{1-\lambda},\dfrac{\kappa c}{1-\lambda}])=1$. The formula for
the characteristic function is a consequence of the other
statements, and the proof is complete.

\begin{rem}
Note that if the characteristic function of $\mu_*$ is integrable, then $\mu_*$ has a continuous and bounded
density which is given by
\[
f_*(x)=\frac{1}{2\pi}\int_{-\infty}^{\infty}\exp(-i x r)\phi_*(r)dr,\;\;x\in\realnos.
\]
On the other hand if $\mu_*$ has a density then $\phi_*(r)\to 0$ as $|r|\to\infty$.

Note also that if a density exists then it must be zero outside
the interval $[-\dfrac{\kappa c}{1-\lambda},\dfrac{\kappa
c}{1-\lambda}]$.
\end{rem}


Let $Y=[a,b]$ be an interval  and $h(y) =y$ for $y\in Y$. Thus $h$
is bounded and for this choice of $h$ all moments of the
corresponding limiting distribution $\mu_*$ exists by Proposition
\ref{p:mom}. However, the calculation might be quite tedious. We
are going to determine the measure $\mu_*$ for a specific example
of the transformation $T$ by using a different method.

Let $h_n:Y\to \realnos$ be defined by
 \be
 h_n(y)=\sum_{i=0}^{n}\lambda^{n-i}T^i(y),\;\;y\in Y, n\ge 0.
 \label{hn}
 \ee
Then $v(n\tau)= \kappa h_{n}(\xi_0)$ and $v_n= \kappa \lambda h_{n-1}(\xi_0)$. Thus knowing the limiting
distribution for these sequences is equivalent to knowing the limiting distribution for $h_n(\xi_0)$.

\subsection{Dyadic map}  To give a concrete example for which much of the preceding considerations can  be completely illustrated,
consider the generalized dyadic map
defined by Equation \ref{dyadic}:
\[
T(y)= \left\{
\begin{array}{ll}
2y +1, & y \in \left [-1,0\right ]\\ 2y -1, & y \in \left (0, 1\right ].
\end{array}
\right.
\]


\begin{prop}\label{propdyadic}
Let $\xi,\xi_0$ be random variables uniformly distributed on $[-1,1].$ Let $(\alpha_k)$ be a sequence of
independent random variables taking values drawn from $\{-1,1\}$ with  equal probability. Assume that $\xi$ is
statistically independent of the sequence $(\alpha_k)$. Then for every $\lambda\in(0,1)$
 \be
h_n(\xi_0)\to^d\frac{1}{2-\lambda}\left(\xi+\sum_{k=0}^{\infty}\lambda^{k}\alpha_{k+1}\right).
 \label{limitdyadic}
 \ee
\end{prop}
{\bf Proof.} The random variable
\[\xi_0=\sum_{k=1}^{\infty}\frac{\alpha_k}{2^k}\] is uniformly distributed
on $[-1,1]$. It is easily seen that for the dyadic map \be
\xi_i=T^i(\xi_0)=\sum_{k=1}^{\infty}\frac{\alpha_{k+i}}{2^k}\;\;\mathrm{for}\;\;i\ge
1 .\label{shiftproperty}\ee Using this representation we obtain
\[
\sum_{i=0}^{n-1}\lambda^{n-1-i}\xi_i= \sum_{i=0}^{n-1}\lambda^{n-1-i} \sum_{k=1}^{n-i}\frac{\alpha_{k+i}}{2^k}+
\sum_{i=0}^{n-1}\lambda^{n-1-i}\sum_{k=n-i+1}^{\infty}\frac{\alpha_{k+i}}{2^{k}}.
\]
Changing the order of summation leads to
\begin{eqnarray*}
\sum_{k=1}^{n}\left(\sum_{i=1}^{k}\frac{\lambda^{n-1-k+i}}{2^i}\right)\alpha_k
&+&
\sum_{i=0}^{n-1}\frac{\lambda^{n-1-i}}{2^{n-i}}\sum_{k=1}^{\infty}\frac{\alpha_{k+n}}{2^{k}}=\\
&&\frac{1}{2-\lambda}\sum_{k=1}^{n}\lambda^{n-k}\left(1-\left(\frac{\lambda}{2}\right)^k\right)\alpha_k
+\frac{1-\left(\frac{\lambda}{2}\right)^n}{2-\lambda}\xi_n.
\end{eqnarray*}
Consequently
\[
\sum_{i=0}^{n-1}\lambda^{n-1-i}\xi_i=\frac{1}{2-\lambda}\sum_{k=1}^{n}\lambda^{n-k}\alpha_k +
\frac{1}{2-\lambda}\xi_n - \lambda^n w_n,
\]
where
\[
w_n=\frac{1}{2-\lambda}\left[\sum_{k=1}^{n}\left(\frac{1}{2}\right)^k \alpha_k +
\left(\frac{1}{2}\right)^n\xi_n\right].
\]
This gives
 \be
 h_{n-1}(\xi_0)+\lambda^n
w_n=\frac{1}{2-\lambda}\left(\sum_{k=1}^{n}\lambda^{n-k}\alpha_k +
\xi_n\right).\label{xndyadic}
 \ee
Note that for every $n$ we have $|w_n|\le 2$. Therefore $\lambda^n
w_n$ is a.s. convergent to $0$ as $n\to\infty$. Since
$h_{n}(\xi_0)$ converges in distribution, say to
$\widetilde{\mu}_*,$ we have $h_{n-1}(\xi_0)+\lambda^n
w_n\to^d\widetilde{\mu}_*$ and the random variables on the
right-hand side of Equation \ref{xndyadic} converge in
distribution to $\widetilde{\mu}_*$. Since the random variables
$\alpha_k$ are independent, the random variables
$\sum_{k=1}^{n}\lambda^{n-k}\alpha_k$ and $\xi_n$ are also
independent for every $n$. The same is true for
$\sum_{k=0}^{n-1}\lambda^{k}\alpha_{k+1}$ and $\xi$. Moreover,
$\xi_n+\sum_{k=1}^{n}\lambda^{n-k}\alpha_k$ and
$\xi+\sum_{k=0}^{n-1}\lambda^{k}\alpha_{k+1}$ have identical
distributions. Thus
\[\frac{1}{2-\lambda}\left(\xi+\sum_{k=0}^{n-1}\lambda^{k}\alpha_{k+1}\right)\to^d\mu_*.\]
On the other hand $\sum_{k=0}^{n-1}\lambda^{k}\alpha_{k+1}\to\sum_{k=0}^{\infty}\lambda^{k}\alpha_{k+1}$ almost
surely as $n\to\infty$, but this implies convergence in distribution. The proof is complete.


Before stating our next result, we review some of the known properties of the random variable which appears in Equation
\ref{limitdyadic}. For every $\lambda\in(0,1)$  let \be
 \zeta_\lambda=\sum_{k=0}^{\infty}\lambda^{k}\alpha_{k+1},\label{thezetas}
 \ee
and let $\varrho_\lambda$ be the distribution function of $\zeta_\lambda$,
$\varrho_\lambda(x)=\Pr\{\zeta_\lambda\le x\}$ for $x\in\realnos.$  Explicit expressions for $\varrho_\lambda$
are, in general, not known. The measure induced by the distribution $\varrho_\lambda$ is called an {\it
infinitely convolved Bernoulli measure} (see \cite{peres} for the historical background and recent advances).

It is known \citep{jessen} that $\varrho_\lambda$ is continuous
and it is either absolutely continuous or singular. Recall that
$x$ is a {\it point of increase} of $\varrho_\lambda$ if
$\varrho_\lambda(x-\epsilon)<\varrho_\lambda(x+\epsilon)$ for all
$\epsilon>0$. The set of points of increase of $\varrho_\lambda$
\citep{kershner} is either the interval $[-\frac{1}{1-\lambda},
\frac{1}{1-\lambda}]$ when $\lambda\ge \frac{1}{2}$ or a Cantor
set $K_\lambda$ of zero Lebesgue measure contained in this
interval when $\lambda<\frac{1}{2}$, $\varrho_\lambda$ is always
singular for $\lambda\in (0,\frac{1}{2})$ and the Cantor set
$K_\lambda$ satisfies $K_\lambda=(\lambda K_\lambda+1)\cup(\lambda
K_\lambda-1)$ and $\frac{1}{1-\lambda},-\frac{1}{1-\lambda}\in
K_\lambda$. \cite{wintner} noted that $\varrho_\lambda$ has the
uniform density $\rho_\lambda(x)=\frac{1}{4}1_{[-2,2]}(x)$ for
$\lambda=\frac{1}{2}$ and that it is absolutely continuous for the
$k$th roots of $\frac{1}{2}$. Thus it was suspected that
$\varrho_\lambda$ is absolutely continuous for all $\lambda \in
[\frac{1}{2},1)$. However, \cite{erdos39} showed that
$\varrho_\lambda$ is singular for $\lambda=\dfrac{\sqrt{5}-1}{2}$
and for the reciprocal of the so called Pisot numbers in $(1,2)$.
Later \cite{erdos40} showed that there is a $\beta<1$ such that
for almost all $\lambda\in (\beta,1)$ the measure
$\varrho_\lambda$ is absolutely continuous. Only recently,
\cite{solomyak} showed that $\beta=\frac{1}{2}$.

\begin{prop}\label{prop:halflambda}
For every $\lambda\in(0,1)$ the density $f_*^\lambda$ of the limiting measure $\mu_*^\lambda$ of $v(n\tau)$
satisfies
\[
f_*^\lambda(v)=0\;\;\mathrm{if\; and\; only \;if}\;\; |v|\ge \frac{
\kappa}{1-\lambda}.
\]
Moreover, on the interval $\left(-\frac{\kappa
}{1-\lambda},\frac{\kappa }{1-\lambda}\right)$ we have
\[
f_*^\lambda(v)= \left\{
 \begin{array}{ll}
\frac{2-\lambda}{2\kappa
}\varrho_\lambda\left(\frac{2-\lambda}{\kappa}v+1\right),
& -\frac{\kappa }{1-\lambda}< v <-\frac{\kappa \lambda}{(2-\lambda)(1-\lambda)}\\
\frac{2-\lambda}{2 \kappa}\left[\varrho_\lambda\left(
\frac{2-\lambda}{\kappa}v+1\right) -\varrho_\lambda\left(
\frac{2-\lambda}{\kappa}v-1\right)\right],
&|v|\le \frac{\kappa \lambda}{(2-\lambda)(1-\lambda)}\\
\frac{2-\lambda}{2 \kappa }\left[1-\varrho_\lambda\left(
\frac{2-\lambda}{\kappa}v-1\right)\right], &\frac{\kappa
\lambda}{(2-\lambda)(1-\lambda)}< v < \frac{\kappa }{1-\lambda},
\end{array}
\right.
\]
where $\varrho_\lambda$ is the distribution function of $\zeta_\lambda$ defined by
Equation \ref{thezetas}.
\end{prop}

\noindent {\bf Proof.} Recall that we have $v(n\tau)= h_n(\xi_0)$. By Proposition \ref{propdyadic}, the sequence
$(2-\lambda)h_n(\xi_0)$ converges in distribution to $\xi+\zeta_\lambda$ and the random variables $\xi$ and
$\zeta_\lambda$ are statistically independent. Since $\xi$ has the uniform density on $[-1,1]$ and
$\varrho_\lambda$ is continuous, the density of $\xi + \zeta_\lambda$ is given by
\begin{eqnarray*}
 \int_{-\infty}^\infty\frac 12 1_{[-1,1]}(x-z)d\rho_\lambda(z) &=& \int_{-\infty}^\infty\frac 12
1_{[x-1,x+1]}(z)d\rho_\lambda(z) \\
   &=& \int_{x-1}^{x+1}\frac 12 d\rho_\lambda(z)=
   \frac
12(\varrho_\lambda(x+1)-\rho_\lambda(x-1)).
\end{eqnarray*}
Since $v(n\tau)$ converges in distribution to $\dfrac{\kappa
}{2-\lambda}(\xi+\zeta_\lambda)$,  it follows that $\mu_*^\lambda$
is the distribution of $\dfrac{\kappa
}{2-\lambda}(\xi+\zeta_\lambda)$. Thus $\mu_*^\lambda$ has a density
given by
\[
f_*^\lambda(v)=\dfrac{2-\lambda}{2 \kappa }\left(\varrho_\lambda\left( \dfrac{2-\lambda}{\kappa}v+1\right)
-\varrho_\lambda\left( \dfrac{2-\lambda}{\kappa}v-1\right)\right),\;\;v\in\realnos.
\]
Consequently, $f_*^\lambda(v)=0$ if and only if
\[
\varrho_\lambda\left( \dfrac{2-\lambda}{\kappa}v+1\right)
=\varrho_\lambda\left( \dfrac{2-\lambda}{\kappa}v-1\right).
\]
Since $\varrho_\lambda$ is nondecreasing, it must
be constant outside the set of points on which it is increasing,
which is contained in the interval $[-\frac{ 1
}{1-\lambda},\frac{1}{1-\lambda}]$. Hence $\varrho_\lambda(x)=0$ for
$x\le -\frac{1}{1-\lambda}$ and $\varrho_\lambda(x)=1$ for
$x\ge\frac{1}{1-\lambda}$. Therefore if $|v|\ge \frac{ \kappa
}{1-\lambda}$, then $f_*^\lambda(v)=0$.
If $\lambda\ge \frac{1}{2}$ the
function $\varrho_\lambda$ is increasing on the interval
$[-\frac{1}{1-\lambda},\frac{1}{1-\lambda}]$, thus $f_*^\lambda$ is
positive on $[-\frac{\kappa }{1-\lambda},\frac{\kappa
}{1-\lambda}]$. Now let $\lambda<\frac{1}{2}$. Since
$\frac{1}{1-\lambda},-\frac{1}{1-\lambda}\in K_\lambda$, we also
have $\frac{2\lambda-1}{1-\lambda}, \frac{1-2\lambda}{1-\lambda}\in
K_\lambda$ and they divide the interval
$[-\frac{1}{1-\lambda},\frac{1}{1-\lambda}]$ into three intervals of
length $2$, $2\frac{1-2\lambda}{1-\lambda}$, and $ 2$ respectively.
Since the middle interval has length less than $2$ and the distance
between the points $\frac{2-\lambda}{\kappa}v+1$ and
$\frac{2-\lambda}{\kappa}v-1$ is always $2$, the result follows in
this case as well.

\begin{rem}
If $\lambda=\frac{1}{2}$ then $\varrho_\lambda(x)=\frac{1}{2}(x+2)$, $2-\lambda=\frac{3}{2}$
 and the density is equal to \be f_*^\lambda(v)= \left\{
 \begin{array}{ll}
\vspace{.1in} \dfrac{9}{32 \kappa }(v+2  ), &-2 \kappa < v <-\dfrac{2}{3} \kappa \\
 \vspace{.1in} \dfrac{3}{8 \kappa},  &|v|\le \dfrac{2}{3} \kappa \\
\vspace{.1in} \dfrac{9}{32 \kappa}(2  -v),&\dfrac{2}{3} \kappa < v < 2 \kappa
\end{array}
\right. \label{halflambda} \ee 
\end{rem}


\begin{rem}\label{r:fatbaker}
The invariant measure for the baker transformation $S_\beta$ of Example \ref{fatbaker} is the product of the
distribution of $(1-\lambda)\zeta_\lambda$ and the normalized Lebesgue measure. Thus, in this example the
limiting measure for $v_n$ is the distribution of $(1-\lambda)\zeta_\lambda$, which may be either singular or
absolutely continuous.
\end{rem}

\subsection{Graphical illustration of the velocity density evolution with
dyadic map perturbations}

What is the probability density function of $h_n(\xi_0)$ defined by Equation \ref{hn} when $\xi_0$ is
distributed according to $\nu$?  For many
maps, including the dyadic map example being considered here, this can be calculated analytically
which is the subject of this section.

Let $Y$ be an interval and let $\nu$ have a density $g_*$ with respect to Lebesque measure. Then the
distribution of $h_n(\xi_0)$ is given by
\[
\Pr\{h_n(\xi_0)\in A\}=\Pr\{\xi_0\in h^{-1}_n(A)\}=\int_{h^{-1}_n(A)}g_*(y)dy, \;\;A\in\B(\realnos).
\]
To obtain the density of $h_n(\xi_0)$ with respect to the Lebesgue measure, one has to write the last integral
as $\int_A g_n(x)dx$ for some nonnegative function $g_n$. If the map $h_n:Y\to\realnos$ is nonsingular with
respect to Lebesgue measure, then the Frobenius-Perron operator $P_{h_n}:L^1([-1,1])\to L^1(\realnos)$ for $h_n$
exists and $g_n=P_{h_n}g_*$.

Let $Y=[-1,1]$ and let $T$ be the dyadic map. Remember that $P_T$
has the uniform invariant density $g_*(y)= \frac 12
1_{[-1,1]}(y)$. Since $h_n$ is a linear function on each interval
$\left(\frac{k}{2^n},\frac{k+1}{2^n}\right]$ with constant
derivative, say $h_n'$, we have \be g_n(v) =
P_{h_n}g_*(v)=\frac{1}{2h_n^{'}}\sum_{k=-2^n}^{2^n-1}1_{h_n((\frac{k}{2^n},\frac{k+1}{2^n}])}(v),
\;\; v\in \realnos. \label{dyadicden} \ee The derivative $h_n^{'}$
satisfies the recurrence equation $h_n'=\lambda^{n-1}+2
h_{n-1}^{'}$, $n\ge 1$ and $h_0^{'}=0$ and is equal to
$\dfrac{(\lambda-3)2^n +\lambda^n}{\lambda-2}$ for each $n$.

\begin{figure}
\begin{center} 
  \includegraphics[width=3in]{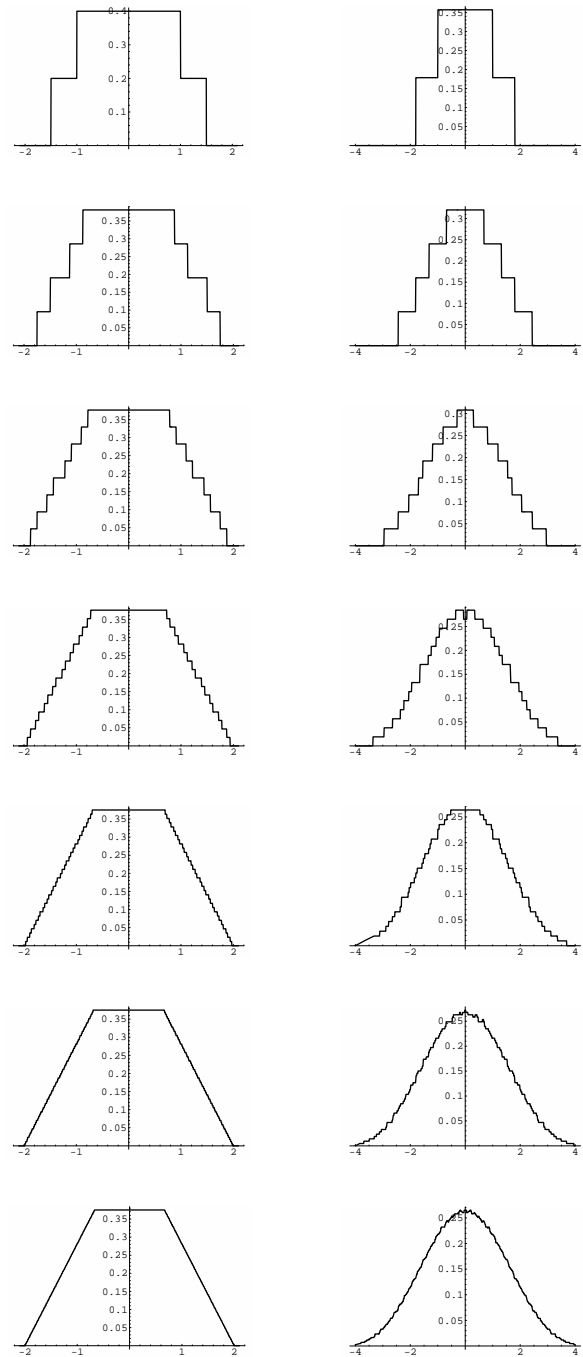}\\
\end{center}
  \caption{This figure shows the successive evolution of the densities $g_n$, $n=2, \ldots, 8$ of the velocity under perturbations from the dyadic map.
  In the left hand series of
  densities, $\lambda = \frac 12$, while on the
  right $\lambda = 0.8$.  The densities $g_n$ when $\lambda = \frac 12$ rapidly approach the
  limiting analytic form $f_*^{\frac 12}$ given in
  Equation  \ref{halflambda}. In both cases, $  \kappa =
  1$.}\label{twolambdas}
\end{figure}

\begin{figure}
\begin{center}  
   \includegraphics[width=5in]{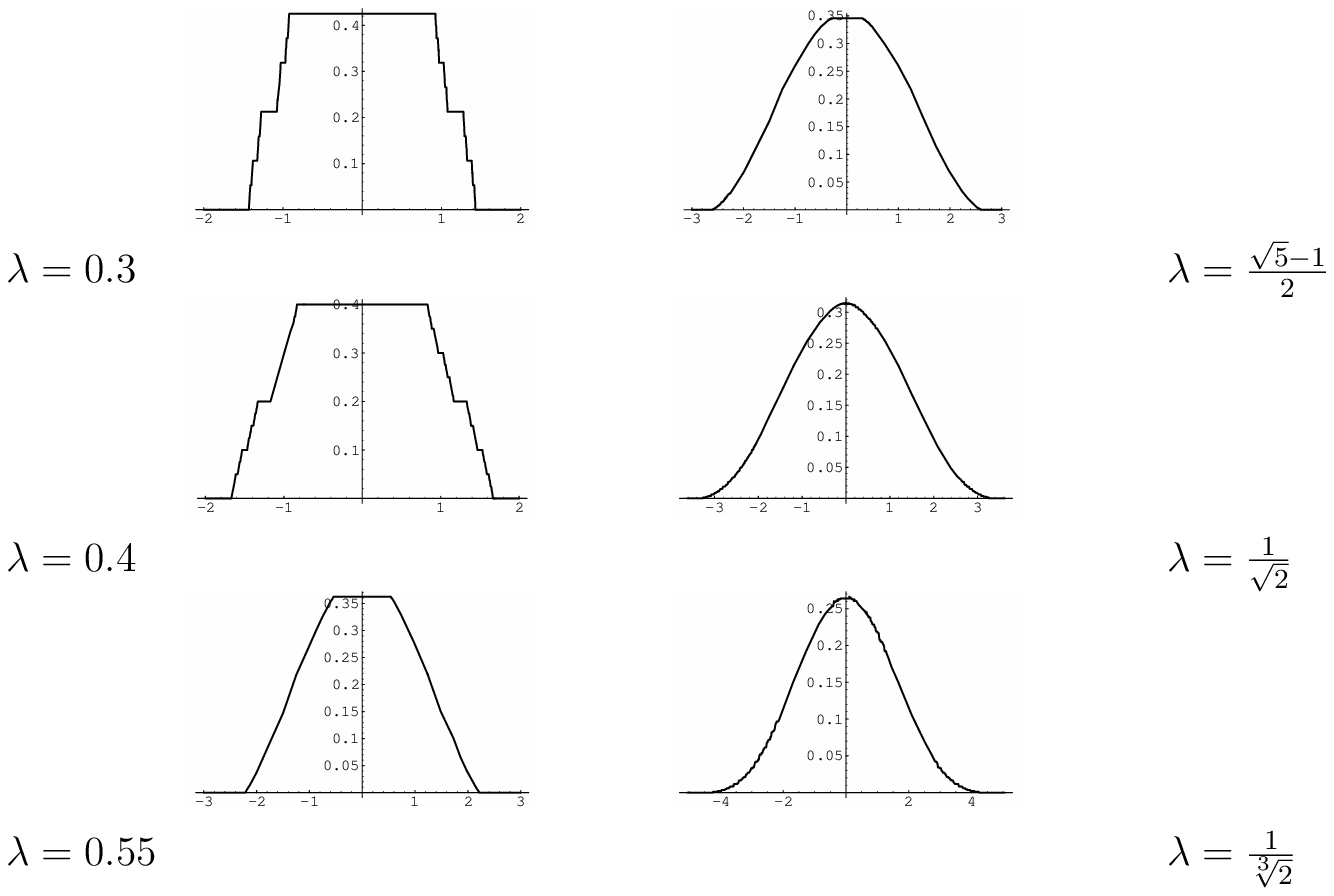}\\
\end{center}
\caption{This figure illustrates the form of the density $g_8(v)$
for perturbations coming from the dyadic map, as computed from
Equation \ref{dyadicden}, and various values of $\lambda$ as
indicated, with $\kappa  = 1$. In every case the initial velocity
density was the uniform invariant density of the dyadic
map.}\label{difflambdas}
\end{figure}

In Figure \ref{twolambdas} we show the evolution of the velocity densities $g_n$ when $T$ is the dyadic map for
two different values of $\lambda$. For $\lambda = \frac 12$ (left hand panels) the density rapidly (by $n=8$)
approaches the analytic form given in Equation \ref{halflambda}.  On the right hand side, for $\lambda = 0.8$
the velocity densities have, by $n=8$, approached a Gaussian-like form but supported on a finite interval.  In
both cases the support of the limiting densities is in agreement with Proposition \ref{prop:halflambda}. Figure
\ref{difflambdas} shows $g_8(v)$ for six different values of
$\lambda$.  

\subsection{r-dyadic map} Let $r\ge 2$ be an integer. Consider
the $r$-dyadic transformation on the interval $[0,1]$
\[
{T}(y)=r y \;(\mathrm{mod}\; 1), \;\;y\in[0,1].
\]
The proof of Proposition \ref{propdyadic} carries over to this transformation when $(\alpha_k)$ is a sequence of
independent random variables taking values in $\{0,1,\ldots,r-1\}$ with  equal probabilities, {\it i.e.}
$\Pr(\alpha_k=i)=\frac{1}{r}$, $i=0,1,\ldots,r-1$, and  $\xi$ is a random variable uniformly distributed on
[0,1] and independent of the sequence of random variables $(\alpha_k)$. Then the limiting measure for $v_n$ is
the distribution of the random variable
\[
\frac{\kappa }{r-\lambda}\left({\xi}+\sum_{k=0}^{\infty}\lambda^{k}\alpha_{k+1}\right).
\]

\setcounter{equation}{0} \setcounter{figure}{0}
\section{Gaussian Behaviour in the limit $\tau\to
0$.}\label{tau}

In a series of papers \citet{beck87}, \citet{beck90},
\citet{beck90a}, \citet{beck96} and \citet{beck99}, motivated by
questions related to alternative interpretations of Brownian
motion, have numerically examined the dynamic character of the
iterates of dynamical systems of the form \bea u_{n+1} &=& \lambda
u_n + \sqrt{\tau} y_n \qquad \lambda \equiv e^{-\gamma\tau}
\label{becksys1} \\ y_{n+1} &=& T(y_n),\label{becksys2}
 \eea in which $T$ is a `chaotic' mapping and
$\tau$ is a small temporal scaling parameter. They refer to these
systems as linear Langevin systems, and point out that they arise
from the following
\[
\dot{u}=-\gamma u+ \sqrt{\tau}\sum_{i=1}^{\infty}y_{i-1}\delta(t-i\tau).
\]
Integrating this equation one obtains Equation \ref{becksys1} with $u_n=u(n\tau)$.

For situations in which the map $T$ is selected from the class of Chebyshev maps [c.f. Equation \ref{chebyeqn}
 and \citet{adler64}],
 \citet{beck2001} have provided abundant numerical evidence that the density of the distribution of the sequence of iterates $\{u_i\}_{i=0}^N$,
for $N$ quite large, may be approximately normal, or Gaussian, as $\lambda \to 1$, and \citet{beck99} have
provided some of the same type of numerical evidence for perturbations coming from the dyadic and hat maps.  Our
results provide the analytic basis for these observations.

In this section we consider and answer the question when one can obtain
 Gaussian processes by studying appropriate scaling limits of the
velocity and position variables. We first recall what is meant by a Gaussian process.

An $\realnos$-valued stochastic process $\{\zeta(t);
t\in(0,\infty)\}$ is called {\it Gaussian} if, for every integer
$l\ge 1$ and real numbers $0< t_1<t_2<\ldots <t_l<\infty$ the random
vector $(\zeta(t_1),\ldots,\zeta(t_l))$ has a joint normal
distribution or equivalently, for all $d_j\in\realnos$,
$j=1,\ldots,k$, the random variable $\sum_{j=1}^l d_j\zeta(t_j)$ is
Gaussian. The finite dimensional distributions of a Gaussian process
are completely determined by its first moment $m(t)=E(\zeta(t))$ and
its covariance function
\[K_\zeta(t,s)=E(\zeta(t)-m(t))(\zeta(s)-m(s)), \;\;s,t> 0.\] If
$m(t)\equiv 0$, $t> 0$ we say that $\zeta$ is a zero-mean Gaussian
process. The initial random variable $\zeta(0)$ can be either
identically equal to zero or can be any other random variable
independent of the process $\{\zeta(t); t\in(0,\infty)\}$.

Now we recall the Ornstein-Uhlenbeck theory of Brownian motion for a free particle. The Ornstein-Uhlenbeck
velocity process is a solution of the stochastic differential equation
\[
dV(t)=-\gamma V(t)dt + \sigma_0dw(t),
\]
where $w$ is a standard Wiener process, and the solution of this equation is
\[
V(t)=e^{-\gamma t}V(0) + \sigma_0\int_{0}^t e^{-\gamma(t-s)}dw(s).
\]
In other words, $V$ is an Ornstein-Uhlenbeck velocity process if
$\zeta$ defined by $\zeta(t)=V(t)-e^{-\gamma t}V(0)$, $t\ge 0$, is a
zero-mean Gaussian process with covariance function \be\label{covOU}
K_\zeta(t,s)=\frac{\sigma_0^2}{2\gamma}(e^{2\gamma
\min(t,s)}-1)e^{-\gamma(t+s)}.\ee If the initial random variable
$V(0)$ has a normal distribution with mean zero and variance
$\frac{\sigma_0^2}{2\gamma}$, then $V$ itself is a stationary,
zero-mean Gaussian process with covariance function
\[
K_V(t,s)=\frac{\sigma_0^2}{2\gamma}e^{-\gamma|t-s|}.
\]
Let $X(t)$ denote the position of a Brownian particle at time $t$.
Then
\[
X(t)=X(0)+\int_{0}^tV(s)ds.
\]
In other words, $X$ is an Ornstein-Uhlenbeck position process if $\eta$ defined by
$\eta(t)=X(t)-X(0)-\frac{1-e^{-\gamma t}}{\gamma}V(0)$ is a zero-mean Gaussian process with covariance function
\[
K_\eta(t,s)=\frac{\sigma_0^2}{2\gamma^3}\left(2\gamma\min(t,s)
-2+2e^{-\gamma t}+2e^{-\gamma s}-e^{-\gamma
|t-s|}-e^{-\gamma(t+s)}\right).
\]
In particular the variance of $\eta(t)$ is equal to $\frac{\sigma_0^2}{2\gamma^3}(2\gamma t -3+4e^{-\gamma
t}-e^{-2\gamma t})$.

Let $h\in L^2(Y,\B,\nu)$ be such that $\int h(y)\nu(dy)=0$. Assume
that $x_0$, $v_0$, and $\xi_0$ are independent random variables on
$(Y,\B,\nu)$ and $\xi_0$ is distributed according to $\nu$. The
solution of Equation \ref{perturbed} is of the form
\[
v(t)=e^{-\gamma (t-n\tau)}v(n\tau),\;\;t\in[n\tau,(n+1)\tau),\;n\ge 0.
\]
We indicate the dependence of $x(t)$ and $v(t)$ on $\tau$ by
writing $x_\tau(t)$ and $v_\tau(t)$ respectively. Let
$n=[\frac{t}{\tau}]$, where the notation $[\cdot]$ indicates the
integer value of the argument, for $t\in[n\tau,(n+1)\tau)$,
substitute $\lambda=e^{-\gamma \tau}$,  and use Equation
\ref{calvel} to obtain \be v_\tau(t)=e^{-\gamma t}v_0 + \kappa
e^{-\gamma t}\sum_{j=0}^{[\frac{t}{\tau}]}e^{\gamma \tau
j}h(T^j(\xi_0)),\;\;t\ge 0\label{contvel}, \ee and Equation
\ref{posxn} to obtain \be x_\tau(t)=x_0+\frac{1-e^{-\gamma
t}}{\gamma} v_0+\frac{\kappa
}{\gamma}\left(\sum_{j=0}^{[\frac{t}{\tau}]}h(T^j(\xi_0))-e^{-\gamma
t}\sum_{j=0}^{[\frac{t}{\tau}]}e^{\gamma \tau
j}h(T^j(\xi_0))\right).\label{contpos}\ee

Observe that the first moment of $v_\tau(t)$ is equal to
\[
\int v_\tau(t)\nu(dy)=e^{-\gamma t}\int v_0(y)\nu(dy),
\]
since the random variables $h(T^j(\xi_0))$ have  a first moment equal to $0$.  Assume for simplicity that $\int
h(y)h( T^j(y))\nu(dy)=0$ for $j \geq 1$ and set $\sigma^2=\int h^2(y)\nu(dy)$. Since the random variables $v_0$
and $h(T^j(\xi_0))$ are independent, the second moment of $v_\tau(t)$ takes the form
\begin{eqnarray*}
\int v_\tau(t)^2\nu(dy)&=&e^{-2\gamma t}\int v_0^2(y)\nu(dy) +\kappa^2 e^{-2\gamma
t}\sum_{j=0}^{[\frac{t}{\tau}]}e^{2\gamma \tau j}\int
h^2(T^j(y))\nu(dy)\\
 &=& e^{-2\gamma t}\int v_0^2(y)\nu(dy)+\sigma^2 \kappa^2
e^{-2\gamma t}\frac{1-e^{2\gamma\tau([\frac{t}{\tau}]+1)}}{1-e^{2\gamma \tau}}.
\end{eqnarray*}
If $\sigma$ and $\gamma$ do not depend on $\tau$, we have
\[
\lim_{\tau\to 0} \sigma^2e^{-2\gamma
t}\left(e^{2\gamma\tau([\frac{t}{\tau}]+1)}-1\right)\frac{\tau}{e^{2\gamma
\tau}-1}=\frac{\sigma^2}{2\gamma}(1-e^{-2\gamma t}).
\]
Hence the limit of the second moment of $v_\tau(t)$ as $\tau\to 0$ is finite and positive if and only if $\kappa
$ depends  on $\tau$ in such a way that
\[
\lim_{\tau\to 0}\frac{\kappa_\tau^2}{\tau}
\]
is finite and positive.

\cite{beck87} take $\kappa_\tau=\sqrt{\tau}$ from the outset, and claim that in the limit $\tau\to 0$ the
process $v_\tau(t)$ converges to the Ornstein-Uhlenbeck velocity process when the sequence $(h\circ T^j)$ has a
so called $\phi$-mixing property\footnote{A sequence of random variables $\{\xi_j:j\ge 0\}$ is called
$\phi$-mixing if
\[
\lim_{n\to\infty}\sup\{\frac{|\Pr(A\cap B)-\Pr(A)\Pr(B)|}{\Pr(A)}: A\in \F_{1}^{k}, b\in \F_{k+n}^\infty, k\ge
1\}= 0
\]
where $\F_{1}^{k}$ and $\F_{k+n}^\infty$ denote the $\sigma$-algebra generated by the random variables
$\xi_1,\ldots,\xi_k$ and $\xi_{k+n},\xi_{k+n+1},\ldots$ respectively.
  }
on the probability space $(Y,\B,\nu)$.  In fact, the following result can be proved.

\begin{thm}\label{OUvelocity}
Let $(Y,\B,\nu)$ be a normalized measure space and $T:Y\to Y$ be
ergodic with respect to $\nu$. Let $h\in L^2(Y,\B,\nu)$  be such
that $ \sum_{n=0}^\infty ||\FP_{T,\nu}^n h||_2<\infty $ and let
\[\sigma=\left(\int h(y)^2\nu(dy)+2\sum_{n=1}^\infty\int h(y)h( T^n(y))\nu(dy)\right)^{1/2}\] be positive.
Assume that $\gamma>0$ and
\[
\lim_{\tau\to 0}\frac{\kappa_\tau^2}{\tau}=1.
\]
Then for each $v_0$ the finite dimensional distributions of the velocity process $v_\tau$ given by Equation
\ref{contvel} converge weakly as $\tau\to 0$ to the finite dimensional distributions of the Ornstein-Uhlenbeck
velocity process $V$ for which $V(0)=v_0$ and $\sigma_0=\sigma$.
\end{thm}
{\noindent \bf Proof.} By Theorem \ref{FCLT1} we have
$\FP_{T,\nu}(\tilde{h})=0$ where $\tilde{h}=h+f- f\circ T$ and $
f=\sum_{n=1}^\infty \FP_{T,\nu}^n h$. 
For $t\ge 0$ and $\tau>0$ define
\[
\zeta_\tau(t)=\kappa_\tau e^{-\gamma
t}\sum_{j=0}^{[\frac{t}{\tau}]}e^{\gamma \tau j}\tilde{h}\circ
T^j,\;\;\tilde{\zeta}_\tau(t)=\kappa_\tau e^{-\gamma
t}\sum_{j=0}^{[\frac{t}{\tau}]}e^{\gamma \tau j}(f\circ
T^{j+1}-f\circ T^j).
\]
Then
\[
v_\tau(t)=e^{-\gamma t}v_0 + \zeta_\tau(t)+ \tilde{\zeta}_\tau(t).
\]
Observe that
\[
\tilde{\zeta}_\tau(t)=\kappa_\tau e^{-\gamma t} (e^{\gamma \tau [\frac{t}{\tau}]}f\circ
T^{[\frac{t}{\tau}]+1}-f)+\kappa_\tau e^{-\gamma t}(e^{-\gamma \tau}-1)\sum_{j=1}^{[\frac{t}{\tau}]}e^{\gamma
\tau j}f\circ T^j.
\]
Hence
\[
||\tilde{\zeta}_\tau(t)||_2\le 2 |\kappa_\tau| e^{-\gamma t} (e^{\gamma \tau [\frac{t}{\tau}]}+1)||f||_2,
\]
and consequently
\[
||\tilde{\zeta}_\tau(t)||_2\le 4|\kappa_\tau|||f||_2,\;\;t\ge 0,\;\;\tau>0.
\]
This and Lemma \ref{t:Gaussian} imply that the finite dimensional distributions of $v_\tau(t)-e^{-\gamma t}v_0$
converge weakly to the corresponding finite dimensional distributions of a zero mean Gaussian process $\zeta$
with $\zeta(0)=0$ and the covariance function $K_{\zeta}(t,s)$ given by Equation \ref{covOU} where $\sigma_0^2=
||\tilde{h}||_2^2$, 
which completes the proof.

For the corresponding position process we have the following.

\begin{thm}
Under the assumptions of Theorem \ref{OUvelocity}, let $V(0)=v_0$. Then for each $x_0$ the finite dimensional
distributions of the position process $x_\tau$ given by Equation \ref{contpos} converge weakly as $\tau\to 0$ to
the finite dimensional distributions of the Ornstein-Uhlenbeck position process $X$ for which $X(0)=x_0$.
\end{thm}

{\noindent\bf Proof.} This follows from Lemma \ref{t:Gaussianpos} similarly as the preceding theorem follows
from Lemma \ref{t:Gaussian}.

\begin{exmp}
{\em Let us apply Theorem \ref{OUvelocity} to a transformation $T:[-1,1]\to [-1,1]$ and $h(y)=y$. 
We have $\FP_{T,\nu}h=0$ when $T$ is the hat map, Equation
\ref{hat}. Then $\sigma^2=1/3$. Thus all assumptions of Theorem
\ref{OUvelocity} are satisfied. When $T$ is one of the Chebyshev
maps (\ref{chebyeqn}) $S_N$ we also have $\FP_{T,\nu}h=0$ by Example
\ref{exp:sol} and $\sigma^2=1/2$.  For the dyadic map
(\ref{dyadic}), the series $\sum_{n=1}^\infty \FP_{T,\nu}^nh$ is
absolutely convergent in $L^2([-1,1],\B([-1,1]),\nu)$ and is equal
to $h$. This implies that $\sigma=||2 h-h\circ T||_2$, thus in this
case $\sigma=1$, which can be easily calculated.  Thus all of the
numerical examples and studies of Beck and co-workers cited above
are covered by this example.}
\end{exmp}

\setcounter{equation}{0} \setcounter{figure}{0}
\section{Discussion}\label{disc}

In this paper we were motivated by the strong statistical
properties of discrete dynamical systems to consider  when
Brownian motion like behaviour could emerge in a simple toy
system.  To do this, we have reviewed and significantly extended a
class of central limit theorems for discrete time maps.  These new
results, presented primarily in Sections \ref{s:brownian} and
\ref{technical}, were then applied in Section  \ref{anal} to the
Langevin-like equations
\begin{eqnarray}\nonumber \dfrac{dx (t)}{dt}&=&v(t),\\
\nonumber\frac{dv(t)}{dt} & = & -\gamma v(t) + \kappa_\tau \eta(t)
\end{eqnarray} in which the underlying noise $\eta(t)$ need not be a
Gaussian noise but may be substituted by
\[
\eta(t)=\sum_{n=0}^\infty h(\xi(t)) \delta(t-n\tau)
\]
with a highly irregular deterministic function $\xi(n\tau)$. When
the variables $h(\xi(n\tau))$ are uncorrelated Gaussian
distributed (thus in fact independent) random variables then the
limiting distribution of $v(n\tau)$ is  Gaussian. This is need not
be the case for the deterministic noise produced by perturbations
derived from highly chaotic semi-dynamical systems. However
(Section \ref{tau}), in the  limit $\tau\to 0$ both types of noise
produce the same stochastic process in this limit, the
Ornstein-Uhlenbeck process. Finally, in Section \ref{identify} we
have illustrated all of our results of Section \ref{anal} for the
specific case of perturbations derived from the {\it exact} dyadic
map.

The significance of these considerations is rather broad.  It is
the norm in experimental observations that any experimental
variable that is recorded will be ``contaminated" by ``noise".
Sometime the distribution of this noise is approximately Gaussian,
sometimes not.  The considerations  here illustrate quite
specifically that the origins of the noise observed experimentally
need not be due to the operation of a random process (random in
the sense that there is no underlying physical law allowing one to
predict exactly the future of the process based on the past).
Rather, the results we present strongly suggest, as an
alternative, that the fluctuations observed experimentally might
well be the signature of an underlying deterministically chaotic
process.

\section*{Acknowledgments}
This work was supported by MITACS (Canada) and the Natural Sciences
and Engineering Research Council (NSERC grant OGP-0036920, Canada).
The first draft of this paper was written while the second author
was visiting McGill University, whose hospitality and support are
gratefully acknowledged.  We thank Rados{\l}aw Kami\'nski for
providing the computations on which Figure \ref{s0009} was based.
\appendix

\setcounter{equation}{0} \setcounter{figure}{0}
\section{Appendix: Limit Theorems for Dependent Random
Variables}\label{triangular}

This Appendix reviews  known general central limit theorems from
probability theory which we can use  directly in the context of
noninvertible dynamical systems. With their help we then prove
several results which we have used  in the preceding Sections.

Consider random variables arranged in a double array
\begin{eqnarray}
 & \zeta_{1,1}, \zeta_{1,2}, \ldots, \zeta_{1,k_1}  \nonumber \\
 & \zeta_{2,1}, \zeta_{2,2}, \ldots, \zeta_{2,k_2} \\
 \label{array}
 &\vdots  \nonumber\\
& \zeta_{n,1}, \zeta_{n,2}, \ldots, \zeta_{n,k_n}\nonumber  \\
&\vdots\nonumber
  \end{eqnarray}
with $k_n\to\infty$ as $n\to \infty$. We shall give conditions
which imply that the row sums converge in distribution to a
Gaussian random variable $\sigma N(0,1)$, that is \be
\sum_{i=1}^{k_n} \zeta_{n,i}\to^d \sigma N(0,1).
\label{limitarray}\ee We shall require the {\it Lindeberg
condition} \be \lim_{n\to\infty}\sum_{i=1}^{k_n}E(\zeta_{n,i}^2
1_{\{|\zeta_{n,i}|>\epsilon\}})= 0 \;\;\mbox{for
every}\;\;\epsilon>0.\;\;\label{lindp}\ee


If independence in each row is allowed, then we have the classical
Lindeberg-Feller theorem.
\begin{thm}(\citet[Theorem 7.2.1]{chung})\label{lfthm}
Let  the random variables $\zeta_{n,1},\ldots,\zeta_{n,k_n}$ be
independent for each $n$. Assume that $E(\zeta_{n,i})=0$ and
$\sum_{i=1}^{k_n}E(\zeta_{n,i}^2)=1$, $n\ge 1$.  Then  the
Lindeberg condition holds if and only if
\[
\sum_{i=1}^{k_n} \zeta_{n,i}\to^d  N(0,1),
\]
and  \be \mbox{for all}\;\; \delta>0,\;\;\max\limits_{1\le i\le
k_n} \Pr\{|\zeta_{n,i}|>\delta\}\to 0 \label{neg}.\ee
\end{thm}
If one considers  a map $T$ on a probability space $(Y,\B,\nu)$
which preserves the measure $\nu$, and defines $\zeta_{n,i}$ to be
$\frac{1}{\sqrt{n}} f\circ T^{i-1}$ for $i=1,\ldots, n$ with $f$
measurable, then Condition \ref{neg} holds.  This is because
\[
\Pr\{|\zeta_{n,i}|>\delta\}=\nu (\{y\in Y: |f(T^{i-1}(y))|>\delta
\sqrt{n}\}),
\]
and $\{y\in Y: |f(T^{i-1}(y))|>\delta \sqrt{n}\}=T^{-i+1}(\{y\in
Y: |f(y)|>\delta \sqrt{n}\})$. Thus by the invariance of $\nu$
this leads to
\[
\max\limits_{1\le i\le k_n}\Pr\{|\zeta_{n,i}|>\delta\}=\nu(\{y\in
Y: |f(y)|>\delta \sqrt{n}\})\to 0.
\]
Similarly, if one takes a square integrable $f$, then the
Lindeberg condition \ref{lindp} is satisfied. Indeed, we have
\[
E(|\zeta_{n,i}|^{2}1_{\{|\zeta_{n,i}|>\epsilon\}}) =
\frac{1}{\sqrt{n}^{2}}\int_{\{z:|f(T^{i-1}(z))|\ge
\sqrt{n}\epsilon\}} f^2(T^{i-1}(y))\nu(dy) \] and by the change of
variables applied to $T^{i-1}$ this reduces to
\[
\frac{1}{{n}}\int_{\{|f|\ge \sqrt{n}\epsilon\}} f^2(y) \nu(dy).
\]
Hence
\[
\sum_{i=1}^{n}E(|\zeta_{n,i}|^{2}1_{\{|\zeta_{n,i}|>\epsilon\}})=\int_{\{|f|\ge
\sqrt{n}\epsilon\}} f^2(y) \nu(dy),
\]
which converges to $0$ by the Dominated Convergence Theorem under
our assumption that $f^{2}$ is integrable.

Since our random variables are dependent, we cannot apply the
above theorem. Instead, we use the notion of martingale
differences for which there is a natural generalization of Theorem
\ref{lfthm}. Moreover, additional assumptions are needed, as one
can easily check that if $T$ is the identity map, then
\[
\frac{1}{\sqrt{n}} \sum_{i=1}^n f\circ T^{i-1}=\frac{n}{\sqrt{n}}f
\]
which can not be convergent to a Gaussian random variable.

We first recall the definition of conditional expectation. Let a
probability space $(\Omega,\F,\Pr)$ be given and let
$\mathcal{G}$ be a sub-$\sigma$-algebra of $\F$. 
For $\zeta\in L^1(\Omega,\F,\Pr)$ there exists a random variable
$E(\zeta|\mathcal{G})$, called the {\it conditional expected value
} of $\zeta$ given $\mathcal{G}$, having the following properties:
it is $\mathcal{G}$ measurable, integrable and satisfies the
equation
\[
\int_{A} E(\zeta|\mathcal{G})(\omega)\Pr (d\omega)= \int_{A}
\zeta(\omega) \Pr(d\omega),\;\;A\in\mathcal{G}.
\]
The existence and uniqueness of $E(\zeta|\mathcal{G})$ for a given
$\zeta$ follows from the Radon-Nikodym theorem. The transformation
$\zeta\mapsto E(\zeta|\mathcal{G})$ is a linear operator between
the spaces $L^1(\Omega,\F,\Pr)$ and $L^1(\Omega,\mathcal{G},\Pr)$,
so sometimes it is called an {\it operator of conditional
expectation}.

Let $\{\zeta_{n,i}: 1\le i\le k_n, n\ge 1\}$ be a family of random
variables defined on a probability space $(\Omega,\F,\Pr)$. For
 each $n\ge 1$, let  a family $\{\F_{n,i}: i\ge 0\}$ of
sub-$\sigma$-algebras of $\F$ be given. Consider the following set
of conditions \bit \item[(i)] $E(\zeta_{n,i})=0$ and
$E(\zeta_{n,i}^2)<\infty$, \item[(ii)] $\F_{n,i-1}\subseteq
\F_{n,i}$, \item[(iii)] $\zeta_{n,i}$ is $\F_{n,i}$ measurable,
\item[(iv)] $E(\zeta_{n,i}|\F_{n,i-1})=0$ for each $1\le i\le
k_n$, $n\ge 1$. \eit A family $\{\F_{n,i},\zeta_{n,i}:1\le i\le
k_n, n\ge 1\}$ satisfying conditions (i)-(iv) is called a {\it
(square integrable) martingale differences array}.

The next theorem is  from \cite{billingsley95}.

\begin{thm}(\citet[Theorem 35.12]{billingsley95})\label{CLGen}
Let $\{\zeta_{n,i}:1\le i\le k_n, n\ge 1\}$ be a martingale
difference array satisfying the Lindeberg condition \ref{lindp}.
If \be \sum_{i=1}^{k_n}E(\zeta_{n,i}^2|\F_{n,i-1})\to^{P}
\sigma^2, \label{varianceb}\ee
 where $\sigma$ is a nonnegative
constant, then \be \sum_{i=1}^{k_n} \zeta_{n,i}\to^d \sigma
N(0,1). \ee
\end{thm}

If the limit in Condition \ref{varianceb} is a random variable
instead of the constant $\sigma^2$, we obtain convergence to
mixtures of normal distributions \citet[Corollary p.
561]{eagleson75}.

\begin{thm}\label{CLEgl}
Let $\{\zeta_{n,i}:1\le i\le k_n, n\ge 1\}$ be a martingale
difference array satisfying the Lindeberg condition \ref{lindp}.
If there exists an
$\F_\infty=\bigcap_{n=1}^{\infty}\F_{n,0}$-measurable, a.s.
positive and finite random variable $\eta$ such that \be
\sum_{i=1}^{k_n}E(\zeta_{n,i}^2|\F_{n,i-1})\to^{P} \eta,
\label{condvar}\ee then $\sum_{i=1}^{k_n} \zeta_{n,i}$ is
convergent in distribution to a measure whose characteristic
function is $\varphi(r)=E(\exp(-\frac{1}{2}r^2\eta)).$
\end{thm}

The above result shows that to obtain a normal distribution in the
limit a specific normalization is needed and we have the following

\begin{thm}(\citet[Theorem 3.6]{gaenslerjoos})\label{CLTRN}
Let $\{\zeta_{n,i}:1\le i\le k_n, n\ge 1\}$ be a martingale
difference array and let $\eta$ be a real-valued random variable
such that $\Pr(0<\eta<\infty)=1$. Suppose that \be
\lim_{n\to\infty}E(\max_{1\le i\le
k_n}|\zeta_{n,i}|)=0\label{llin}\ee and \be
\sum_{i=1}^{k_n}\zeta_{n,i}^2\to^{P} \eta. \label{variancevar}\ee If
$\eta$ is $\F_{n,i}$-measurable for each $n$ and for each $1\le i\le
k_n$, then \[ \frac{\sum_{i=1}^{k_n}
\zeta_{n,i}}{\sqrt{\sum_{i=1}^{k_n}\zeta_{n,i}^2}}\to^d N(0,1).
\]
\end{thm}

\begin{rem}
If $\eta$ in Condition \ref{variancevar} is constant,
$\eta=\sigma^2$, then the conclusion of Theorem \ref{CLTRN} is
equivalent to
\[
\sum_{i=1}^{k_n} \zeta_{n,i}\to^d \sigma N(0,1).
\]
Note also that Condition \ref{llin} is implied by the Lindeberg
condition.
\end{rem}

The next result gives conditions for moment convergence in Theorem
\ref{CLGen}.

\begin{thm}(\citet[Theorem]{hall78},\citet[Theorem 3]{teicher})
Let $\{\zeta_{n,i}:1\le i\le k_n, n\ge 1\}$ be a martingale
difference array with $\sum_{i=1}^{k_n}E(\zeta_{n,i}^2)=\sigma^2$
where $\sigma>0$. Suppose that for $p>1$ \be
\sum_{i=1}^{k_n}E|\zeta_{n,i}|^{2p}\to 0\qquad\mbox{and}\qquad
E|\sum_{i=1}^{k_n}E(\zeta_{n,i}^2|\F_{n,i-1})-\sigma^2|^{p}\to
0\label{c:moments}. \ee Then
\[
\lim_{n\to\infty}E|\sum_{i=1}^{k_n}\zeta_{n,i}|^{2p}=E|N(0,\sigma^2)|^{2p}.
\]
\end{thm}

Let $(Y,\B,\nu)$ be a normalized measure space and $T:Y\to Y$ be a
measurable map such that $T$ preserves the measure $\nu$. Recall
from Section \ref{sdsys} the relation \ref{fpcond} between the
transfer operator $\FP_{T,\nu}$, the Koopman operator and the
operator of conditional expectation which gives
\[
\FP_{T,\nu}\circ U_T f=f\;\;\mbox{and}\;\;U_T\circ \FP_{T,\nu}
f=E(f|T^{-1}(\B)),\;\;f\in L^1(Y,\B,\nu).
\]

\begin{lem}\label{mdarr} Let $(Y,\B,\nu)$ be a normalized measure space, and $T:Y\to Y$ be a
measurable map such that $T$ preserves the measure $\nu$. Let
$\{c_{n,i}: 1\le i\le k_n, n\ge 1\}$ be a family of real numbers and
$h\in L^2(Y,\B,\nu)$. Suppose that $\FP_{T,\nu}h=0$. Then
\[
\zeta_{n,i}=c_{n,i}\,h\circ T^{k_n-i},\;\;1\le i\le k_n,
\;\;\zeta_{n,i}=0,\;\;i> k_n,
\]
with
\[
\F_{n,i}=T^{-k_n + i}(\B),\;\;0\le i\le k_n, \;\;\mbox{and}\;\;
\F_{n,i}=\B,\;\; i> k_n,\;\;n\ge 1
\]
is a martingale difference array and if
$c_{n,i}=\frac{1}{\sqrt{k_n}}$, $1\le i\le k_n$ then the following
hold
\begin{description}
    \item[(i)]  Lindeberg condition \ref{lindp};
    \item[(ii)] Conditions \ref{condvar} and \ref{variancevar} with $\eta=E(h^2|\mathcal{I})$ where $\mathcal{I}$ is the
$\sigma$-algebra of all $T$-invariant sets;
    \item[(iii)] Condition \ref{varianceb}
with $\sigma^2=\int h^2d\nu$ provided that $T$ is ergodic;
    \item[(iv)] Condition \ref{c:moments} for every $p>1$ provided that $T$ is ergodic and $h\in L^{\infty}(Y,\B,\nu)$.
\end{description}
%
\end{lem}
{\noindent\bf Proof.} To check conditions (ii), and (iii) of the
definition of a martingale difference array, observe that
$T^{-j-1}(\B)\subset T^{-j}(\B)$ and $h\circ T^j$ is $T^{-j}(\B)$
measurable. The Koopman and transfer operators for the iterated map
$T^j$ are just the j$^{th}$ iterates of the operators $U_T$ and
$\FP_{T,\nu}$. From this and Equation \ref{fpcond} we have
$\FP_{T,\nu}^j U_T^j h=h$ and
\[
E(h\circ T^j|T^{-j-1}(\B))=U_T^{j+1} \FP_{T,\nu}^{j+1}(h\circ
T^j)=U_T^{j+1} \FP_{T,\nu} h.
\]
Since $\FP_{T,\nu}h=0$, we see that $E(h\circ T^j|T^{-j-1}(\B))=0$
for $j\ge 0$ which proves condition (iv).

The Lindeberg condition reduces, through a change of variables, to
\[
\sum_{i=1}^{k_n}E(\zeta_{n,i}^{2}1_{\{|\zeta_{n,i}|>\epsilon\}})=\int_{\{|h|\ge
\sqrt{k_n}\epsilon\}} h^{2} \nu(dy),
\]
but $h^2$ is integrable and the Lindeberg condition follows. To
obtain Condition \ref{condvar} use Equation \ref{fpcond}, change
the order of summation
\begin{eqnarray*}
\sum_{i=1}^{k_n}E(\zeta_{n,i}^2|\F_{n,i-1})&=&
\frac{1}{k_n}\sum_{i=1}^{k_n}E(h^2\circ
T^{k_n-i}|T^{-k_n+i-1}(\B))\\
&=& \frac{1}{k_n}\sum_{i=1}^{k_n}U_T^{k_n-i+1} \FP_{T,\nu}^{k_n-i+1}U_T^{k_n-i} (h^2)\\
&=&\frac{1}{k_n}\sum_{i=1}^{k_n}U_T^{k_n-i+1}
\FP_{T,\nu}(h^2)=\frac{1}{k_n}\sum_{i=0}^{k_n-1}U_T^{i+1}
\FP_{T,\nu} (h^2)
\end{eqnarray*}
and apply Birkhoff's ergodic theorem to the integrable function
$U_T\FP_{T,\nu}(h^2)$ to conclude that this sequence is convergent
to  $E( U_T\FP_{T,\nu} (h^2)|\mathcal{I})$  almost everywhere (with
respect to $\nu$), and consequently in probability. Since
$U_T\FP_{T,\nu}(h^2)=E(h^2|T^{-1}(\B))$ and
$\mathcal{I}\subseteq{T^{-1}(\B)}$, we have $E( U_T\FP_{T,\nu}
(h^2)|\mathcal{I})=E(h^2|\mathcal{I})$. Similarly, Condition
\ref{variancevar} follows from the Birkhof ergodic theorem. In
addition, if $T$ is ergodic, then $\eta$ is constant a.e. and is
equal to $\int h^2 d\nu$. Since $h^2\in L^p(Y,\B,\nu)$ and $p>1$, we
have
\[
\sum_{i=1}^{k_n}E|\zeta_{n,i}|^{2p}=\frac{1}{n^p}\int
h^{2p}(y)\nu(dy)\to 0
\]
and $U_T\FP_{T,\nu} (h^2)\in L^p(Y,\B,\nu)$. By the ergodic theorem
in $L^p$ spaces we  get
\[\sum_{i=1}^{k_n}E(\zeta_{n,i}^2|\F_{n,i-1})=\frac{1}{k_n}\sum_{i=0}^{k_n-1}U_T^{i}
U_T\FP_{T,\nu} (h^2)\to \int U_T\FP_{T,\nu} (h^2)(y)\nu(dy)
\]
in $L^p(Y,\B,\nu)$, but $\int U_T\FP_{T,\nu} (h^2)(y)\nu(dy)=\int
h^2(y)\nu(dy)$ and the proof is complete.

We now turn to the FCLT for $(h\circ T^i)_{i\ge 0}$. Let
$(Y,\B,\nu)$ be a normalized measure space and $T:Y\to Y$ be
ergodic with respect to $\nu$. Let $h\in L^2(Y,\B,\nu)$ and
$\sigma=||h||_2>0$. We define a random function
\[
\psi_n(t)=\frac{1}{\sigma \sqrt{n}}\sum_{i=0}^{[nt]-1}h\circ T^{i}
\;\;\mbox{for}\;\;t\in [0,1]
\]
(where the sum from $0$ to $-1$ is set to be $0$). Note that
$\psi_n$ is a right continuous step function, a random variable of
$D[0,1]$ and $\psi_n(0)=0$.


\begin{lem}\label{finitedim}
If $\FP_{T,\nu}h=0$, then the finite dimensional distributions of
$\psi_n$ converge to those of the Wiener process $w$.
\end{lem}
{\noindent \bf Proof.} To show that the finite dimensional
distributions of $\psi_n$ converge to the corresponding finite
dimensional distributions of $w$ we use the Cram\'er-Wold
technique. If the $c_1,\ldots,c_k$ are arbitrary real numbers and
$t_0=0<t_1<\ldots <t_k\le 1$,  we put 
\[
\zeta_{n,i}= \left\{
 \begin{array}{ll}
\dfrac{c_j}{\sigma\sqrt{n}}h\circ T^{n-i}, &n-[nt_j]<i\le
n-[nt_{j-1}],\;j=1,\ldots,k
\\
 0, &1\le i\le n-[nt_k]\;\;\mbox{and}\;\;t_k<1\\
 0,&i>n
\end{array}
\right. \] Observe that
\[
\sum_{i=1}^{n}\zeta_{n,i}=\sum_{j=1}^k
c_j(\psi_n(t_j)-\psi_n(t_{j-1})).
\]
By Lemma \ref{mdarr} $\zeta_{n,i}$ 
is a martingale differences array and we will verify the
conditions of Theorem \ref{CLGen}. For the Lindeberg condition
note that
\[
\sum_{i=1}^nE(\zeta_{n,i}^21_{\{|\zeta_{n,i}|>\epsilon\}})=\sum_{j=1}^{k}\frac{c_j^2}{\sigma^2
n}\left([nt_j]-[nt_{j-1}])E(h^21_{\{|h|>\epsilon\sigma
\sqrt{n}c_j^{-1}\}}\right)
\]
and as a finite sum of sequences converging to $0$ it is
convergent to $0$. For Condition \ref{varianceb}, observe that
\[
\sum_{i=1}^nE(\zeta_{n,i}^2|T^{-n+i-1}(\B))=\sum_{j=1}^{k}\frac{c_j^2}{\sigma^2
n}\sum_{i=[nt_{j-1}]+1}^{[nt_j]}U_T^iP_T(h^2)\to^P \sum_{j=1}^k
c_j^2(t_j-t_{j-1})
\]
by the ergodicity of $T$, and the fact that $\sigma^2=\int
h^2d\nu$. Therefore, by Theorem \ref{CLGen},
\[
\sum_{i=1}^{n}\zeta_{n,i}\to^d \sqrt{\sum_{j=1}^k
c_j^2(t_j-t_{j-1})}\,\,N(0,1).
\]
Thus $\sum_{j=1}^k c_j(\psi_n(t_j)-\psi_n(t_{j-1}))$ converges to
the Gaussian distributed random variable with mean $0$ and
variance $\sum_{j=1}^k c_j^2(t_j-t_{j-1})$, but this is the
distribution of  $\sum_{j=1}^k c_j(w(t_j)-w(t_{j-1}))$ which
completes the proof.

%
%
%
%

\begin{lem}\label{FCLGen}
If $\FP_{T,\nu}h=0$, then Condition \ref{tight} holds for each
positive $\epsilon$.
\end{lem}
{\noindent\bf Proof.} Define a martingale difference array
\[
\zeta_{n,i}=\left\{
 \begin{array}{ll}\frac{1}{\sigma\sqrt{n}}
h\circ T^{n-i}, & 1\le i\le n\\ 0, & i>n ,\;\;n\ge 1.
\end{array}\right.
\]
Let also $\zeta_{n,0}=0$ and
$\widetilde{\psi}_n(t)=\sum_{i=0}^{[nt]} \zeta_{n,i}$,
$t\in[0,1]$, $n\ge 1$. We have
\[
\psi_n(t)=\widetilde{\psi}_n(1)-\widetilde{\psi}_n(1-t).
\]
We first observe that
\begin{eqnarray*}
  \sup_{|t-s|\le\delta}|\psi_n(s)-\psi_n(t)| &\le & \sup_{|t-s|\le\delta}|\widetilde{\psi}_n(s)-\widetilde{\psi}_n(t)| \\
  &\le & 4\sup_{k}\sup_{k\delta< t\le
(k+1)\delta}|\widetilde{\psi}_n(t)-\widetilde{\psi}_n(k\delta)|.
\end{eqnarray*}
This gives
\[
\nu(\sup_{|t-s|\le\delta}|\psi_n(s)-\psi_n(t)|>\epsilon)\le
\sum_{k\delta<1}\nu(\sup_{k\delta< t\le
(k+1)\delta}|\widetilde{\psi}_n(t)-\widetilde{\psi}_n(k\delta)|>\frac{\epsilon}{4}).
\]
Now applying Lemma \ref{finitedim} and arguments similar to those
of \citet[pp. 64-65]{brown71}, one can complete the proof.

For the next results we need the following

\begin{lem}\label{naverage}
Let $(z_i)_{i\ge 1}$ be a sequence of real numbers such that
\[
\lim_{n\to\infty}\dfrac{\sum_{i=1}^n z_i}{n}=z.
\]
Then
\[
\lim_{n\to\infty}\dfrac{\sum_{i=1}^{k_n}
a_n^{i}z_i}{\sum_{i=1}^{k_n} a_n^{i}}=z
\]
for every sequence of integers $k_n\ge 1$ and every sequence of
real numbers $a_n$ satisfying \be
\lim_{n\to\infty}k_n=\infty,\;\;\lim_{n\to\infty}a_n=1,\;\;\mbox{and}\;\;\lim_{n\to\infty}a_n^{k_n}\neq
1 \label{scallim}\ee and either $a_n> 1$ or $0<a_n< 1$ for all
$n\ge 1$.
\end{lem}
{\noindent\bf Proof.} The Abel method of summation,
\[
\sum_{i=k}^m c_ib_i=c_m\sum_{i=k}^mb_i
-\sum_{i=k}^{m-1}(c_{i+1}-c_i)\sum_{j=k}^ib_i
\]
can be used to write
\[
\sum_{i=1}^{k_n} a_n^{i}z_i-z\sum_{i=1}^{k_n}
a_n^{i}=-\sum_{i=1}^{k_n-1}(a_n^{i+1}-a_n^{i})(\sum_{j=1}^i
 z_j -iz)+ a_n^{k_n} (\sum_{i=1}^{k_n} z_i - k_n z).
\]
Fix $\epsilon>0$ and let $n_0$ be such that
\[
\left|\frac{\sum_{i=1}^{m}z_i }{m}-z\right|\le
\epsilon\;\;\mbox{for}\;\;m\ge n_0.
\]
Suppose that $a_n>1$ for all $n\ge 1$. The other case is proved
analogously.  Combining these yields, for $k_n>n_0$,
\[
\left|\frac{\sum_{i=1}^{k_n} a_n^{i}z_i}{\sum_{i=1}^{k_n}
a_n^{i}}-z\right|\le
\frac{\sum_{i=1}^{n_0-1}(a_n^{i+1}-a_n^{i})\left|{\sum_{j=1}^i
 z_j} -iz\right|}{\sum_{i=1}^{k_n}
a_n^{i}}+ \epsilon\frac{\sum_{i=n_0}^{k_n-1}(a_n^{i+1}-a_n^{i})i
+a_n^{k_n} k_n}{\sum_{i=1}^{k_n} a_n^{i}}.
\]
Letting $n\to\infty$ we see that the first term on the right goes
to zero, while the second term goes to $\epsilon$ times a constant
not depending on $\epsilon$, which completes the proof.

\begin{thm}\label{t:ndifferent}
Let $(Y,\B,\nu)$ be a normalized measure space and $T:Y\to Y$ be
ergodic with respect to $\nu$. Let $(k_n)$, $(a_n)$ be sequences
satisfying Condition \ref{scallim}. Let $c\in\realnos$ and let
$(c_n)$ be a sequence of real numbers such that
\[\lim_{n\to\infty}k_nc_n^2=c^2.\]
If $h\in L^2(Y,\B,\nu)$ is such that $\FP_{T,\nu}h=0$, then
\[
c_n\sum_{i=1}^{k_n}a_n^{i} h\circ T^{i}\to^d \sigma N(0,1),
\]
where $\sigma=\sqrt{\dfrac{c^2(a^2-1)}{\ln a^2}}||h||_2$ and
$a=\lim\limits_{n\to\infty}a_n^{k_n}$.
\end{thm}
{\noindent \bf Proof.} From Lemma \ref{mdarr} it follows that
$\zeta_{n,i}=c_n a_n^{k_n+1-i}h\circ T^{k_n+1-i}$ is a martingale
difference array and that
\[
\lim_{n\to\infty}\frac{1}{\sqrt{k_n}}E(\max_{1\le j\le k_n}|h\circ
T^j|)=0.
\]
We shall apply Theorem \ref{CLTRN}. We have
\[
\max_{1\le i\le k_n} |\zeta_{n,i}|\le |c_n|
\max(a_n^{k_n},1)\max_{1\le i\le k_n}|h\circ T^{i}|,
\]
so
\[
E(\max_{1\le i\le k_n} |\zeta_{n,i}|)\le \sqrt{k_n}|c_n|
\max(a_n^{k_n},1)E(\frac{1}{\sqrt{k_n}}\max_{0\le i\le k_n}|h\circ
T^{i}|).
\]
Letting $n\to\infty$ we see that Condition \ref{llin} holds. To
verify Condition \ref{variancevar} note that
\begin{eqnarray*}
\sum_{i=1}^{k_n}\zeta_{n,i}^2=c_n^2 \sum_{i=1}^{k_n}
a_n^{2i}h^2\circ T^{i}=c_n^2 \sum_{i=1}^{k_n}
a_n^{2i}\frac{\sum_{i=1}^{k_n} a_n^{2i} h^2\circ
T^{i}}{\sum_{i=1}^{k_n} a_n^{2i}}.
\end{eqnarray*}
Therefore Birkhoff's ergodic theorem,  Lemma \ref{naverage}, and
the fact that
\[
\lim_{n\to\infty}c_n^2 \sum_{i=1}^{k_n} a_n^{2i}= \frac{c^2
(a^2-1)}{\ln a^2}
\]
complete the proof.

\begin{cor}\label{ndifferent}
Under the assumptions of Theorem \ref{t:ndifferent}, if for $h\in
L^2(Y,\B,\nu)$ the series $\sum_{n=0}^\infty \FP_{T,\nu}^n h $ is
convergent in $L^2(Y,\B,\nu)$, then
\[
c_n\sum_{i=1}^{k_n}a_n^{i} h\circ T^{i}\to^d \sigma N(0,1),
\]
where $\sigma=\sqrt{\dfrac{c^2(a^2-1)}{\ln a^2}}||h+f-f\circ T||_2$
and $ f=\sum_{n=1}^\infty \FP_{T,\nu}^n h$.
\end{cor}
{\noindent\bf Proof.} Theorem \ref{FCLT1} implies
$\FP_{T,\nu}(h+f-f\circ T)=0$. Thus
\[
c_n\sum_{i=1}^{k_n}a_n^{i} (h+f-f\circ T)\circ T^{i}\to^d \sigma
N(0,1)
\]
by Theorem \ref{t:ndifferent}. Therefore it remains to prove that
\be c_n\sum_{i=1}^{k_n}a_n^{i} (f\circ T-f)\circ T^{i}\to^P
0.\label{e:prolim} \ee Observe that the left-hand side of Equation
\ref{e:prolim} is equal to
\[
c_n (a_n^{k_n} f\circ T^{k_n+1}- f\circ T) + c_n (a_n^{-1}-1)
\sum_{i=1}^{k_n}a_n^{i} f\circ T^{i}.
\]
Since $c_n\to 0$ as $n\to\infty$, the first term converges in
probability to $0$.  From Lemma \ref{naverage} and Birkhoff's
ergodic theorem it follows that
\[
\frac{\sum_{i=1}^{k_n}a_n^{i} f\circ
T^{i}}{\sum_{i=1}^{k_n}a_n^{i}}\to^P \int f(y)\nu(dy).
\]
Therefore the sequence
\[
 c_n (a_n^{-1}-1)
\sum_{i=1}^{k_n}a_n^{i} f\circ T^{i}=c_n
(1-a_n^{k_n})\frac{\sum_{i=1}^{k_n}a_n^{i} f\circ
T^{i}}{\sum_{i=1}^{k_n}a_n^{i}}
\]
is also convergent in probability to $0$, which completes the
proof.

\begin{rem}
Note that we can conclude from Theorem \ref{t:ndifferent} 
that
\[
c_n\sum_{i=m}^{k_n}a_n^{i} h\circ T^{i}\to^d \sigma N(0,1),
\]
where $m\ge 0$ is any fixed integer, because  $c_n$ goes to zero
and $a_n$ to $1$ as $n\to \infty$, so the difference
\[
c_n\sum_{i=1}^{k_n}a_n^{i} h\circ T^{i}-c_n\sum_{i=m}^{k_n}a_n^{i}
h\circ T^{i},
\]
which is either equal to $c_nh$ or $c_n\sum_{i=1}^{m}a_n^{i}
h\circ T^{i}$, converges in probability to zero.
\end{rem}

\begin{lem}\label{t:Gaussian}
Let $(Y,\B,\nu)$ be a normalized measure space, $T:Y\to Y$ be
ergodic with respect to $\nu$, and $\gamma\neq 0$ be a constant.
Let $\kappa_\tau$, $\tau>0$, be such that
\[
\lim_{\tau\to 0}\frac{\kappa_\tau^2}{\tau}=1.
\]
If $h\in L^2(Y,\B,\nu)$ is such that $\FP_{T,\nu}h=0$, then the
finite dimensional distributions of the process $\zeta_\tau$ defined
by
\[
\zeta_\tau(t)=\kappa_\tau e^{-\gamma
t}\sum_{j=0}^{[\frac{t}{\tau}]}e^{\gamma \tau j}h\circ
T^j,\;\;t\ge 0,\;\;\tau>0
\]
converge weakly as $\tau\to 0$ to the corresponding finite
dimensional distributions of the zero-mean Gaussian process $\zeta$
for which $\zeta(0)=0$ and
\[E\zeta(t)\zeta(s)=\frac{||h||^2_2}{2\gamma}(e^{2\gamma
\min(t,s)}-1)e^{-\gamma(t+s)},\;t,s>0.\]
\end{lem}
{\noindent \bf Proof.} To prove the convergence of the finite
dimensional distributions of $\zeta_\tau$ to the corresponding
finite dimensional distributions of the Gaussian process $\zeta$,
it is enough to prove that for any $l\ge 1$, real numbers
$0<t_1<...<t_l<\infty$ and $d_1,...,d_l$ the distribution of
$\sum_{j=1}^{l}d_j\zeta(t_j)$ is Gaussian and that
$\sum_{j=1}^{l}d_j\zeta_\tau(t_j)$ converges in distribution
as $\tau\to 0$ to $\sum_{j=1}^{l}d_j\zeta(t_j)$. 

We consider first the case of $l=1$. It follows from Theorem
\ref{t:ndifferent} that for $t>0$ the distribution of
$\zeta_\tau(t)$ converges weakly as $\tau\to 0$ to a Gaussian
random variable.  To see this let $\tau_n$ be a sequence going to
zero as $n\to\infty$. Take $k_n=[\frac{t}{\tau_n}]$,
$a_n=e^{\gamma \tau_n}$, $c_n=\kappa_{\tau_n}e^{-\gamma t}$, and
observe that
\[
\lim_{n\to\infty}k_n=\infty,\;\;\lim_{n\to\infty}a_n=1,\;\;\lim_{n\to\infty}a_n^{k_n}=e^{\gamma
t}\;\;\mbox{and}\;\;\lim_{n\to\infty}k_nc_n^2=te^{-2\gamma t},
\]
and $\frac{te^{-2\gamma t}(e^{2\gamma t}-1)}{\ln e^{2\gamma
t}}=\frac{1-e^{-2\gamma t}}{2\gamma}$. The theorem then implies
that
\[
\kappa_{\tau_n} e^{-\gamma
t}\sum_{j=0}^{[\frac{t}{\tau_n}]}e^{\gamma \tau_n
j}h(T^j(\xi_0))\to
N\left(0,\frac{\sigma_0^2}{2\gamma}(1-e^{-2\gamma t})\right)
\]
where $\sigma_0^2=\int h^2(y)\nu(dy)$ and $\xi_0$ is distributed
according to $\nu$. Consequently $\zeta_\tau(t)\to^d \zeta(t)$ as
$\tau\to 0$, where $\zeta(t)$ is a Gaussian distributed random
variable with mean $0$ and variance given by
\[
\frac{||h||^2_2}{2\gamma}(1-e^{-2\gamma t}), \;\;t>0.
\]
Note that $\zeta_{\tau}(0)=\kappa_\tau h$. Since $\lim_{\tau\to
0}\kappa_\tau=0$, we also have $\zeta_{\tau}(0)\to 0$ as $\tau\to
0$.

We next consider the case of $l=2$.  The case of arbitrary $l$ is
deduced analogously from Theorem \ref{CLTRN}. Let $t_1<t_2$ and
$d_1,d_2$ be given. Let $\tau_n$ be a sequence going to zero as
$n\to\infty$. Set $k_{n,1}=[\frac{t_1}{\tau_n}]$,
$k_{n,2}=[\frac{t_2}{\tau_n}]$, $k_n=k_{n,2}+1$, and observe that
$k_{n,1}<k_{n,2}$ for all $n$ sufficiently large. Define
\[
\eta_{n,j}=\left\{
\begin{array}{ll}
d_2e^{-\gamma t_2}\kappa_{\tau_n}e^{\gamma\tau_n (k_n-j)}h\circ T^{k_n-j}, & 0<j\le k_{n,2}-k_{n,1}, \\
(d_2e^{-\gamma t_2}+d_1e^{-\gamma
t_1})\kappa_{\tau_n}e^{\gamma\tau_n (k_n-j)}h\circ
T^{k_n-j}, & k_{n,2}-k_{n,1}<j\le k_n,\\
0, & \mathrm{otherwise}.
\end{array}\right.
\]
Then we have
\[
d_1\zeta_{\tau_n}(t_1)+d_2\zeta_{\tau_n}(t_2)=\sum_{j=1}^{k_n}\eta_{n,j}.
\]
Observe that
\begin{eqnarray*}
\sum_{j=1}^{k_n}\eta_{n,j}^2 &=& d_2^2e^{-2\gamma
t_2}\kappa_{\tau_n}^2\sum_{j=0}^{k_{n,2}}e^{2\gamma\tau_n j}h^2\circ
T^{j} \\
   & &+ (2d_2d_1e^{-\gamma (t_2+ t_1)}+d_1^2e^{-2\gamma
t_1})\kappa_{\tau_n}^2\sum_{j=0}^{k_{n,1}}e^{2\gamma\tau_n
j}h^2\circ T^{j}.
\end{eqnarray*}
As in the proof of Theorem \ref{t:ndifferent}, we check that Theorem
\ref{CLTRN} applies to $\{\eta_{n,j}:1\le j\le k_n, n\ge 1\}$ and
conclude that
\[
d_1\zeta_{\tau_n}(t_1)+d_2\zeta_{\tau_n}(t_2)\to^d \sigma N(0,1)
\]
where $\sigma^2=\frac{||h||^2_2}{2\gamma}(d_2^2(1-e^{-2\gamma
t_2})+2d_2d_1e^{-\gamma (t_2+ t_1)}(e^{2\gamma
t_1}-1)+d_1^2(1-e^{-2\gamma t_1}))$. Since $\sigma N(0,1)$ is the
distribution of $d_1\zeta(t_1)+d_2\zeta(t_2)$ and
$E(\zeta(t_1)\zeta(t_2))=\frac{||h||^2_2}{2\gamma}e^{-\gamma (t_2+
t_1)}(e^{2\gamma t_1}-1)$, the proof of the lemma is complete.

\begin{lem}\label{t:Gaussianpos}
Let $(Y,\B,\nu)$ be a normalized measure space, $T:Y\to Y$ be
ergodic with respect to $\nu$, and $\gamma\neq 0$ be a constant.
Let $\kappa_\tau$, $\tau>0$, be such that
\[
\lim_{\tau\to 0}\frac{\kappa_\tau^2}{\tau}=1.
\]
If $h\in L^2(Y,\B,\nu)$ is such that $\FP_{T,\nu}h=0$, then the
finite dimensional distributions of the process $\eta_\tau$ defined
by
\[
\eta_\tau(t)=\frac{\kappa_\tau}{\gamma}\sum_{j=0}^{[\frac{t}{\tau}]}(1-e^{\gamma
(\tau j-t)})h\circ T^j,\;\;t\ge 0,\;\;\tau>0
\]
converge weakly as $\tau\to 0$ to the corresponding finite
dimensional distributions of the zero-mean Gaussian process $\eta$
for which $\eta(0)=0$ and
\[E\eta(t)\eta(s)=\frac{||h||^2_2}{\gamma^3}(2\gamma\min(t,s)
-2+2e^{-\gamma t}+2e^{-\gamma s}-e^{-\gamma
|t-s|}-e^{-\gamma(t+s)})\] for $t,s>0$.
\end{lem}
The lemma follows from Theorem \ref{CLTRN} in a similar fashion as
the preceding lemma.

\bibliographystyle{apalike}
\bibliography{zpf}

\begin{thebibliography}{}

\bibitem[Adler and Rivlin, 1964]{adler64}
Adler, R. and Rivlin, T. (1964).
\newblock Ergodic and mixing properties of {C}hebyshev polynomials.
\newblock {\em Proc. Amer. Math. Soc.}, 15:794--796.

\bibitem[Alexander and Yorke, 1984]{alexyorke}
Alexander, J.~C. and Yorke, J.~A. (1984).
\newblock Fat baker's transformations.
\newblock {\em Erg. Theory Dyn. Syst.}, 4:1--23.

\bibitem[an~der Heiden and Mackey, 1982]{adhmcm82}
an~der Heiden, U. and Mackey, M.~C. (1982).
\newblock The dynamics of production and destruction: Analytic insight into
  complex behaviour.
\newblock {\em J. Math. Biol.}, 16:75--101.

\bibitem[Beck, 1990a]{beck90a}
Beck, C. (1990a).
\newblock Brownian motion from deterministic dynamics.
\newblock {\em Physica A}, 169:324--336.

\bibitem[Beck, 1990b]{beck90}
Beck, C. (1990b).
\newblock Ergodic properties of a kicked damped particle.
\newblock {\em Commun. Math. Phys.}, 130:51--60.

\bibitem[Beck, 1996]{beck96}
Beck, C. (1996).
\newblock Dynamical systems of {L}angevin type.
\newblock {\em Physica A}, 233:419--440.

\bibitem[Beck and Roepstorff, 1987]{beck87}
Beck, C. and Roepstorff, G. (1987).
\newblock From dynamical systems to the {L}angevin equation.
\newblock {\em Physica A}, 145:1--14.

\bibitem[Benedics and Carleson, 1985]{benedics}
Benedics, M. and Carleson, L. (1985).
\newblock On iterations of $1-a x^2$ on $(-1,1)$.
\newblock {\em Ann. Math.}, 122:1--25.

\bibitem[Billingsley, 1968]{billingsley68}
Billingsley, P. (1968).
\newblock {\em Convergence of Probablility Measures}.
\newblock John Wiley \& Sons, New York.

\bibitem[Billingsley, 1995]{billingsley95}
Billingsley, P. (1995).
\newblock {\em Probablility and Measure}.
\newblock John Wiley \& Sons, New York.

\bibitem[Boltzmann, 1995]{boltzmann96}
Boltzmann, L. (1995).
\newblock {\em Lectures on {G}as {T}heory}.
\newblock Dover, Mineola, N.Y.

\bibitem[Bowen, 1975]{bowen75}
Bowen, R. (1975).
\newblock Equilibrium states and the ergodic theory of {A}nosov
  diffeomorphisms.
\newblock volume 470 of {\em Springer Lecture Notes in Math.}, New York.

\bibitem[Boyarsky and Scarowsky, 1979]{boyarsky79}
Boyarsky, A. and Scarowsky, M. (1979).
\newblock On a class of transformations which have unique absolutely continuous
  invariant measures.
\newblock {\em Trans. Amer. Math. Soc.}, 255:243--262.

\bibitem[Briggs et~al., 2001]{briggs01}
Briggs, M., Sengers, J., Francis, M., Gaspard, P., Gammon, R., Dorfman, J., and
  Calabrese, R. (2001).
\newblock Tracking a colloidal particle for the measurement of dynamic
  entropies.
\newblock {\em Physica A}, 296:42--59.

\bibitem[Brown, 1971]{brown71}
Brown, B.~M. (1971).
\newblock Martingale central limit theorems.
\newblock {\em Ann. Math. Statist.}, 42:59--66.

\bibitem[Chew and Ting, 2002]{chew}
Chew, L. and Ting, C. (2002).
\newblock Microscopic chaos and {G}aussian diffusion processes.
\newblock {\em Physica A}, 307:275--296.

\bibitem[Chung, 2001]{chung}
Chung, K.~L. (2001).
\newblock {\em A course in probability theory}.
\newblock Academic Press, 3 edition.

\bibitem[Denker, 1989]{denker89}
Denker, M. (1989).
\newblock The central limit theorem for dynamical systems.
\newblock In {\em Dynamical systems and ergodic theory (Warsaw, 1986)},
  volume~23 of {\em Banach Center Publ.}, pages 33--62. PWN, Warsaw.

\bibitem[Dorfman, 1999]{dorfman99}
Dorfman, J. (1999).
\newblock {\em An Introduction to Chaos in Nonequilibrium Statistical
  Mechanics}, volume~14 of {\em Cambridge Lecture Notes in Physics}.
\newblock Cambridge University Press, Cambridge, New York.

\bibitem[Dudley, 1989]{dudley}
Dudley, R.~M. (1989).
\newblock {\em Real Analysis and Probability}.
\newblock Wadsworth, Belmont, California.

\bibitem[Eagleson, 1975]{eagleson75}
Eagleson, G.~K. (1975).
\newblock Martingale convergence to mixtures of infinitely divisible laws.
\newblock {\em Ann. Probab.}, 3:557--562.

\bibitem[Eckmann and Ruelle, 1985]{eckmannruelle85}
Eckmann, J.-P. and Ruelle, D. (1985).
\newblock Ergodic theory of chaos and strange attractors.
\newblock {\em Rev. Modern Phys.}, 57(3, part 1):617--656.

\bibitem[Einstein, 1905]{einstein05}
Einstein, A. (1905).
\newblock {\"U}ber die von der molekularkinetischen {T}heorie der {W}{\"a}rme
  geforderte {B}ewegung von in ruhenden {F}l{\"u}ssigkeiten suspendierten
  {T}eilchen.
\newblock {\em Ann. d. Physik}, 17:549--560.

\bibitem[Erd{\"o}s, 1939a]{erdos39}
Erd{\"o}s, P. (1939a).
\newblock On a family of symmetric {B}ernoulli convolutions.
\newblock {\em Amer. J. Math.}, 61:974--976.

\bibitem[Erd{\"o}s, 1939b]{erdos40}
Erd{\"o}s, P. (1939b).
\newblock On the smoothness properties of a family of {B}ernoulli convolutions.
\newblock {\em Amer. J. Math.}, 62:180--186.

\bibitem[F{\"u}rth, 1956]{furth}
F{\"u}rth, R., editor (1956).
\newblock {\em Investigations on the Theory of the Brownian Movement}, New
  York. Dover.

\bibitem[Gaenssler and Joos, 1992]{gaenslerjoos}
Gaenssler, P. and Joos, K. (1992).
\newblock Another view on martingale central limit theorems.
\newblock {\em Stochastic Processes Appl.}, 40(2):181--197.

\bibitem[Gallavotti, 1999]{gallavotti}
Gallavotti, G. (1999).
\newblock {\em Statistical Mechanics: A Short Treatise}.
\newblock Springer Verlag, Berlin, New York.

\bibitem[Gaspard et~al., 1998]{gaspard98}
Gaspard, P., Briggs, M., Francis, M., Sengers, J., Gammon, R., Dorfman, J., and
  Calabrese, R. (1998).
\newblock Experimental evidence for microscopic chaos.
\newblock {\em Nature}, 394:865--868.

\bibitem[Gordin, 1969]{gordin69}
Gordin, M.~I. (1969).
\newblock The central limit theorem for stationary processes.
\newblock {\em Dokl. Akad. Nauk SSSR}, 188:739--741.

\bibitem[Hall, 1978]{hall78}
Hall, P. (1978).
\newblock The convergence of moments in the martingale central limit theorem.
\newblock {\em Z. Wahrsch. Verw. Gebiete}, 44(3):253--260.

\bibitem[Hilgers and Beck, 1999]{beck99}
Hilgers, A. and Beck, C. (1999).
\newblock Approach to {G}aussian stochastic behavior for systems driven by
  deterministic chaotic forces.
\newblock {\em Phys. Rev. E.}, 60:5385--5393.

\bibitem[Hilgers and Beck, 2001]{beck2001}
Hilgers, A. and Beck, C. (2001).
\newblock Higher-order correlations of {T}chebyscheff maps.
\newblock {\em Physica D}, 156:1--18.

\bibitem[Hunt et~al., 2002]{hunt02}
Hunt, B.~R., Kennedy, J.~A., Li, T.-Y., and Nusse, H.~E. (2002).
\newblock S{LYRB} measures: natural invariant measures for chaotic systems.
\newblock {\em Phys. D}, 170(1):50--71.

\bibitem[Iosifescu, 1992]{iosif92}
Iosifescu, M. (1992).
\newblock A very simple proof of a generalization of the
  {G}auss-{K}uzmin-{L}evy theorem on continued fractions, and related
  questions.
\newblock {\em Rev. Roumaine Math. Pures Apll.}, 37:901--914.

\bibitem[Jab{\l}o{\'n}ski, 1991]{jab91}
Jab{\l}o{\'n}ski, M. (1991).
\newblock A central limit theorem for processes generated by a family of
  transformations.
\newblock {\em Dissertationes Mathematicae}, 307:1--62.

\bibitem[Jab{\l}o{\'n}ski et~al., 1985]{jabkow85}
Jab{\l}o{\'n}ski, M., Kowalski, Z., and Malczak, J. (1985).
\newblock The rate of convergence of iterates of the {F}robenius-{P}erron
  operator for {L}asota-{Y}orke transformations.
\newblock {\em Univ. Jagel. Acta Math.}, 25:189--193.

\bibitem[Jab{\l}o{\'n}ski and Malczak, 1983a]{jab83}
Jab{\l}o{\'n}ski, M. and Malczak, J. (1983a).
\newblock A central limit theorem for piecewise convex mappings of the unit
  interval.
\newblock {\em T\^{o}hoku Math. J.}, 35:173--180.

\bibitem[Jab{\l}o{\'n}ski and Malczak, 1983b]{jabmal83}
Jab{\l}o{\'n}ski, M. and Malczak, J. (1983b).
\newblock The rate of convergence of iterates of the {F}robenius-{P}erron
  operator for piecewise convex transformations of the unit interval.
\newblock {\em Bull. Pol. Ac.: Math.}, 31:249--254.

\bibitem[Jakobson, 1981]{jakobson}
Jakobson, M. (1981).
\newblock Absolutely continuous invariant measure for one-parameter families of
  one-dimensional maps.
\newblock {\em Commun. Math. Phys.}, 81:39--88.

\bibitem[Jessen and Wintner, 1935]{jessen}
Jessen, B. and Wintner, A. (1935).
\newblock Distribution functions and the {R}iemann zeta function.
\newblock {\em Trans. Amer. Math. Soc.}, 38:48--88.

\bibitem[Kappler, 1931]{kappler}
Kappler, E. (1931).
\newblock Versuche zur {M}essung der {A}vogadro-{L}oschmidtschen {Z}eit aus der
  {B}rownschen bewegung einer {D}rehwaage.
\newblock {\em Ann. Physik}, 11:233--256.

\bibitem[Keller, 1980]{keller80}
Keller, G. (1980).
\newblock Un th\'eor\`eme de la limite centrale pour une classe de
  transformations monotones par morceaux.
\newblock {\em C.R. Acad. Sc. Paris}, 291:155--158.

\bibitem[Kershner and Wintner, 1935]{kershner}
Kershner, R. and Wintner, A. (1935).
\newblock On symmetric {B}ernoulli convolutions.
\newblock {\em Amer. J. Math.}, 57:541--548.

\bibitem[Lasota and Mackey, 1994]{almcmbk94}
Lasota, A. and Mackey, M.~C. (1994).
\newblock {\em Chaos, Fractals and Noise: Stochastic Aspects of Dynamics}.
\newblock Springer-Verlag, Berlin, New York, Heidelberg.

\bibitem[Liverani, 1996]{liverani}
Liverani, C. (1996).
\newblock Central limit theorem for deterministic systems.
\newblock In {\em International Conference on Dynamical Systems (Montevideo,
  1995)}, volume 362 of {\em Pitman Res. Notes Math. Ser.}, pages 56--75.
  Longman, Harlow.

\bibitem[Mackey, 1989]{mcm89rmp}
Mackey, M.~C. (1989).
\newblock The dynamic origin of increasing entropy.
\newblock {\em Rev. Mod. Phys.}, 61:981--1016.

\bibitem[Mackey, 1992]{mcmtdbk}
Mackey, M.~C. (1992).
\newblock {\em Time's Arrow: The Origins of Thermodynamic Behaviour}.
\newblock Springer-Verlag, Berlin, New York, Heidelberg.

\bibitem[Mackey and Glass, 1977]{mcmlg77sci}
Mackey, M.~C. and Glass, L. (1977).
\newblock Oscillation and chaos in physiological control systems.
\newblock {\em Science}, 197:287--289.

\bibitem[Mazo, 2002]{mazo}
Mazo, R. (2002).
\newblock {\em Brownian {M}otion: {F}luctuations, {D}ynamics, and
  {A}pplications}.
\newblock Claredon Press, Oxford.

\bibitem[Peres et~al., 2000]{peres}
Peres, Y., Schlag, W., and Solomyak, B. (2000).
\newblock Sixty years of {B}ernoulli convolutions.
\newblock In {\em Fractal geometry and stochastics, II (Greifswald/Koserow,
  1998)}, Progr. Probab. 46, pages 39--65, Basel. Birkh{\"a}user.

\bibitem[Pollicott and Sharp, 2002]{pollicottsharp}
Pollicott, M. and Sharp, R. (2002).
\newblock Invariance principles for interval maps with an indifferent fixed
  point.
\newblock {\em Comm. Math. Phys.}, 229:337--346.

\bibitem[Ratner, 1973]{ratner73}
Ratner, M. (1973).
\newblock The central limit theorem for geodesic flows on n-dimensional
  manifolds of negative cuorvature.
\newblock {\em Israel J. Math.}, 16:181--197.

\bibitem[Ruelle, 1976]{ruelle76}
Ruelle, D. (1976).
\newblock A measure associated with {A}xiom {A} attractors.
\newblock {\em Am. J. Math.}, 98:619--654.

\bibitem[Ruelle, 1978]{ruelle78}
Ruelle, D. (1978).
\newblock Sensitive dependence on initial condition and turbulent behavior of
  dynamical systems.
\newblock {\em Ann. N.Y. Acad. Sci.}, 316:408--416.

\bibitem[Ruelle, 1979]{ruelle79}
Ruelle, D. (1979).
\newblock Microscopic fluctuations and turbulence.
\newblock {\em Phys. Let.}, 72A:81--82.

\bibitem[Ruelle, 1980]{ruelle80}
Ruelle, D. (1980).
\newblock Measures describing a turbulent flow.
\newblock {\em Ann. N.Y. Acad. Sci.}, 357:1--9.

\bibitem[Schulman, 1997]{schulman97}
Schulman, L.~S. (1997).
\newblock {\em Time's Arrows and Quantum Measurement}.
\newblock Cambridge University Press, Cambridge.

\bibitem[Sinai, 1972]{sinai72}
Sinai, Y.~G. (1972).
\newblock Gibbs measure in ergodic theory.
\newblock {\em Russian Math. Surveys}, 27:21--69.

\bibitem[Solomyak, 1995]{solomyak}
Solomyak, B. (1995).
\newblock On the random series $\sum \pm \lambda^n$ (an {E}rd{\"o}s problem).
\newblock {\em Annals of Math.}, 142:611--625.

\bibitem[Teicher, 1988]{teicher}
Teicher, H. (1988).
\newblock Distribution and moment convergence of martingales.
\newblock {\em Probab. Theory Related Fields}, 79(2):303--316.

\bibitem[Thaler, 1980]{thaler}
Thaler, M. (1980).
\newblock Estimates of the invariant densities of endomorphisms with
  indifferent fixed points.
\newblock {\em Israel J. Math.}, 37:303--314.

\bibitem[Tsujii, 1996]{tsujii96}
Tsujii, M. (1996).
\newblock On continuity of {B}owen-{R}uelle-{S}inai measures in families of
  one-dimensional maps.
\newblock {\em Comm. Math. Phys.}, 177(1):1--11.

\bibitem[Tyran-Kami\'nska, 2004]{tyran}
Tyran-Kami\'nska, M. (2004).
\newblock An invariance principle for maps with polynomial decay of
  correlations.
\newblock {\em Preprint}.

\bibitem[Viana, 1997]{viana}
Viana, M. (1997).
\newblock Stochastic dynamics of deterministic systems.
\newblock {\em Col. Bras. de Matem{\'a}tica}, 21:197.

\bibitem[Wintner, 1935]{wintner}
Wintner, A. (1935).
\newblock On convergent {P}oisson convolutions.
\newblock {\em Amer. J. Math.}, 57:827--838.

\bibitem[Wong, 1979]{wong79}
Wong, S. (1979).
\newblock A central limit theorem for piecewise monotonic mappings of the unit
  interval.
\newblock {\em Ann. Prob.}, 7:500--514.

\bibitem[Young, 1992]{young}
Young, L.-S. (1992).
\newblock Decay of correlations for certain quadratic maps.
\newblock {\em Commun. Math. Phys.}, 146:123--138.

\bibitem[Young, 1999]{young99}
Young, L.-S. (1999).
\newblock Recurrence times and rates of mixing.
\newblock {\em Israel J. Math.}, 110:153--188.

\bibitem[Young, 2002]{young02}
Young, L.-S. (2002).
\newblock What are {SRB} measures, and which dynamical systems have them?
\newblock {\em J. Statist. Phys.}, 108:733--754.

\end{thebibliography}
\end{document}